\documentclass[11pt,a4paper]{article}
\pdfoutput = 1

\usepackage{jheppub}
\usepackage[latin1]{inputenc} 
\usepackage{bm}
\usepackage{color}
\usepackage{graphicx}
\usepackage{cancel}
\usepackage{xspace}
\usepackage{latexsym} 
\usepackage{amsfonts}
\usepackage{amssymb} 
\usepackage{mathrsfs}
\usepackage{parskip}
\usepackage{enumerate}

\DeclareMathAlphabet{\mathpzc}{OT1}{pzc}{m}{it}

\newcommand{\stau}{\tilde{\tau}_1^-}
\newcommand{\snu}{\tilde{\nu}}
\newcommand{\neu}{\tilde{\chi}^0}
\newcommand{\cha}{\tilde{\chi}^+}
\newcommand{\slr}{\tilde{l}_R}

\newcommand{\der}{{\rm d}}

\title{Multi-lepton signatures at LHC from sneutrino dark matter}

\author[1,2]{Chiara Arina} 
\author[1]{and Maria Eugenia Cabrera}
\affiliation[1]{GRAPPA Institute, University of Amsterdam, Science Park 904, 1090 GL Amsterdam (Netherlands)}
\affiliation[2]{Institut d'Astrophysique de Paris, 98bis boulevard Arago, 75014 Paris (France)}

\abstract{We investigate multi-lepton LHC signals arising from an extension at
  the grand unification scale of the minimal supersymmetric
      standard model (MSSM) involving right-handed neutrino superfields. In
  this framework neutrinos have Dirac masses and the mixed sneutrinos are the
  lightest supersymmetric particles and hence the dark matter candidates. We
  analyze the model parameter space in which the sneutrino is a good dark
  matter particle and has a direct detection cross-section compatible with the
  LUX bound. Studying the supersymmetric mass spectrum of this region, we find
  several signatures relevant for LHC, which are distinct from the predictions
  of the MSSM with neutralino dark matter. For instance two opposite sign and
  different flavor leptons, three uncorrelated leptons and long-lived staus
  are the most representative. Simulating both the signal and expected
  background, we find that the multi-lepton signatures and the long-lived stau
  are in the reach of the future run of LHC with a luminosity of 100/fb. We
  point out that if one of these signatures is detected, it might be an
  indication of sneutrino dark matter.}


\begin{document}

\maketitle   
\flushbottom


\section{Introduction}\label{sec:intro}

The Standard Model of particle physics has encountered a tremendous success, however it leaves many questions unanswered, such as the hierarchy problem, the origin of neutrino masses and the origin of a non-baryonic dark matter candidate. Several of these hints point toward the existence of new physics around the TeV scale and a very well motivated theoretical scenario is supersymmetry (SUSY)~\cite{Nilles:1983ge,Haber:1984rc,Martin:1997ns}.

The constrained minimal supersymmetric standard model (MSSM) however has come to a critical point, firstly because of the null results of ATLAS and CMS on searches for supersymmetric partners of the standard model particles. Besides setting strong limits on the mass of the colored supersymmetric sector~\cite{Chatrchyan:2013xna,Aad:2013wta} the non observation of squark and gluinos has moved the experimentalist attention to the electroweak gaugino sector, where the LHC collaborations have now set constraints stronger than LEP searches~\cite{CMS:2013dea,TheATLAScollaboration:2013hha}. Secondly, the recent discovery of the Higgs boson, with a mass around 126 GeV~\cite{CMS:yva,ATLAS:2013sla} has crucial importance for the MSSM,  as such value requires large radiative corrections, which scales as the logarithm of the supersymmetric masses, in particular with the stop masses. Consequently, the latter must be rather high, well above 1 TeV unless the stop sector has maximal mixing, thus suggesting that the mass scale of SUSY particles could be substantially higher than expected from fine-tuning arguments. This would also make very challenging, if not impossible, to detect SUSY at LHC in a direct or indirect way~\cite{Farina:2011bh,Balazs:2012qc,Akula:2012kk,Buchmueller:2012hv,Arbey:2012dq,Strege:2012bt,Cabrera:2012vu}. Prospects for detection remain interesting if some supersymmetric states are still sufficiently light, which in general implies to go beyond the constrained MSSM. There are several possibilities, such as non universal Higgs masses, non universal gaugino masses, which all ends up to some sort of split SUSY scenarios where a part of the spectrum is heavy and the rest is still in the reach of LHC. In general the electroweak SUSY fermions are at TeV scale, while the scalars are at much higher scale. 

The particle physics model we analyze here is a far less investigated extension of the MSSM focused on the sneutrino, the scalar super-partner of the left-handed neutrino, which plays the role of dark matter candidate instead of the usual neutralino. Since neutrinos have masses, as is now clearly understood by a host of independent and very robust experimental results and theoretical analyses~\cite{Fogli:2012ua,Schwetz:2008er}, the motivation for considering this model is to study a natural and direct extension of the MSSM which contains terms in the supersymmetric lagrangian which can drive neutrino masses.  The model we consider incorporates at the same time the new physics required to explain two basic problems of astroparticle physics: the origin of neutrino masses and the nature of dark matter. We do not attempt to be totally exhaustive on the type of supersymmetric model and we focus on the dark matter phenomenology of the sneutrino within the particular framework described below. Connections between neutrino physics and the phenomenology of sneutrino dark matter arise in general in supersymmetric see-saw models and have been discussed in {\it e.g.}~\cite{Hirsch:1997is,Grossman:1997is,Hall:1997ah,Borzumati:2000mc,Dedes:2007ef,Arina:2007tm,Arina:2008bb,MarchRussell:2009aq,An:2011uq,DeRomeri:2012qd,BhupalDev:2012ru,Guo:2013sna,Banerjee:2013fga}. The sneutrino as dark matter candidate is excluded in the MSSM, because it has a non-zero hypercharge. Indeed its couplings to the $Z$ boson makes it annihilate too efficiently in the early Universe, hence its final relic abundance is lower than the value $\Omega_{\rm DM} h^2$ measured by the Planck satellite~\cite{Ade:2013zuv}. Besides the problem of being under abundant, the direct detection is the most stringent limit for this candidate: the scattering cross-section off nucleus is mediated by $Z$ boson exchange on $t$-channel, giving order to spin-independent (SI) cross-section of about $10^{-39} {\rm cm^2}$, value excluded already a decade ago for dark matter particles heavier than 10 GeV. The picture changes if we include in the MSSM a right-handed neutrino superfield (MSSM+RN from here on), which gives rise to Dirac neutrino masses. Being the theory supersymmetric then, the superfield contains as well a scalar right-handed field, the scalar neutrino right $\tilde{N}$. This field, if at TeV scale, can mix with the left-handed partner $\tilde{\nu}_L$ and make the sneutrino, mostly right-handed, a viable dark matter candidate~\cite{Arkani-Hamed:2000bq,Arina:2007tm,Thomas:2007bu,Belanger:2010cd}. Pure right-handed sterile sneutrinos are viable dark matter candidates as well, as discussed {\it e.g.} in~\cite{Asaka:2005cn,Cerdeno:2009dv,Khalil:2011tb,Choi:2012ap}.

The phenomenology of the MSSM+RN model has been investigated in the framework of LHC constraints and direct detection in~\cite{Biswas:2009rba,Dumont:2012ee} and for indirect detection and cosmology in~\cite{Hooper:2004dc,Arina:2007tm,Choi:2012ap}. In this paper we review the status of the sneutrino as dark matter after the Higgs boson mass measurements, by exploring the SUSY parameter space with the soft breaking terms fixed at the grand unification (GUT) scale, however allowing for non universal slepton and gaugino masses. We also assess the impact of the most recent exclusion bound from LUX~\cite{Akerib:2013tjd} for dark matter direct searches. In this framework the colored particles will be mostly heavier than the scalar leptons and gauginos, so that we can satisfy the requirement of having a Higgs at 125 GeV.  

A scenario where the dark matter candidate is different than the standard neutralino and is linked to the neutrino physics is plausible and has interesting motivations. Therefore it would be worthy to improve the study on this model and the analysis of the distinctive signatures expected at colliders,  which is the main motivation of our paper.  Signatures at LHC from sneutrinos, arising form the strong production of squarks and gluinos, have been investigated in~\cite{Thomas:2007bu,Belanger:2011ny,Lee:2011ti}. By exploiting the tight connection with the lepton sector, we instead focus on multi-lepton signatures that can arise from the sneutrino dark matter. 
We consider three peculiar signatures, which can be disentangled from MSSM standard scenario, based mainly on these signals: two opposite sign leptons with different flavor and three uncorrelated leptons. An efficient way of probing these signatures is via direct chargino production~\cite{Cabrera:2012gf}, as we will discuss in details.  We run Monte Carlo simulated events followed by detector simulations for representative  benchmarks that can arise in the MSSM+RN parameter space, in which the sneutrino is a good dark matter candidate. We discuss how these signatures can be  detected and eventually distinguished with respect to the standard MSSM picture, whenever possible. Besides the multi-lepton final states, we consider signals coming from long-lived charged particles, in particular the lightest scalar tau mass eigenstate ($\stau$). In a corner of the MSSM+RN parameter space such particles can have life-time long enough to decay outside the detector volume or in the hadronic calorimeter, giving for instance disappearing tracks as signature. Interestingly we will show that all these signatures are connected to the dark matter relic density constraint: the annihilation processes to get $\Omega_{\rm DM} h^2$ will fix the SUSY mass spectrum and hence the signals at collider.

The rest of the paper is organized as follows. The supersymmetric model under investigation is described in section~\ref{sec:model}, while in section~\ref{sec:num} we define our numerical analysis. In section~\ref{sec:DM} the parameter space of the MSSM+RN that leads to good dark matter candidates is detailed. Section~\ref{sec:cs} provides an in-depth discussion of the relevant signatures at collider from sneutrino dark matter and differences/similarities with respect to the standard MSSM framework, together with section~\ref{sec:disc}. Finally we summarize our findings in section~\ref{sec:concl}. In Appendix~\ref{sec:appA} we discuss the prior choice and show the marginalized one dimensional posterior probability density functions for the parameters relevant for the dark matter phenomenology.

\section{Supersymmetric framework}\label{sec:model}

The MSSM+RN model we use has been defined in~\cite{Arkani-Hamed:2000bq,Borzumati:2000mc,Arina:2007tm} and is similar to~\cite{Belanger:2010cd}.  The superpotential for Dirac right-handed neutrino superfield , with the lepton violating term absent, is given by
\begin{equation}\label{lrsuppot}
W = \epsilon_{ij} (\mu \hat H^{u}_{i} \hat H^{d}_{j} - Y_{l}^{IJ} \hat H^{d}_{i} \hat L^{I}_{j} \hat R^{J}
+ Y_{\nu}^{IJ} \hat H^{u}_{i} \hat L^{I}_{j} \hat N^{J} )\,,
\end{equation}
where $Y_{\nu}^{IJ}$ is a matrix in flavor space (which we choose to be real and diagonal), from which the mass of neutrinos, of Dirac nature, are obtained, $m_D^{I} = v_{u}Y_{\nu}^{II}$. In the soft-breaking potential there are additional contributions due to the new scalar fields
\begin{equation}\label{lrsoftpot}
\hspace{-0.5cm}
V_{\rm soft} = (M_{L}^{2})^{IJ} \, \tilde L_{i}^{I \ast} \tilde L_{i}^{J} + 
(M_{N}^{2})^{IJ} \, \tilde N^{I \ast} \tilde N^{J} - 
 [\epsilon_{ij}(\Lambda_{l}^{IJ} H^{d}_{i} \tilde L^{I}_{j} \tilde R^{J} + 
\Lambda_{\nu}^{IJ} H^{u}_{i} \tilde L^{I}_{j} \tilde N^{J})  + \mbox{h.c.}]\,,
\end{equation}
where both matrices $M_{N}^{2}$ and $\Lambda_{\nu}^{IJ}$ are real and diagonal, with common entries $m_{N^2_k}$ and $A_{\snu}^k$ respectively ($k$ is the flavor index). Defining the sneutrino interaction basis by the vector $\Phi^\dag=(\tilde{\nu}_L^\ast,\, \tilde N^\ast)$, the sneutrino mass potential is
\begin{eqnarray}
V_{\rm mass}^k =\frac{1}{2}\, \Phi^{\dag}_{LR}\, \mathcal{M}^2_{LR}\, \Phi_{LR}\,,
\end{eqnarray}
where the squared--mass matrix $\mathcal{M}^2_{LR}$ reads
\begin{equation}
\mathcal{M}^2_{LR}  =   \begin{pmatrix}
m^2_{L_k} + \frac{1}{2} m_{Z}^{2} \cos(2\beta) + m_D^2  & \, \, \, \frac{v}{\sqrt{2}} A_{\snu}^k \sin\beta - \mu m_D\rm cotg\beta  \cr\\
                 \frac{v}{\sqrt{2}} A_{\snu}^k \sin\beta - \mu m_D\rm cotg\beta   & m^2_{N_k} + m_D^2
                 \end{pmatrix}  \,.
  \label{eq:masslr}
\end{equation}
The Dirac neutrino mass is small, and can be safely neglected. Here $m^2_{L_k}$ is the soft mass term for the three SU(2) leptonic doublets, $\tan\beta = v_u/v_d$ and $v^2=v_u^2+v_d^2=(246\,  {\rm GeV})^2$, with $v_{u,d}$ being the Higgs vacuum expectation values.

The off diagonal term is relevant for the mixing among the mass eigenstates: {\it i.e.} for $A_{\snu}^k = \eta Y_\nu$, namely the trilinear term aligned to the neutrino Yukawa, this term is necessarily very small as compared to the diagonal entries and is therefore negligible. However, $A_{\snu}^k$ can be in general a free parameter and may naturally be of the order of the other entries of the matrix, and induce a sizable mixing of the lightest sneutrino in terms of left-handed and right-handed fields. We define the mixing as follows
\begin{eqnarray}\label{lr_eigenstates}
\left\{
\begin{array}{l}
 \snu_1^k = -\sin\theta_{\snu}\;\snu_L + \cos\theta_{\snu}\;\tilde{N}\,,\\
 \snu_2^k = +\cos\theta_{\snu}\;\snu_L + \sin\theta_{\snu}\;\tilde{N}\,,
\end{array}
\right.
\end{eqnarray}
where $\theta_{\snu}$ is the mixing angle. Sizeable mixings reduce the coupling to the $Z$ boson, which
couples only to left-handed fields, and therefore have relevant impact on all the sneutrino phenomenology,
as recognized in refs.~\cite{Arkani-Hamed:2000bq,Smith:2001hy,Grossman:1997is,Arina:2007tm,Belanger:2010cd}.

With the inclusion of $\hat{N}$, the renormalization group equations (RGEs) are modified as
\begin{eqnarray}
\frac{\der m^2_{N_k}}{\der \ln\mu} & = & \frac{4}{16 \pi^2} \left(A^{k}_{\snu}\right)^2\,,\\ \nonumber
\frac{\der m^2_{L_k}}{\der \ln\mu} & = & \left( \rm MSSM\,  terms \right) + \frac{2}{16 \pi^2} \left(A^{k}_{\snu}\right)^2\,,\\  \nonumber
\frac{\der A_{\snu}^k}{\der \ln\mu} & = & \frac{2}{16 \pi^2} \left(- \frac{3}{2} g^2_2 - \frac{3}{10} g_1^2 + \frac{3}{2} Y_t^2 + \frac{1}{2} Y_\tau^2 \right) A_{\snu}^k\,,\\ \nonumber
\frac{\der m^2_{H_u}}{\der \ln\mu} & = & \left( \rm MSSM\,  terms \right) +\sum_{k=1,3} \frac{2}{16 \pi^2} \left(A^{k}_{\snu}\right)^2\,, \nonumber
\end{eqnarray}
with $\mu$ being the renormalization scale, $g_2$ and $g_1$ the SU(2) and U(1) gauge couplings, $Y_{t,\tau}$ the top and $\tau$ Yukawa respectively. Notice that the right-handed soft mass receives corrections only from the trilinear term, which affects as well the running of the left-handed part. This was already recognized in~\cite{Belanger:2010cd,Belanger:2011ny}, but we report it here as well to set the basis of our analysis. 

Assuming negligible electron and muon Yukawas and keeping only the $\tau$ Yukawa $Y_\tau$ in the RGEs leads to $\snu_{1e}=\snu_{1\mu}$ and  $\snu_{1\tau}$ to be the lightest sneutrino mass eigenstate and hence the LSP. Similarly for the heavier states we have $\tilde{\nu}_{e2}=\tilde{\nu}_{\mu 2}$ and $\tilde{\nu}_{\tau 2} \equiv \snu_2$. From here on we drop the flavor index and consider the sneutrino dark matter to be constituted by $\snu_{1\tau}\equiv \snu_1$, unless stated otherwise. The relevant parameters at electroweak (EW) scale for the sneutrino sector are the two mass eigenstates $m_{\snu_1}$ and $m_{\snu_2}$ and the mixing angle $\theta_{\snu}$, related to the $A_{\snu}$ term via $ \sin 2\theta_{\snu} = \sqrt{2} A_{\snu} v \sin \beta/\left(m^2_{\snu_2}-m^2_{\snu_1}\right)$.

\section{Set up of the numerical analysis}\label{sec:num}

\subsection{Parameters and methodology}\label{sec:par}

We study the MSSM+RN in the framework in which the soft parameters are considered non-universal at a high scale $M_X$, where supersymmetry breaking is transmitted to the observable sector via gravity mediated mechanism. The model is defined by the following free parameters, whose initial values are understood to be fixed at the scale $M_X$
\begin{equation}
\{ \theta_i \}  = \{ M_1, M_2, M_3, m_L, m_R, m_N, m_Q, m_H, A_L,A_{\snu}, A_Q, B, \mu\} \,,
\label{eq:param}
\end{equation}
where the $M_i$ are the gaugino masses and $m_H$ denote the common entry for the two Higgs doublet masses, taken to be equal ($m_{H_u} = m_{H_d}$). The $A_L$  and $A_Q$ are the scalar trilinear couplings for the sleptons and squarks respectively. Finally $\mu$ and $B$ are the mass term for the Higgs fields in the superpotential, equation~\ref{lrsuppot}, and the coefficient of the bilinear term in the soft scalar potential, equation~\ref{lrsoftpot}.

Since there are many free parameters, a random scan would turn out quite inefficient in exploring the parameter space, as it scales as $n^2$, where $n$ is the number of random variables. To accomplish an efficient sampling we adopt an approach based on Bayes' theorem 
\begin{equation}
p(\theta_i | d) \propto \mathcal{L}(d|\theta_i) \pi(\theta_i)\,,
\end{equation}
where $d$ are the data under consideration, $\mathcal{L}(d|\theta_i)$  is the likelihood function, encoding how our model describes the data, $\pi(\theta_i)$ is the prior probability distribution function (pdf) associated to each parameter, and $p(\theta_i | d)$ is the posterior pdf. 

The prior pdf is independent on the data and describes our belief on the value
of the theoretical parameters, before the confrontation with the experimental
results. All parameters are soft SUSY breaking terms, except $\tan\beta$:
since they have a common origin it is reasonable to assume that they have
similar size, and the initial conditions are given at the GUT scale, $M_X\sim
10^{16}$ GeV. We assume gaugino masses non-universality, allowing the three
parameters to vary free within a similar range of values.  The scalar masses
are not unified, even though we still assume them to be common for all the
three flavors. The parameter $m_N$ in general does not depend on the other
mass parameters, in particular is not linked to $m_{L}$, which instead is
related to the charged slepton masses. On the same vein we consider $m_R$
independent as well. As the major focus of the model is the slepton sector, we
consider one common soft mass term for all scalar quarks, $m_Q$. Similarly we
let free to vary the trilinear scalar couplings, which are taken to be equal
for all flavors. For the charged sleptons we keep the usual alignment to the
Yukawas, while for the sneutrino sector, the $A_{\snu}$ term is let free, to
provide efficient mixing between the left and right component of the
sneutrino. We use flat prior in the ranges defined in
    table~\ref{tab:priors}. 

\begin{table}[t!]
\caption{Nested sampling parameters and priors for the MSSM+RN framework.\label{tab:priors}}
\begin{center}
\begin{tabular}{ll}
\hline
 NS parameter & Prior range\\
\hline
 $M_1,M_2$  & $(-4000 \to 4000)$ GeV \\
 $\log_{10}(M_3/\rm GeV)$ & $-4 \to 4$  \\ 
 $\log_{10}(m_Q/\rm GeV)$ & $2 \to 5$ \\
$ m_L,m_R$ & $(1 \to 2000)$ GeV \\
 $m_N$ & $1 \to 2000$ GeV\\
 $\log_{10}(A_Q/\rm GeV)$ &  $-5 \to 5$\\
 $A_{L}$ & $(-4000 \to 4000)$ GeV \\
 $A_{\snu}$ & $(-1000 \to 1000)$ GeV \\
 $\log_{10}(m_H/\rm GeV)$ & $1 \to 5$ \\
 $\tan\beta$ &  $3 \to  50$\\
\hline
\end{tabular}
\end{center}
\end{table}

Usually a common choice for MSSM parameters is \{$\tan\beta$, $\rm
sign(\mu)$\}, which replaces the bilinear term and the Higgs mass term in the
superpotential, \{$B$, $\mu$\}, see expression in equation~\ref{eq:param}. To
consistently pass from one set of parameters to the other we follow the
prescription in~\cite{Cabrera:2008tj,Cabrera:2009dm}. This approach consists
in taking $M_Z$ on the same foot as the rest of the
    experimental data and computes the Jacobian of the transformation in the
    parameter space, adding it consistently to the posterior pdf. The
Jacobian factor has a beneficial impact of incorporating a fine-tuning penalization, giving low statistical weight to points with very
    large masses\footnote{It allows to extent the prior range up to values
  close to $M_X$: as these region are suppressed statistically, the results of
  the sampling are essentially independent on the prior range of the soft
  parameters.}. 

\subsection{Constraints and Observables}\label{sec:co}

Here we describe what are the constraints and observables implemented in our likelihood, as summarized in table~\ref{tab:co}. 

The dark matter phenomenology is constrained by requiring that the relic abundance matches the value measured by the Planck satellite~\cite{Ade:2013zuv} and that the sneutrino has a cross-section off nucleus below the LUX exclusion bound~\cite{Akerib:2013tjd}. The sneutrino, being a scalar, has only SI interaction with the nucleus, mediated either by a $Z$ or Higgs boson. Due to the mixing between left and right sneutrino fields, the lightest sneutrino $\snu_1$ coupling with the Z boson is reduced by a factor $\sin\theta_{\snu}$. The averaged cross-section is given by
\begin{equation}
\xi \sigma^{SI}_{ n} = \xi \frac{4 \mu^2_{n}}{\pi} \frac{\left( Z f_p + (A-Z) f_n\right)^2}{A^2} \,,
\end{equation}
where $\xi \equiv \min(\Omega_{\rm DM} h^2, \Omega_{\snu} h^2)$, $\mu_{n}$ is the sneutrino-nucleon reduced mass, $A$ ($Z$) is the mass (atomic number) of the nucleus. The couplings $f_n$, $f_p$ to neutron and proton respectively are computed directly from the model parameters. We do not consider the uncertainties related to the pion nucleon sigma term $\sigma_{\pi n}$, kept fixed to most recent value from lattice simulations~\cite{Junnarkar:2013ac}, and we refer the interested reader to {\it e.g.}~\cite{Koch:1982pu,Gasser:1990ap,Bottino:1999ei,Pavan:2001wz,deAustri:2013saa}. We average on the number of proton and neutron nucleons to extract the cross-section on Xe nucleus, which is then compared to the LUX exclusion limit.

If $\snu_1$ is light enough to be produced in $Z$ decay, its contribution to the $Z$ width is given by:
\begin{equation}
\Delta\Gamma_{Z} = \sin^4\theta_{\snu}\,\frac{\Gamma_{\nu}}{2}
\left[ 1- \left(\frac{2 m_{\snu_1}}{m_{Z}}\right)^{2}\right]^{3/2} \,\;  \theta(m_Z-2\,m_{\snu_1})\,,
\label{eq:Zlr}
\end{equation}
where $\Gamma_\nu=166$ MeV is the $Z$ decay width into neutrinos. The quantity in equation~\ref{eq:Zlr} is well measured and can not be larger than 2 MeV~\cite{ALEPH:2005ab}. On the same vein we require that light sneutrinos below the Higgs resonance, hence produced by Higgs decay, should not contribute more than 65\% to its invisible decay width~\cite{ATLAS:2013pma,CMS:2013yda}.
\begin{table}[t!]
\caption{Summary of the observables and constraints used in the analysis. The left part stands for the observables that have an actual measure, whose corresponding likelihood follows a Gaussian distribution centered on the measured value provided in the table together with the standard deviation.  The right side of the table is for the exclusion bound case, for which the likelihood function encodes the 90\% or 95\% CL with a step function. \label{tab:co}}
\begin{center}
\begin{tabular}{|rl | rl|}
\hline
Observable & Measured & Observable & Limit  \\
\hline
 $\Omega_{\rm DM} h^2$  & $ 0.1186 \pm 0.0031 {\rm (exp)}$  &$ \xi \sigma_n^{SI}  $  &  LUX (90\% CL)  \\
 & $\pm 20\% \rm (theo)$ &  $m_{\tilde{e},\tilde{\mu}}$  &  $> 100$ GeV (LEP 95\% CL)  \\
$m_h$  & $ (125.85 \pm 0.4)$ GeV (exp)  &  $m_{\stau}$  &  $> 85$ GeV (LEP 95\% CL)  \\
  &  $ \pm 4 \rm \, GeV \rm (theo)$ &   $m_{\tilde{\chi}_1^+}$  &  $>  100$ GeV (LEP 95\% CL)    \\
  $\Delta\Gamma_Z^{\rm invisible}$  & $ (166 \pm 2 )$ MeV  &  ${\rm BR}(h \to {\rm invisible})$  & $ > 65\%$ (LHC 95\% CL) \\
 $B \to X_s \gamma$ & $(3.55 \pm 0.24 \pm 0.09) \times 10^{-4} $& & \\
 $B_s \to \mu^+ \mu^-$ & $ 3.2  \times 10^{-9}$& & \\
 &$ (+1.4 - 1.2) \times 10^{-9}$ (stat)& & \\
  & $ (+0.5 - 0.3 )  \times 10^{-9}$ (sys)& & \\

\hline
\end{tabular}
\end{center}
\end{table}

As far as it concerns the particle physics bounds, we require that the mass of the lightest CP-even Higgs satisfies the Higgs boson mass measured by both CMS~\cite{CMS:yva} and ATLAS~\cite{ATLAS:2013sla} collaborations. The value of the Higgs mass we use as observable is a statistical mean of both CMS and ATLAS measurements, as obtained in~\cite{Cabrera:2012vu}. The charged slepton masses, $m_{\tilde{e},\tilde{\mu}}$, should be  compatible with the mass lower bound from LEP, which occurs at $100$ GeV~\cite{sleptons2004}; a similar bound applies as well for the lightest chargino mass eigenstate. The $\stau$ has a somehow lower exclusion bound of $85$ GeV, which comes from LEP measurement on the $W$ boson decay width~\cite{sleptons2004}.  We are aware of the latest bounds on the gluino and squark masses from ATLAS~\cite{TheATLAScollaboration:2013fha} in simplified models, however we do not include them in our analysis due to complications in translating the bounds for a framework with a LSP of different nature than the neutralino. We also consider the constraints coming from the rare decays $B \to X_s\gamma$~\cite{Amhis:2012bh} and $B^0_s \to \mu^+ \mu^-$~\cite{Aaij:2012nna}.

The full likelihood function is the product of the individual likelihoods associated to an experimental result. For the quantities for which positive measurements have been made (as listed in the left part of table~\ref{tab:co}), we assume a Gaussian likelihood function with a variance given by combining the theoretical and experimental variances. For the observables for which only lower or upper limits are available we use a likelihood modelled as step function on the $x\%$ confidence level (CL) of the exclusion limit.

On a practical level, the model has been implemented in
\texttt{FeynRules}~\cite{Duhr:2011se,Degrande:2011ua} (FR), by adding the
appropriate term in the superpotential and in the soft SUSY breaking
potential, following equations~\ref{lrsoftpot} and~\ref{lrsuppot}. We generate
output files compatible with \texttt{CalcHep} in order to use the public code
\texttt{micrOMEGAS\_3.2}~\cite{Belanger:2013oya} for the computation of the
sneutrino relic density and elastic scattering cross-section. The FR package
produces as well outputs compatible with the public code
\texttt{MadGraph5}~\cite{Alwall:2011uj} (MG5), which we use for the collider
analysis at parton level. The input parameters are given in the SUSY Les Houches Accord 2 format~\cite{Allanach:2008qq}. The Monte Carlo simulation of the events make use of
\texttt{Pythia 8}~\cite{Sjostrand:2007gs} (as implemented within
\texttt{MadGraph5}) for hadronization, as well as of the detector simulator
\texttt{Delphes 3}~\cite{deFavereau:2013fsa}, with default ATLAS
specifications. The supersymmetric particle spectrum is computed with the code
\texttt{SoftSusy}, appropriately modified to adapt to
\texttt{micrOMEGAS\_3.2}. Finally the sampling of the parameter space is done
with the code \texttt{MultiNest\_v3.2}~\cite{Feroz:2007kg,Feroz:2008xx}, which
has the tolerance set to 0.5 and the number of live points to
4000\footnote{Technically we run two chains, one of them
      requiring charginos lighter than 900 GeV to have a better
     sampling for the region accessible to LHC.}. The $B$-physics observables are computed by interfacing the program with \texttt{SuperIso}~\cite{Mahmoudi:2008tp}.

\section{Sneutrinos as good dark matter candidates}\label{sec:DM}

Instead of pursuing a full Bayesian analysis based on the posterior pdf, we use the equally weighted posterior sample. This sample contains points drawn randomly from the posterior pdf. More details  about the sampling are given in Appendix~\ref{sec:appA}, where we also comment on the impact of changing priors. Indeed the main interest of our analysis is firstly to find a correlation between the parameter space that leads to the good relic density and SI cross-section with the LHC signatures and secondly to obtain an efficient sampling of the parameter space. 
\begin{figure}[t]
\begin{minipage}[t]{0.49\textwidth}
\centering
\includegraphics[width=1.\columnwidth,trim=1mm 0mm 13mm 10mm, clip]{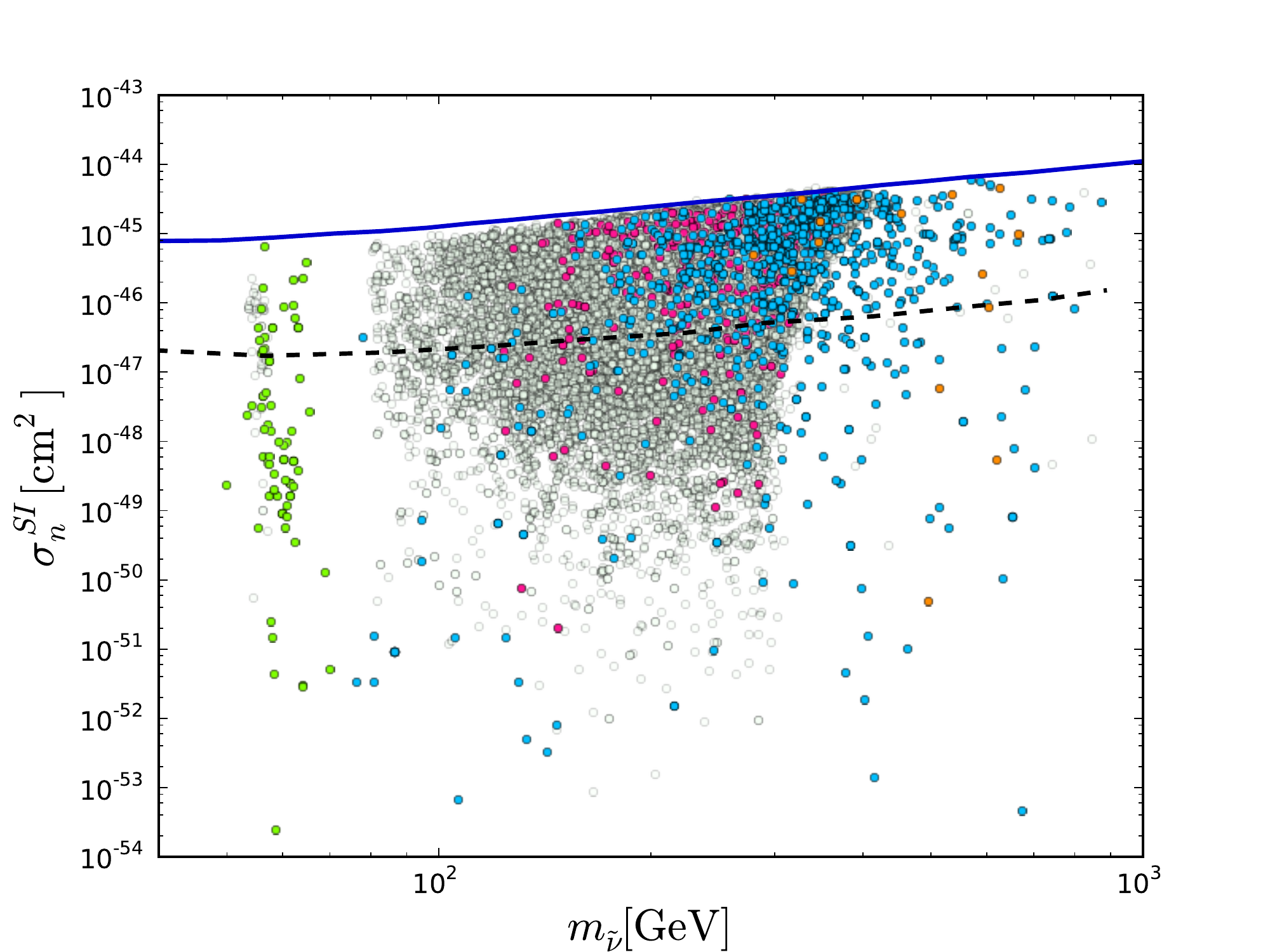}
\end{minipage}
\hspace*{0.2cm}
\begin{minipage}[t]{0.49\textwidth}
\includegraphics[width=1.\columnwidth,trim=1mm 0mm 13mm 10mm, clip]{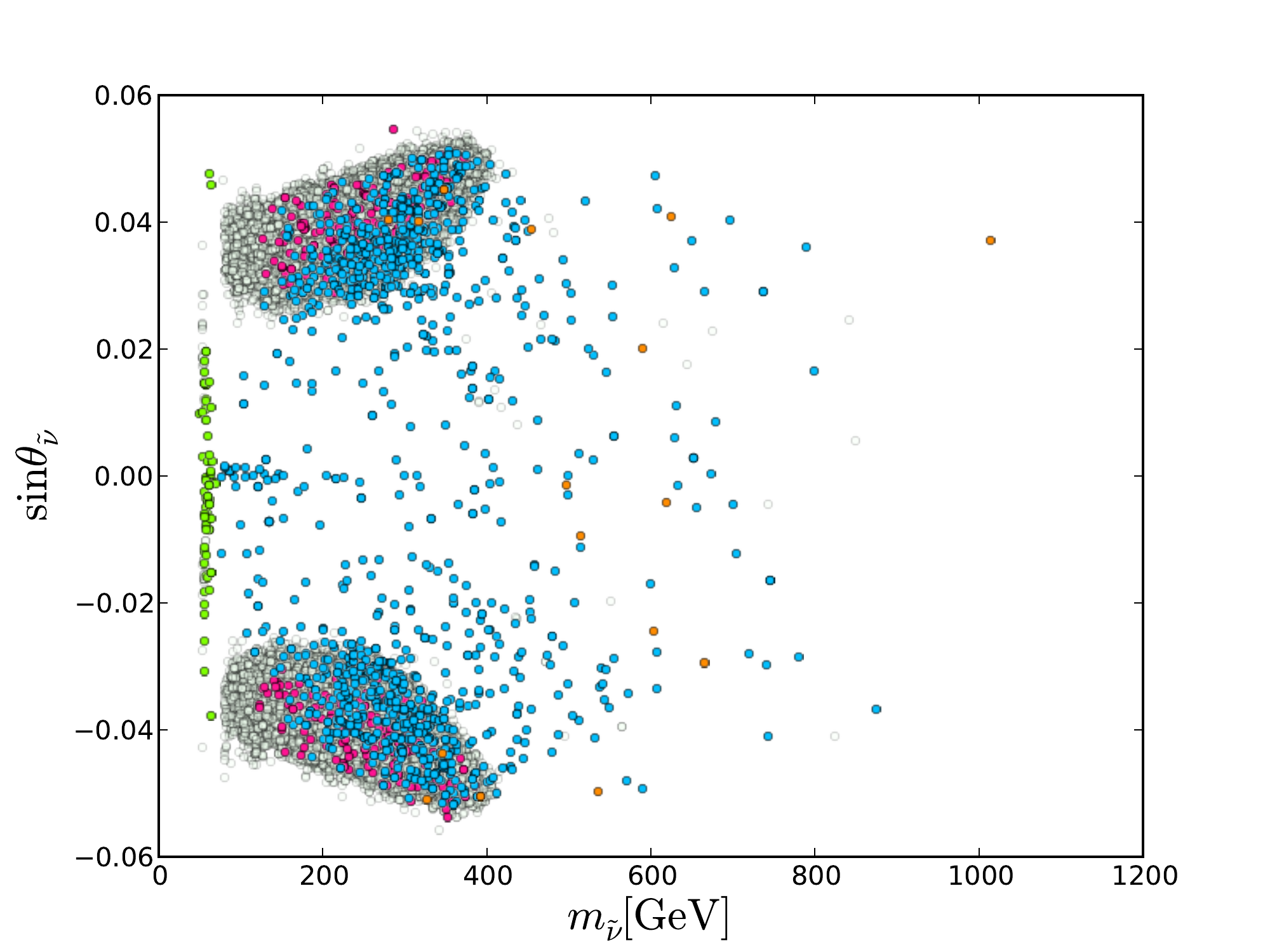}
\end{minipage}
\caption{{\it Left:} Equal weight points from \texttt{MultiNest} chains plotted as a function of the sneutrino mass $m_{\snu}$ and the scattering cross-section $\sigma^{\rm SI}_n$. The green points denote the Higgs resonance region, the magenta points have $(m_{\neu_1}-m_{\snu})/m_{\snu}< 0.10$, the blue points have $(m_{\neu_1}-m_{\snu})/m_{\snu}< 0.10$ and $(m_{\cha_1}-m_{\neu_1})/m_{\neu_1}<0.10$, the orange points denote the long-lived $\stau$, while the gray points do not exhibit a particular pattern in the mass spectrum. The blue solid line is the current exclusion limit by LUX, while the black dashed curve represents the projection for XENON1T.  {\it Right:} Same as left in the $\{ \snu-\sin\theta_{\snu} \}$-plane. All points satisfy the relic abundance constraint. For details on the priors and free parameters see text in section~\ref{sec:num} and~\ref{sec:appA}. \label{fig:goodDM}}
\end{figure}

The result of the \texttt{MultiNest} run is illustrated in figure~\ref{fig:goodDM}: in the left panel the cross-section $\sigma^{\rm SI}_n$ versus the sneutrino mass is shown and all points have a relic density compatible with Planck measurement. We first note that there are not light sneutrinos with masses below the Higgs resonance around 63 GeV. As we adopt boundary conditions for the SUSY parameters at the GUT scale, we did not find light sneutrinos as viable solution for the dark matter candidates, contrary to~\cite{Arina:2007tm,Belanger:2010cd} in which the SUSY parameters are fixed at EW scale. In order to have a very light sneutrino of about 3-10 GeV with good relic density and not excluded by LUX, a very large $A_{\snu}$ is required. In the RGEs the trilinear couplings affect the running of the scalar masses, hence a large value of the scalar trilinear term for the sneutrino induces instabilities and tachyonic solutions\footnote{As already stated, we do not investigate the full posterior pdf of the MSSM+RN, hence it could be that light sneutrinos are viable, however these solutions are difficult to find and might  be in strong tension with the Higgs branching ratio into invisible.}. From the sampling, we highlight four regions, three of which have particular pattern in the SUSY mass spectrum. They all are relevant for LHC physics, giving rise to different signatures, as discussed in the next section, as well as they are characterized by different annihilation channels to achieve the relic density. 

On the Higgs pole (green points), the dominant role for attaining the correct relic abundance is played by the sneutrino itself via  the $\snu_1 \snu^{\ast}_1 \to f\bar{f}$ annihilations mediated by the Higgs boson, as by definition of resonance region. As a consequence, the sneutrino mixing angle, shown as a function of $m_{\snu}$ in the right panel of figure~\ref{fig:goodDM}, is fixed mainly by the requirement of being compatible with the LUX exclusion bound~\cite{Akerib:2013tjd}. In order to suppress sufficiently the $Z$ boson exchange on $t$-channel the sneutrino mixing can not be larger than 0.02. The rest of the SUSY spectrum is not directly constrained by relic density requirements and does not follow a particular pattern. The gaugino sector can be lighter than the scalar lepton sector or vice versa, in other words the Higgs resonance region contains a rich LHC phenomenology. 

Another region, denoted by the orange points, has the characteristic of having long-lived $\stau$. In particular the mass splitting between the sneutrino and the scalar tau, which is the NLSP (next to lightest SUSY particle) is smaller than 1 GeV. We discuss the details of the $\stau$ decay and life-time in section~\ref{sec:s2}. Here we comment on the SUSY spectrum, which is similar to sort of split SUSY scenario, in the sense that the LSP and NLSP are at an energy scale still in reach of LHC, while all the other SUSY particles will remain elusive, above 1 TeV. The correct relic density for $\snu_1$ is achieved in two ways.  When the sneutrinos have a small left-handed component, larger than about 0.02, the dominant annihilation channels are $\snu_1 \snu^{\ast}_1 \to W^+W^-, ZZ, hh, t \bar{t}$ and coannihilation with the $\stau$ is also relevant, such as $\snu_1 \stau \to Z W^-, h W^-$. However the more right-handed the sneutrino becomes, the more the annihilation channels  $\tilde{\tau}_1^+\stau \to W^+W^-, ZZ, hh, t \bar{t}$ dominate for achieving $\Omega_{\rm DM} h^2$. 

The magenta points denote the region where sneutrino and neutralino are close in mass within 10\%, and the neutralino is mostly bino. The relic density in this case is fixed only by sneutrino annihilation, via these dominant processes: $\snu_1 \snu^{\ast}_1 \to W^+W^-, hh, ZZ$ and $\snu_1 \snu^{\ast}_1 \to t\bar{t}$, whenever the top threshold is opened. The mixing angle should be still sizable, around 0.02-0.04, to provide $\Omega_{\snu} h^2$ in accord with the measured value. On the contrary, the blue points denote the parameter space where the sneutrino is degenerate with the neutralino at the level of 10\% and in addition with the chargino, at the same percentage level, that is neutralinos and charginos are either winos or higgsinos. Two possibilities for the relic density can arise. First, if the mass spectrum of the $\neu_1,\neu_2$ and $\cha_1$ is compressed,  coannihilation is crucial for fixing the correct relic density. For instance it can involve a large number of annihilation processes such as $\neu_2 \cha_1 \to q \bar{q}', \bar{l} \nu_l, Z W^+$, $\neu_{1,2} \neu_{1,2} \to q\bar{q},W^+ W^-$ and $\cha_1 \tilde{\chi}^-_1 \to q\bar{q}, W^+ W^-$. The second typology arises if the chargino is almost degenerate with the neutralino within 1-2\%. In this case the relic density is driven only by $\cha_1$ and $\neu_1$ co-annihilation. In both cases the contribution of sneutrinos to $\Omega_{\rm DM} h^2$ is irrelevant. This is the reason why the sneutrinos in this region can be almost purely right-handed and highly elusive for dark matter direct detection. This behavior seems also to be the common one for heavy sneutrinos.

In the rest of the sampling, gray points, there isn't a particular pattern for SUSY mass spectrum. The correct sneutrino relic density is achieved by the sneutrino annihilating mainly via the $s$-channel exchange of a $Z$ boson (from here on called Z-boson region)  into $W^- W^+, t \bar{t}, f\bar{f}$ and via $t$-channel neutralino and chargino  exchange  going into $f\bar{f}$. The mixing angle exhibits a sizable component of left-handed component, as shown in figure~\ref{fig:goodDM}. The values around 0.04 are a good compromise between achieving $\Omega_{\rm DM} h^2$ and being compatible with the direct detection bounds. A large part of the sneutrino dark matter parameter space can be probed by next generation of dark matter experiment, such as XENON1T~\cite{Aprile:2012zx}, denoted by the black dashed line in figure~\ref{fig:goodDM}. 

The effect of the small $Y_\tau$ in the mass spectrum makes the electron and muon sneutrino slightly heavier than $\snu_1$; they will eventually decay into the LSP, with a process mediated most likely by the off-shell lightest neutralino ($\snu_{e,\mu} \to \nu_{e,\mu} \tilde{\chi}^{0 \ast}_1 \to \nu_{e,\mu} \snu_1 \nu_\tau$) and producing only neutrinos plus the LSP. However this process might be very suppressed by the splitting in mass between sneutrino flavors, so that the $\snu_e,\snu_\mu$ can be long-lived, with a life-time that can range from $10^{-4} \rm s$ up to the age of the Universe. We not discuss further the implication of this decay. As far as it concerns the LHC phenomenology all three sneutrino flavors can be produced in the decay chains and are indistinguishable. 

\section{Collider signatures}\label{sec:cs}

We have chosen four  benchmark points, which are representative of the rich LHC phenomenology of sneutrino dark matter. The relevant SUSY breaking parameters at the electroweak scale are summarized in table~\ref{tab:benchmark}. For the analysis we used a center of mass energy of 14 TeV and assumed a  luminosity $\mathcal{L} = 100/{\rm fb}$.
\begin{table}[t]
  \caption{Relevant SUSY breaking parameters at electroweak scale characterizing the four benchmarks used for simulating events at LHC, as labelled. The benchmarks are defined as follows: $B1$ is for the long -lived $\stau$, $B2$ for the two same sign leptons, $B3$ for multi-leptons and finally $B4$ for direct chargino production.}
  \centering
  \begin{tabular}[t]{|c|c|c|c|c|}
    \hline
    & $B1$ & $B2$ & $B3$ & $B4$\\
    \hline
    $M_1$ &  $1009.4$ GeV &  $-604.2$ GeV & $358.2$ GeV&  $207.5$ GeV  \\
    $M_2$& $2519.5$ GeV & $-241.9$ GeV & $746.7$ GeV &   $412.4$ GeV \\
    $\mu$ & $5566.8$ GeV & $393.6$ GeV & $1516.9$  GeV & $1046.6$ GeV   \\
    $m_L$ & $2362.8$  GeV & $2128.1$ GeV &  $683.1$ GeV &  $1289.9$ GeV \\
    $m_{\tau_L}$ & $2248.4$  GeV & $2075.0$ GeV & $656.9$ GeV & $1289.7$ GeV  \\
    $m_N$ & $645.0$ GeV & $110.0$ GeV & $60.7$ GeV & $108.2$ GeV \\
    $A_{\tilde{\nu}}$ & $26.5$ GeV &$753.3$ GeV & $-18.7$ GeV  & $-359.7$ GeV  \\
    $\tan{\beta}$ & $41.1$ & $11.3$ & $9.9$ &  $37.02$ \\
    \hline
  \end{tabular}
  \label{tab:benchmark}
\end{table}

\subsection{Long-lived $\stau$}\label{sec:s2}

Long-lived charged particles at LHC have been studied in the MSSM or in models beyond the standard model from a theoretical point of view (see {\it e.g.}~\cite{Jittoh:2005pq,Biswas:2009rba,DeSimone:2010tf}) and are searched in depth experimentally (see {\it e.g.}~\cite{Chatrchyan:2013oca,TheATLAScollaboration:2013qha}). As anticipated in the previous section, and discussed in~\cite{Gupta:2007ui}, in the MSSM+RN framework the long-lived particle is typically a $\stau$. 
\begin{figure}[t]
\begin{minipage}[t]{0.49\textwidth}
\centering
\includegraphics[width=0.8\columnwidth,trim=5mm 5mm 5mm 10mm, clip]{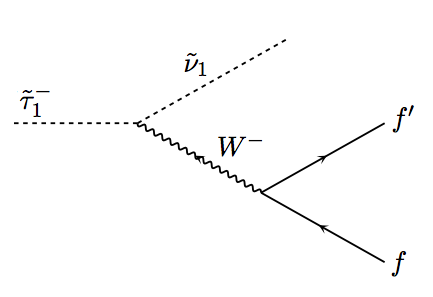}
\end{minipage}
\begin{minipage}[t]{0.49\textwidth}
\includegraphics[width=0.8\columnwidth,trim=5mm 5mm 5mm 8mm, clip]{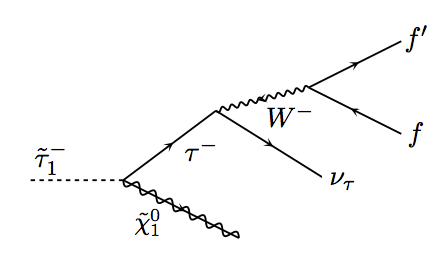}
\end{minipage}
\caption{Three-body decay process for the $\stau$ NLSP, on the left (right) in the MSSM+RN (MSSM) when $\delta m < m_{\tau}$, with $\delta m \equiv m_{\rm NLSP}-m_{\rm LSP}$.\label{fig:dd}}
\end{figure}
The mass matrix of the scalar $\tau$ is 
\begin{eqnarray}\label{massmatrix}
\mathcal{M}^2_{\tilde{\tau}} & = &
\begin{pmatrix}
m^2_{L}+D_{L} + m^2_{f}& m_{\tau}\left(A_{L}+\mu \tan\beta\right)\cr m_{\tau}\left(A_{L}+\mu \tan\beta\right) & m^2_{R}+D_{R}+m^2_{\tau}
\end{pmatrix}\,,
\end{eqnarray}
where the terms $D_{L,R}$ stand for the corrections to the soft masses arising from RGEs. The left-handed soft breaking mass $m_L$ is the only SUSY term common with the sneutrino. As the $\tau$ Yukawa is non negligible, the off-diagonal term is large and induces a mixing between left and right component as
\begin{eqnarray}
\left\{
\begin{array}{l}
 \tilde{\tau}_1 = +\cos\theta_{\tilde{\tau}} \tilde{\tau}_L +\sin\theta_{\tilde{\tau}} \tilde{\tau}_R\,,\\
 \tilde{\tau}_2 = -\sin\theta_{\tilde{\tau}} \tilde{\tau}_L +\cos\theta_{\tilde{\tau}} \tilde{\tau}_R\,.
\end{array}
\right.
\end{eqnarray}
If the $\stau$ is the NSLP and the splitting in mass with the LSP is $\delta m /m_{\snu} < 10\%$ ($\delta m \equiv m_{\stau} - m_{\snu}$), it contributes to the sneutrino relic density as coannihilation processes become relevant. In general the requirement of coannihilation implies that the spitting in mass is much smaller than the $W$ boson mass and the $\stau$ decays into the LSP only via three-body process, illustrated in figure~\ref{fig:dd} (left panel). The smaller the mass splitting the smaller is the decay width because of the suppression in the phase-space, leading eventually to long-lived $\stau$ for $\delta m < 1$ GeV. However the life-time does not only depends on $\delta m$, but  on both the mixing angles $\theta_{\tilde{\tau}}$,  $\theta_{\snu}$ and on the overall mass scale $m_{\snu}$. 

We discuss the long-lived $\stau$ phenomenology using the benchmark point $B1$, given in table~\ref{tab:benchmark}, in which
\begin{eqnarray}
  m_{\stau} = 666.3\,  {\rm GeV}\,, &&  \sin \theta_{\tilde{\tau}} = 0.99\,, \\ \nonumber
  m_{\snu} = 665.5\,  {\rm GeV}\, ,&&  \sin\theta_{\snu} = -0.029\,, \\ \nonumber
  \Gamma_{\tilde{\tau}} = 7.33 \times 10^{-18} \, {\rm GeV}\,,&& \tau_{\tilde{\tau}} = 8.98 \times 10^{-8}\,  {\rm s}\,. 
\label{eq:br1}
\end{eqnarray} 
This is representative of  our sampling  of  long-lived scalar $\tau$ depicted by the orange points (figure~\ref{fig:goodDM}). In $B1$ both $\snu_1$ and $\stau$ are mostly right-handed, but the sneutrino is so sterile that the $\stau$ annihilation alone sets the relic density. The degeneracy between sneutrino and $\stau$ is `accidental': it not entirely due to the left-handed mass\footnote{The neutral component of a SU(2) doublet is always lighter than the charged one due to radiative corrections of the order $\mathcal{O}(100)$ MeV~\cite{Cirelli:2005uq}.} $m_L$ but also to RGEs effects. Searches for long-lived charged particles have excluded $\stau < 300$ GeV~\cite{TheATLAScollaboration:2013qha}, when produced directly in the $pp$ collision, hence we consider only relatively heavy $\stau$. This is the reason why the orange points are present only for heavy sneutrinos with mass larger than about 400 GeV.
\begin{figure}[t]
\centering
\includegraphics[width=0.6\columnwidth,trim=3mm 0mm 19mm 12mm, clip]{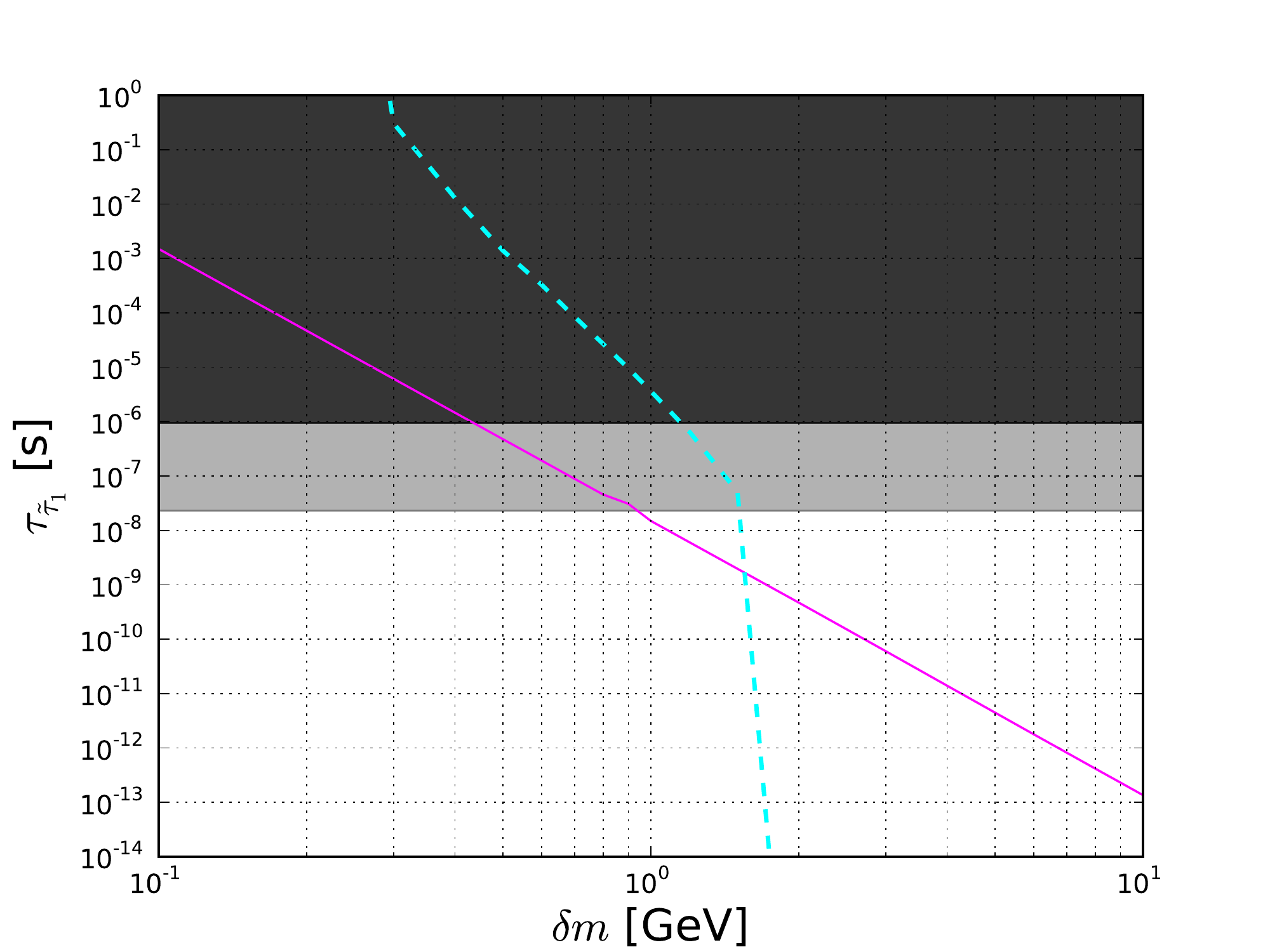}
\caption{$\stau$ life-time as a function of $\delta m \equiv m_{\stau} - m_{\rm LSP}$. The magenta solid (cyan dashed) line stands for the case of the MSSM+RN with the sneutrino LSP (MSSM with $\neu_1$ LSP). The mass of the LSP is fixed at 665 GeV and the $\stau$ is mostly right-handed (see text for details, benchmark $B1$). The light gray area denotes the region with a detectable charged track disappearing in the hadronic calorimeter, while the dark gray area stands for the long-lived particles leaving the detector volume. In both cases a $p_{\rm T}=50$ GeV is assumed. \label{fig:stauchi}}
\end{figure}

Figure~\ref{fig:stauchi} shows the $\stau$ life-time as a function of $\delta m$ (solid magenta line). Below the $W$ threshold, $\tau_{\tilde{\tau}_1}$ is a smoothly increasing function of $\delta m$, because it produces an off-shell $W$, which then decays into on-shell fermions/quarks as $\stau \to W^{-\ast} \snu_1 \to \snu_1 f f'$. As far as $\delta m$ goes below a certain mass threshold, for instance $\tau$ or $\mu$, the decay is suppressed by the reduced number of decay possibilities but is still a three-body decay, hence there are no visible kinks. This is in contrast with the MSSM picture: for a $\stau$ NLSP, its decay into $\neu_1$ is given by a two/three-body process (see figure~\ref{fig:dd}, right panel) until $\delta m < m_{\tau}$. Below $m_{\tau}$ it becomes a three/four-body process, producing a sharp feature in $\tau_{\tilde{\tau}_1}$ (cyan dashed line, figure~\ref{fig:stauchi}): the life-time increases much rapidly than in the MSSM+RN case due to the suppression in phase-space. Conversely, above the $\tau$ mass, the MSSM $\stau$ has a shorter life-time with respect to the MSSM+RN  because the decay is dominated by a two-body process. For details about the long-lived $\stau$ in the MSSM we refer to~\cite{Jittoh:2005pq}. The decay width of the $\stau$ in both scenarios has been computed with MG5 at parton level, up to $\delta m = 0.1$ GeV. Below such value the numerical computation becomes unstable. 
The light and gray band in figure~\ref{fig:stauchi} are a guide to the eye as far as it concerns the relevance of long-lived staus at LHC. The $\stau$ is produced with a certain boost, and travels a distance given by  $d = \tau_{\stau} p_T / m_{\stau}$ in the detector before decaying.  A 666 GeV stau with $p_T =50$ GeV will decay inside the hadronic calorimeter  or inside the detector (gray region) for life-time $\tau_{\tilde{\tau}_1} \sim( 5 \times 10^{-7} \to 10^{-6}) \, {\rm s}$~\cite{ATLAS:2012ab}, implying that the distance travelled is at least 514 mm referring to the ATLAS detector. For life-time larger than $10^{-6}\,  {\rm s}$ (or equivalently decay width $\Gamma_{\tilde{\tau}_1} < 10^{-19}$ GeV) the charged long-lived particle decays outside of the detector volume. For a $p_T=200$ GeV the minimum life-time to decay in the hadronic calorimeter is $\tau_{\tilde{\tau}_1} \sim 10^{-9} {\rm s}$, which corresponds to the change in slope of the $\stau$ life-time in the MSSM scenario. 

There are three ways of detecting long-lived (charged) particles, depending on the distance travelled inside the detector
\begin{enumerate}
\item[(i)] $d \sim \mathcal{O}(1000)\,  \rm mm$: if the heavy long-lived charged particles decay outside of the detector volume, these particles would interact like heavy muons releasing energy by ionization as they travel through the detector. The search is performed by measuring the specific ionization energy loss and the time-of-flight. As these particles travel with velocity $\beta = v/c$ measurably lower than the speed of light, they can be identified and their mass determined via the relation $m = p_T/ (\gamma \beta)$~\cite{Chatrchyan:2013oca,TheATLAScollaboration:2013qha}, with $\gamma$ being the Lorentz factor. 
\item[(ii)] $d \sim \mathcal{O}(100)\,  \rm mm$: if the charged long-lived particles decay inside the hadronic calorimeter, they could be detected by tracks that appear to have few associated hits~\cite{ATLAS:2012ab,Aad:2013yna}. 
\item[(iii)] $d \sim \mathcal{O}(10) \, \rm mm$:  if the long-lived particle is neutral and decays, it gives rise to a displaced vertex. This search characterizes mostly SUSY R-parity violating scenarios and is not relevant for our purposes.
\end{enumerate}

\begin{figure}[t]
\begin{minipage}[t]{0.49\textwidth}
\centering
\includegraphics[width=1.\columnwidth,trim=0mm 0mm 0mm 0mm, clip]{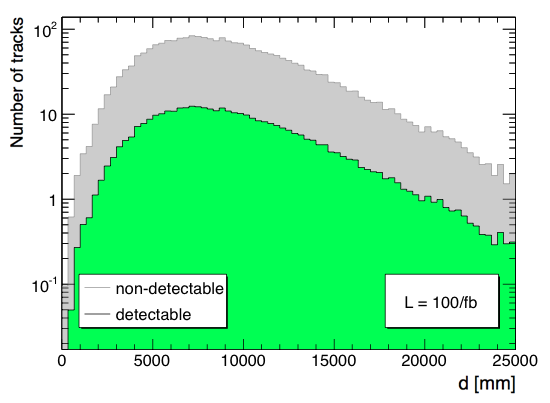}
\end{minipage}
\hspace*{0.2cm}
\begin{minipage}[t]{0.49\textwidth}
\includegraphics[width=1.\columnwidth,trim=0mm 2mm 0mm 0mm, clip]{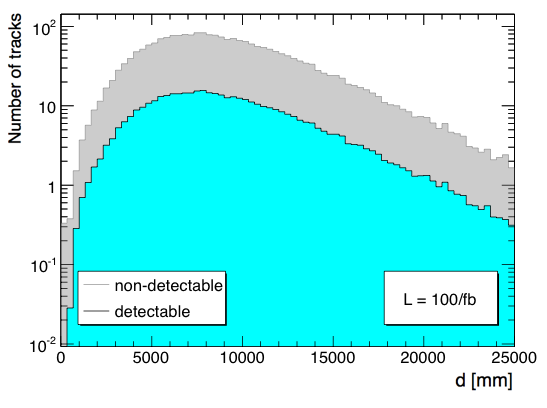}
\end{minipage}
\caption{{\it Left:} Number of events as a function of the distance travelled by the long-lived $\stau$, assuming only one track is identifiable. The green histogram is the expected number of events after applying all cuts (see text), while the gray histogram is the number of events before cuts, as labelled. {\it Right:} Same as left in the case of identification of both charged tracks of the pair produced staus. \label{fig:LLstau}}
\end{figure}

We have simulated Monte Carlo events for the benchmark $B1$, assuming direct production of a pair of $\stau$'s via electroweak Drell-Yan process. The production cross-section is $\sigma = 8.23 \times 10^{-5}$ pb. We operated cuts similar to the search type (ii) to estimate what is the number of long-lived $\stau$ that can be detected. The cuts are  
\begin{enumerate}
\item The track should have no other tracks with $p_T > 0.5$ GeV within a cone of radius $\Delta R = 0.05$; this requirement avoids transverse activity  within a cone centered on the track, namely avoids jets, photons, electrons and muons;
\item The charged particle should travel at least $d=514$ mm, namely $\stau$ should decay inside the hadronic calorimeter;
\item To account for the detector simulation and background subtraction, we convoluted the signal after cuts with the efficiency in detecting charged tracks given in~\cite{TheATLAScollaboration:2013qha}, which is $\epsilon=0.15$ for detecting one track and $\epsilon=0.2$ for detecting both tracks.
\end{enumerate}
The background for escaping charged tracks depends on the type of search. For (i) it is mostly composed of high $p_T$ muons with mis-measured velocity and is data driven, while for (ii), namely disappearing high $p_T$ tracks, it consists of charged hadrons (mostly pions) interacting in the hadronic calorimeter or low $p_T$ charged particles whose $p_T$ is badly measured. We are however not simulating the background and use the criterium 3. to take into account its effect.
The result for $B1$ is illustrated in figure~\ref{fig:LLstau}, where we plot the number of tracks for one detected heavy long-lived $\stau$ (left panel) and for detecting both long-lived charged $\stau$ (right panel), that have been produced in pair. The light-gray region is the number of tracks before applying the cuts described above. Interestingly we see that several events can be measured and most likely they will leave the detector, while only a couple of events are expected to decay inside the hadronic calorimeter. This is due to the fact that the $\stau$ is very massive. Notice that the efficiency for detecting both charged tracks is higher due to less background.

If a long-lived $\stau$ would be detected, could we distinguish between the MSSM and the MSSM+RN scenarios (figure~\ref{fig:LLstau})? It would be tricky to disentangle the two scenarios if the long lived particle decays inside the hadronic calorimeter, however if it decays outside the detector volume and the time-of-flight can be measured, it would be possible to reconstruct its mass by knowing the $p_T$ and have some hints if the LSP is a neutralino or a sneutrino. A long-lived stau is a signature of sneutrino dark matter and can be retrieved in other models, such as pure right-handed sneutrinos~\cite{Cerdeno:2013oya}.

\subsection{Two same sign leptons}
\label{sec:C1toll}
The possibility of having sleptons lighter than neutralinos is an interesting
feature of the phenomenology of MSSM+RN, specially for collider signatures
since this mass hierarchy could lead to potentially powerful
signatures to test sneutrino as LSP at LHC.

Let's start describing the phenomenology of sleptons in the MSSMRN+RN. As it
was mentioned above, the initial parameters are set at the gauge coupling
unification scale $M_X$, where SUSY breaking is transmitted to the observable
sector. When solving the RGEs from $M_X$ to the EW scale, correlations are printed in the mass spectrum. Combining this
effect with the requirement of sneutrinos to be good dark matter candidates, one can be able to understand the typical mass spectrum and therefore study
potential experimental signatures.

In the framework described in section~\ref{sec:model} the stau is the lightest
slepton due to the contribution of the tau Yukawa in the RGE. In addition, depending on the value of $\tan{\beta}$ and the trilinear term, the splitting in mass 
between the lightest stau and the other sleptons could increase. On the other
hand, for the first and second generation of sleptons, typically the left
component is heavier that the lightest neutralino mainly because
its RGE has a term proportional to $M_1^2$ (bino mass) and another one to $M_2^2$
(wino mass). A second and major reason for this mass hierarchy is related
with dark matter, as explained in the previous section: to have a good
sneutrino candidate, $m_L$ is pushed to large values in order to have a mostly right-handed lightest sneutrino. The case
of right-handed slepton is different: its coupling and mass are not constrained by the
requirement of having a sneutrino as a dark matter candidate and its RGE receives contribution from
 $M_1^2$ but not from $M_2^2$. As a consequence, $\tilde{l}_R$ is typically lighter than
the left-handed slepton.
\begin{figure}[t]
\begin{minipage}[t]{0.49\textwidth}
\centering
\includegraphics[width=1.\columnwidth,trim=1mm 0mm 13mm 10mm, clip]{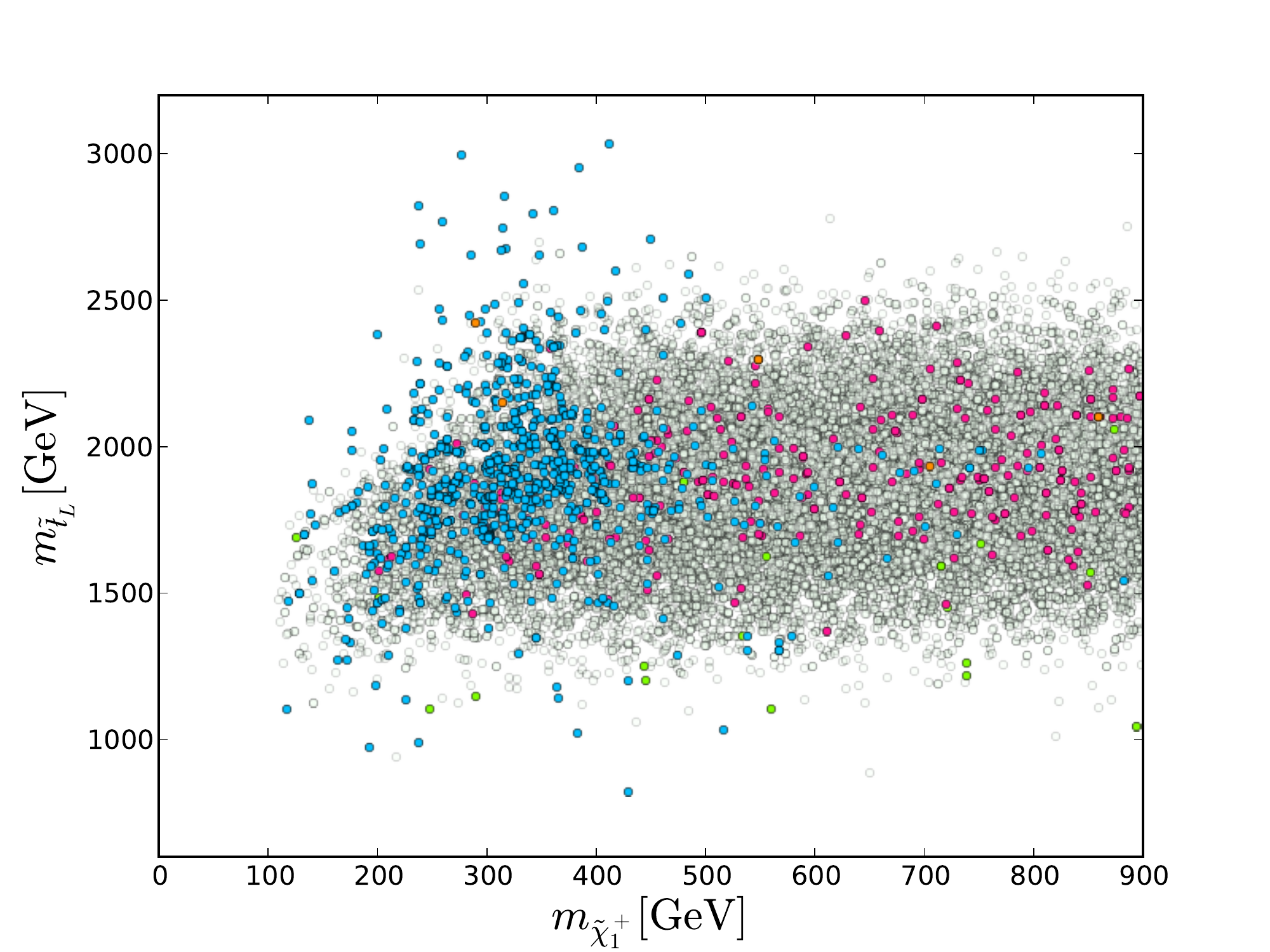}
\end{minipage}
\hspace*{0.2cm}
\begin{minipage}[t]{0.49\textwidth}
\includegraphics[width=1.\columnwidth,trim=1mm 0mm 13mm 10mm, clip]{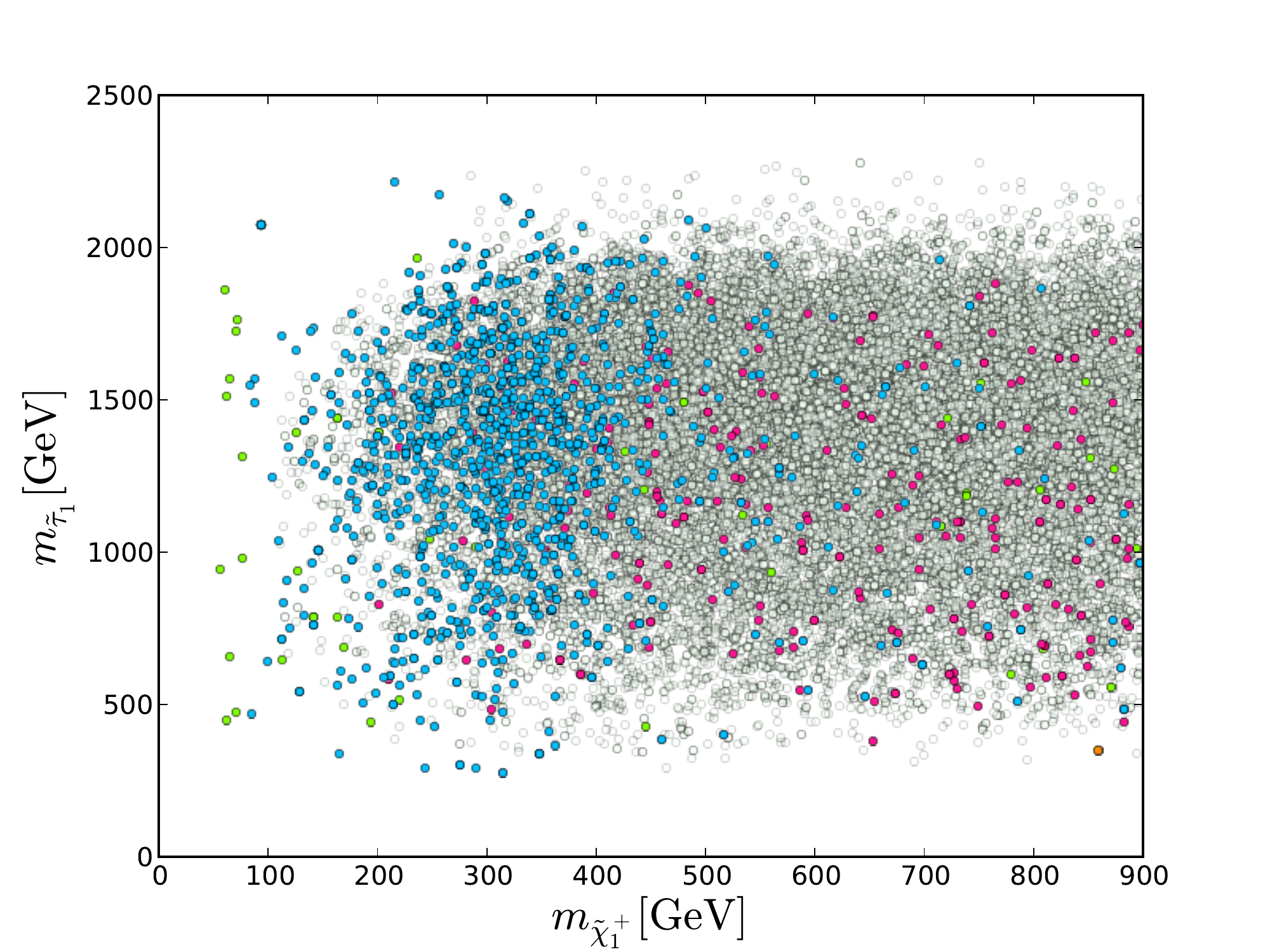}
\end{minipage}
\caption{{\it Left:} Equal weight points in the $\{ \cha_1 - \tilde{l}_L  \}$-space. {\it Right:} Same as left for $\stau$. The color code is as in figure~\ref{fig:goodDM}.}\label{fig:sleptonMasses}
\end{figure}
Figure~\ref{fig:sleptonMasses} shows the relation between the chargino and the
left-handed sleptons and with the staus. As expected, the region with light
$\stau$ is significantly larger than the region accouting for light
left-handed sleptons. We find as well that $\slr$ are lighter than
$\tilde{l}_L$, however we do not show them as we will not use right-handed
scalar leptons in the signatures, as explained more
    in details in section~\ref{sec:gp}. The relation between sleptons and charginos is relevant for collider phenomenology because the dominant production process of electroweak particles is through chargino and neutralino production, via their higgsino and wino components. For that reason we only show the region with charginos lighter than 900 GeV.

It was pointed out by~\cite{Katz:2009qx} that as a consequence of requiring a sneutrino LSP, the chargino-neutralino
production will have a final state with three leptons and missing energy where the
two opposite sign leptons could have different flavors (since sleptons decay through a $W$ boson). This is a very distinctive signature of sneutrino
dark matter with respect to neutralino dark matter. Remember that in the MSSM
the chargino-neutralino production will give a signal of three leptons but the
two opposite sign leptons will have necessarily same flavor, as they are coming from the $Z$ boson or from
the neutralino decay through sleptons.

We instead focus in the region where the NLSP is the lightest stau, $\stau$, so that these signatures are present in all
the regions where a slepton is lighter than the chargino in the gray/magenta points in figure~\ref{fig:goodDM}. We consider the process depicted in figure~\ref{fig:SSleptons}:
\begin{figure}[t]
  \centering
  \includegraphics[width=0.5\linewidth]{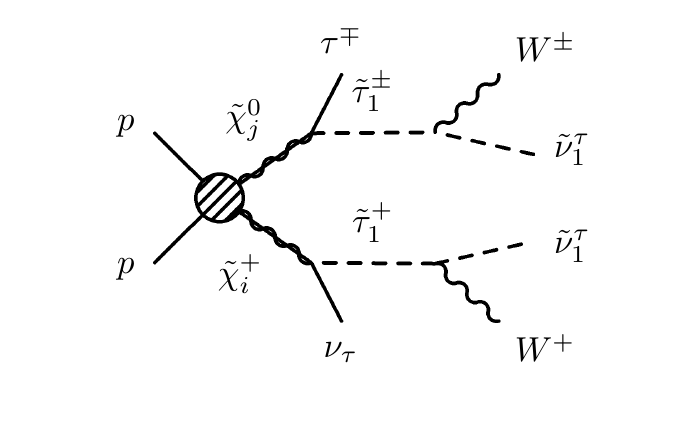}
  \vspace{-0.7cm}
  \caption{Relevant process for the two-lepton signal: chargino-neutralino production and subsequent decay through the lightest stau.}
  \label{fig:SSleptons}
\end{figure}
\begin{eqnarray}\label{eq:b2proc}
  p p &\rightarrow& \tilde{\chi}_i^+\ \tilde{\chi}_j^0 \\ \nonumber
  &\rightarrow& (\ \nu_\tau\ \tilde{\tau}_1^+\ )\ (\ \tau^{\pm}\ \tilde{\tau}_1^\mp
  \ )\\ \nonumber
  &\rightarrow& (\ \nu_\tau\ W^+\ \tilde{\nu}_1^\tau\ )\ (\ \tau^{\pm}\ W^\mp
  \ \tilde{\nu}_1^\tau\ )\,.
\end{eqnarray}
To study this signature in more detail we use benchmark $B2$ described in table~\ref{tab:benchmark}. The relevant masses and mixings are
\begin{eqnarray}
  \label{eq:bench3}
  m_{\tilde{\chi}_1^{\pm}} = 419.3\ \mathrm{GeV},&&
  m_{\tilde{\chi}_2^{0}} = 421.2\ \mathrm{GeV},\\ \nonumber
  m_{\tilde{\nu}_1^{\tau}}=202.6\ \mathrm{GeV},&&
  \sin{\theta_{\tilde{\nu}}} = -0.031\,,\\ \nonumber
  m_{\tilde{\tau}_1}=354.2\ \mathrm{GeV},&&
  \sin{\theta_{\tilde{\tau}}}= -0.00013\,,
\end{eqnarray}
with branching ratios given in table~\ref{tab:BR2}. Notice that the dominant branching ratio is into sneutrinos and leptons.
\begin{table}[t!]
\caption{Relevant branching ratios for chargino and neutralino decays in benchmark $B2$ for the signature of two opposite sign leptons.}
  \centering
  \begin{tabular}[t!]{r c l | c}
   \hline
   &   Process  &  & BR \\
   \hline
    $\tilde{\chi}_1^+ $ & $\rightarrow$ & $\nu_\tau\ \tilde{\tau}_1\ $ & $\ 99.20\%$ \\
    && $ \tau^+ \tilde{\nu}_1\ $ & $\ 0.72\%$ \\
   \hline
    $\tilde{\chi}_2^0 $ & $\rightarrow$ & $\tau^{\pm}\ \tilde{\tau}_1^\mp$ & $\ 99.99\%$ \\
    \hline
  \end{tabular}
  \label{tab:BR2}
\end{table}
The signal we consider here is slightly different from the one assumed in~\cite{Katz:2009qx}, as the final state in~\ref{eq:b2proc} contains two same sign leptons, however the third one is a hadronic $\tau$, which due to the low efficiency in its identification and simulation we are not tagging. 

For the background we consider the production of $W Z\rightarrow W\ l^+\ 
l^-$ and $t\bar{t}W$.  The cross-sections are computed at LO with MG5 and
\texttt{Pythia 6}.

The following cuts are applied,
\begin{enumerate}
\item Two same sign different flavor leptons with $p_T>20$ GeV and pseudo-rapidity $\eta<2.5$;
\item At least one lepton with $p_T > 25$ GeV;
\item $p_T^{\mathrm{miss}}>50$ GeV.
\end{enumerate}
The kinematical variables considered in the analysis are $p_T^\mathrm{miss}$ and the invariant mass of the two selected leptons ($M_{\rm inv}$).
\begin{figure}[t]
  \centering
  \includegraphics[width=0.49\linewidth]{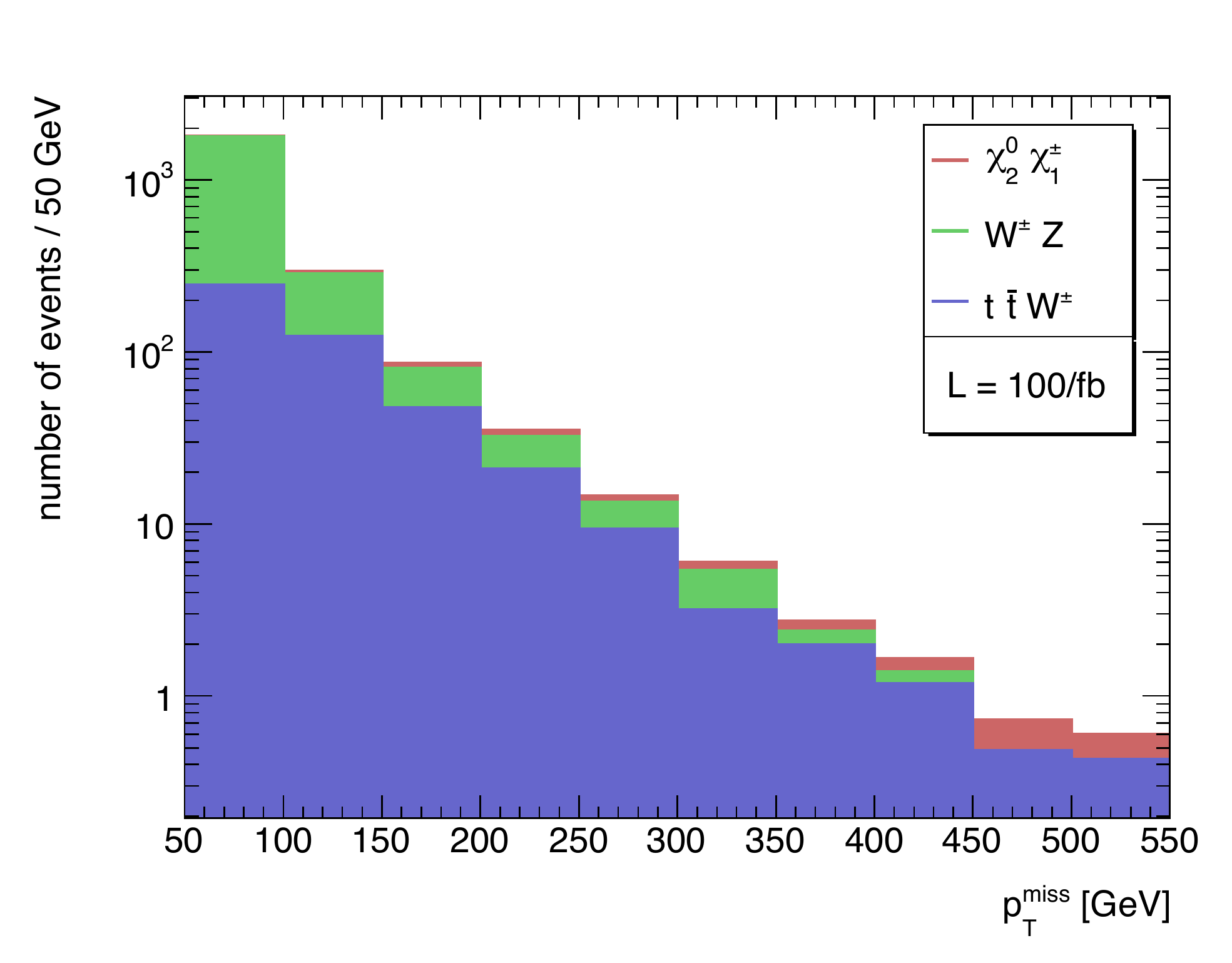}
  \includegraphics[width=0.49\linewidth]{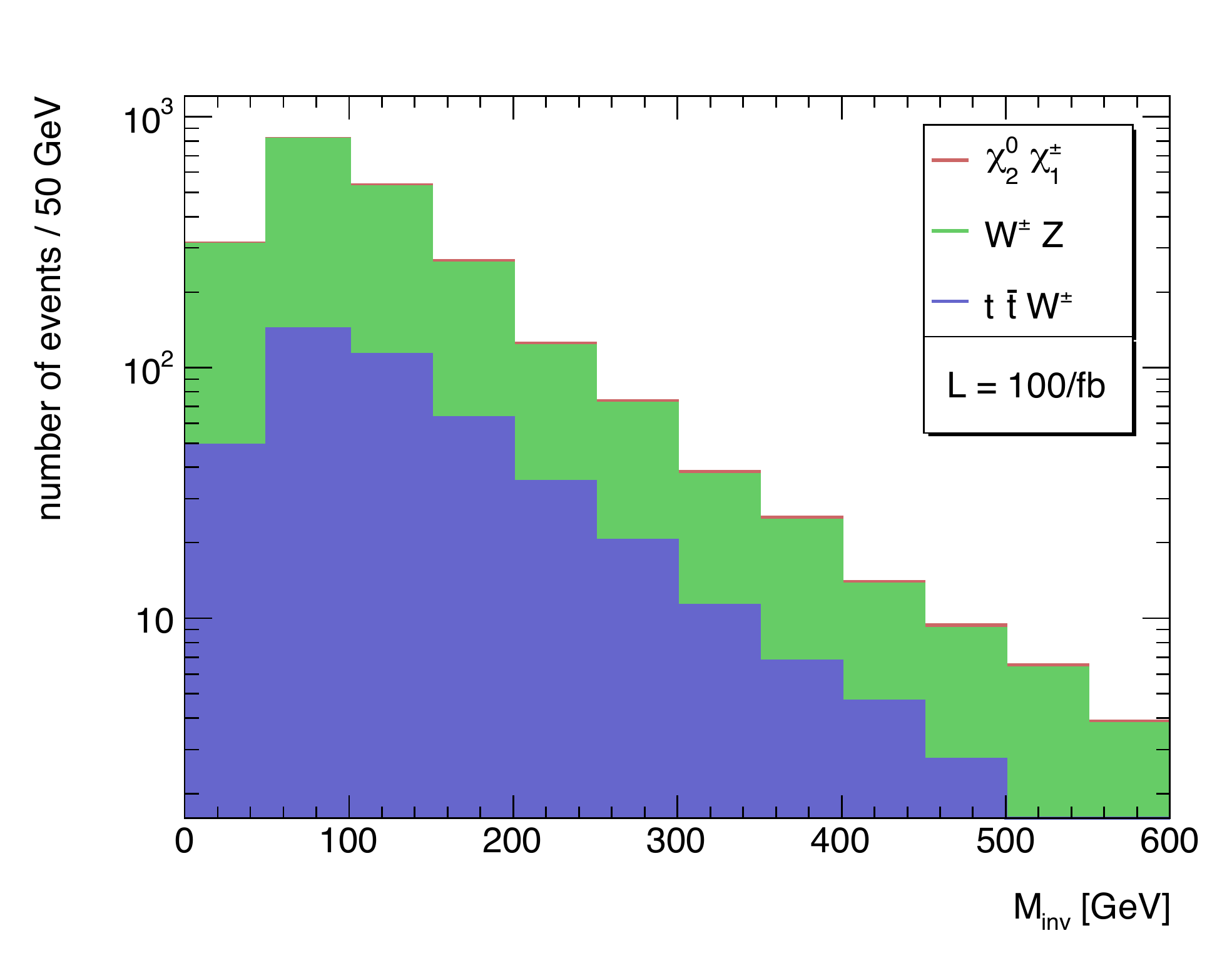}\\
  \includegraphics[width=0.49\linewidth]{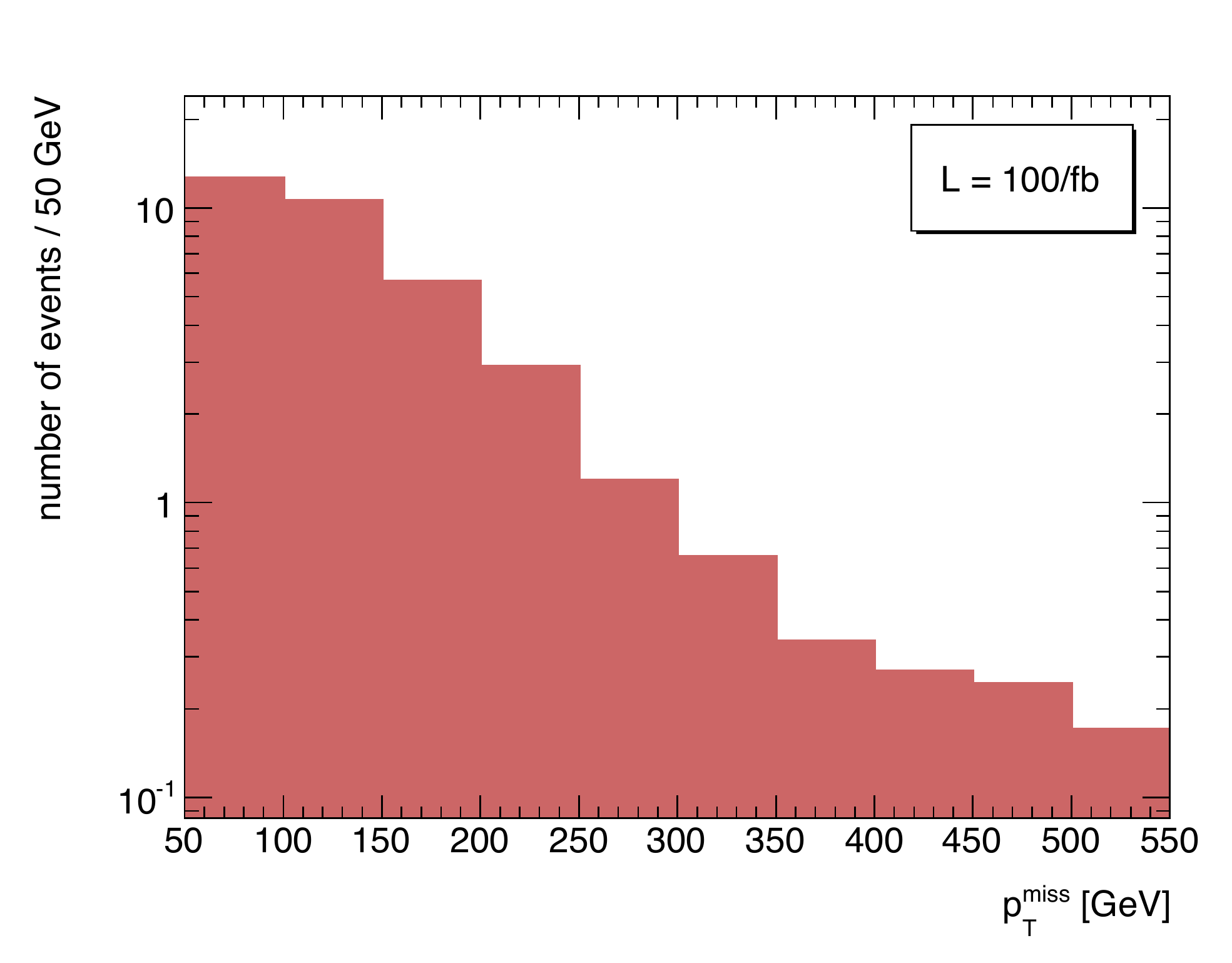}
  \includegraphics[width=0.49\linewidth]{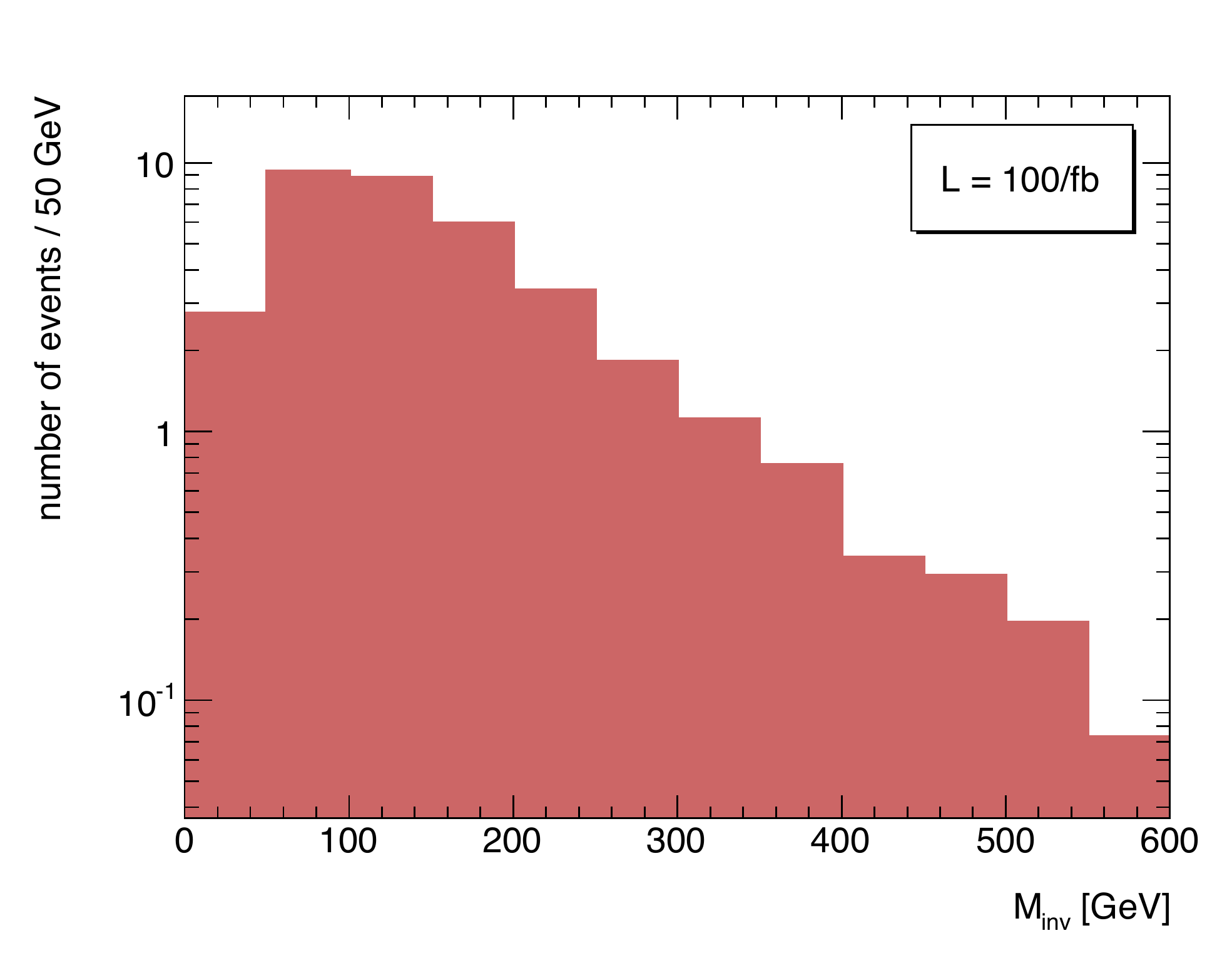}\\
  \includegraphics[width=0.49\linewidth]{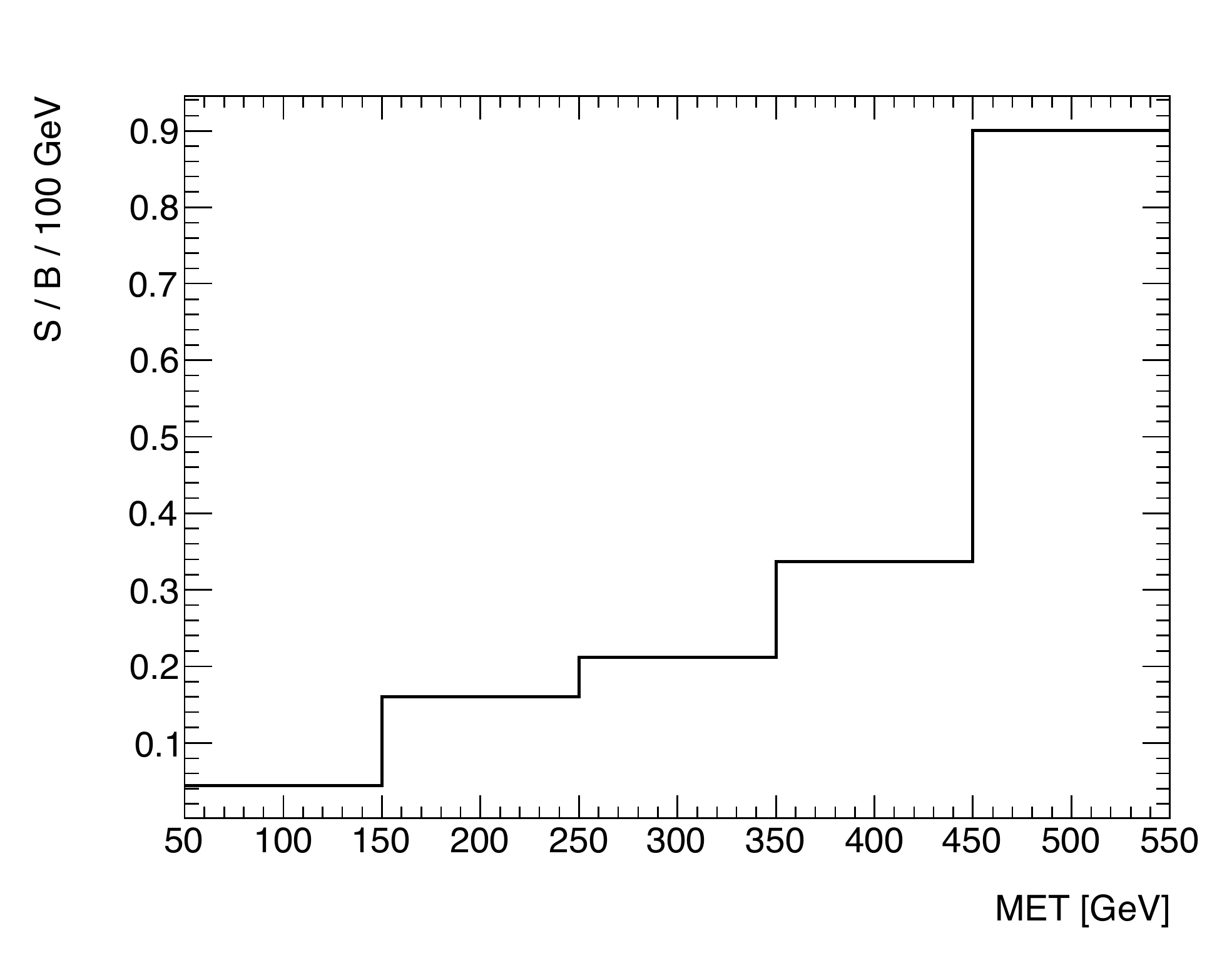}
  \includegraphics[width=0.49\linewidth]{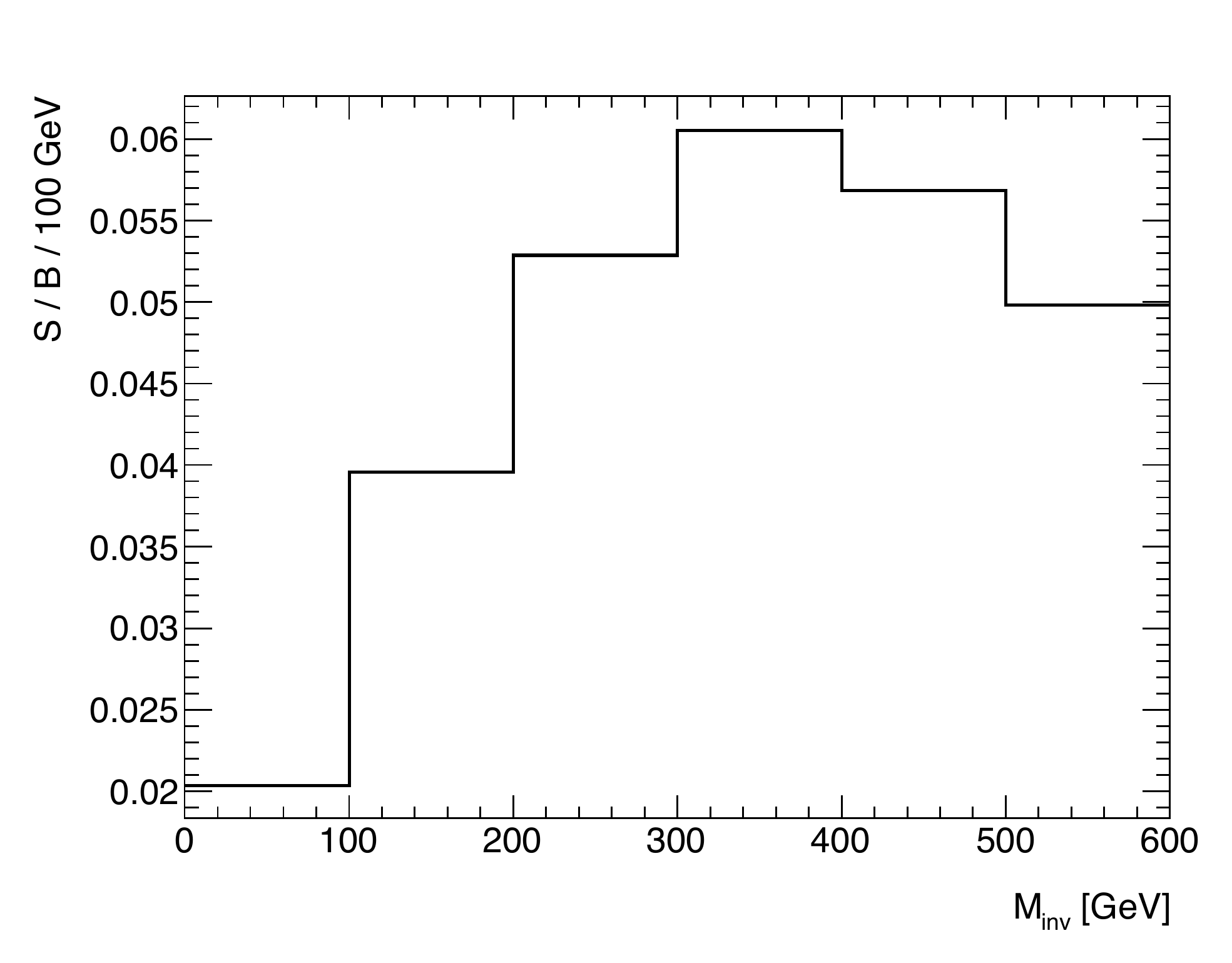}\\
  \caption{{\it Top left:} $p_T^{\mathrm{miss}}$ distribution for chargino-neutralino production of benchmark $B2$. {\it Top right:} Same as left for the $M_\mathrm{inv}$ distribution. The signal and backgrounds are as labelled in the plots. {\it Center (Bottom) left and right:} Same as top for the signal distribution only (signal over background ratio).} 
  \label{fig:NCtoStau}
\end{figure}

Figure~\ref{fig:NCtoStau} shows in the left panel the $p_T^{\mathrm{miss}}$
distribution and in the right panel the $M_{\rm inv}$ distribution. Central panels show
the signal distribution only. As one can see, the signal
    is accumulated at small values of $p_T^{\mathrm{miss}}$ and
    $M_{\mathrm{inv}}$ where the background reaches its maximum value. In
    this region statistical and systematics errors are typically very small and
    therefore even when the ratio between signal and background is small the
    signal could be distinguished. The smallness of the ratio signal over background is illustrated in the lower panels. Of course, NLO contribution have to be
    included and the cuts have to be optimized to get an accurate idea of how
    the signal will stand over the background. Nevertheless, this figure
    gives a good illustration advertising that signals of new physics could be
    hidden at low values of $p_T^{\mathrm{miss}}$ and $M_{\mathrm{inv}}$.

\subsection{Multi-leptons}\label{sec:C1toll}

The phenomenology of the Higgs resonance region is potentially powerful to detect supersymmetry, because of the particular collider signatures that can arise. The composition of the sneutrino requires generally a very small left-handed
component with respect to most
of the points in other regions, see figure~\ref{fig:goodDM}. This implies that the
splitting between $m_L$ and $m_{N}$, and therefore between $m_{\tilde{\nu}_2}$
and $m_{\tilde{\nu}_1}$, tends to be very large. However $\tilde{\nu}_1$ with
mass of about $63$ GeV allows the rest of the mass spectrum to have low mass values as well. For instance $m_L$ does not need to be at about $\mathcal{O}(1)$ TeV to make the $m_{\tilde{\nu}_1}$ to be right enough to be a good dark matter candidate.

On the other hand, since $\tilde{\nu}_1$ and $\tilde{\nu}_2$ are almost pure
right- and left-handed states respectively, charginos and neutralinos couple very weakly to $\tilde{\nu}_1$ and prefer therefore to decay to
$\tilde{\nu}_2$ rather than to the lightest stau or $\tilde{\nu}_1$, as shown in table~\ref{tab:BR3} for the benchmark $B3$ as an example. Let us focus in the process
\begin{eqnarray}
\label{eq:multilepton}
p p &\rightarrow& \tilde{\chi}_1^+\ \tilde{\chi}_2^0 \\ \nonumber
&\rightarrow& (\ l^+\ \tilde{\nu}_2^l\ )\ (\ \nu^{l'}\ \tilde{\nu}_2^{l'}\ )\\
\nonumber
&\rightarrow& (\ l^+\ \nu^l\ \tilde{\chi}_1^0\ )\ (\ \nu^{l'}\  \nu^{l'}\
\tilde{\chi}_1^0\ )\\ \nonumber
&\rightarrow& (\ l^+\ \nu^l\ \tau^\pm\ \tilde{\tau}_1^\mp\ )\ (\ \nu^{l'}\  \nu^{l'}\
\tau^\pm\ \tilde{\tau}_1^\mp\ )\\ \nonumber
&\rightarrow& (\ l^+\ \nu^l\ \tau^\pm\ W^\mp\ \tilde{\nu}_1^\tau\ )\ (\ \nu^{l'}\  \nu^{l'}\
\tau^\pm\ \ W^\mp\ \tilde{\nu}_1^\tau\ ). \nonumber
\end{eqnarray}
\begin{figure}[t]
  \centering
  \includegraphics[width=0.5\linewidth]{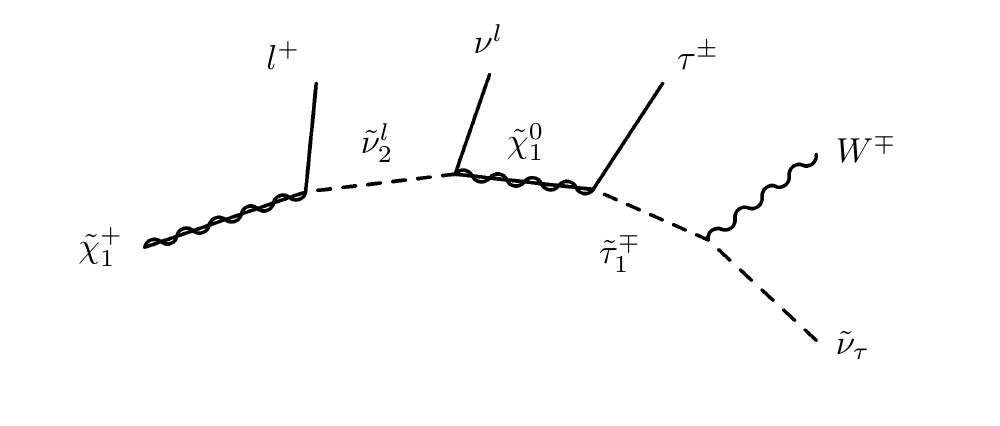} \hspace{-0.5cm}%
  \includegraphics[width=0.5\linewidth]{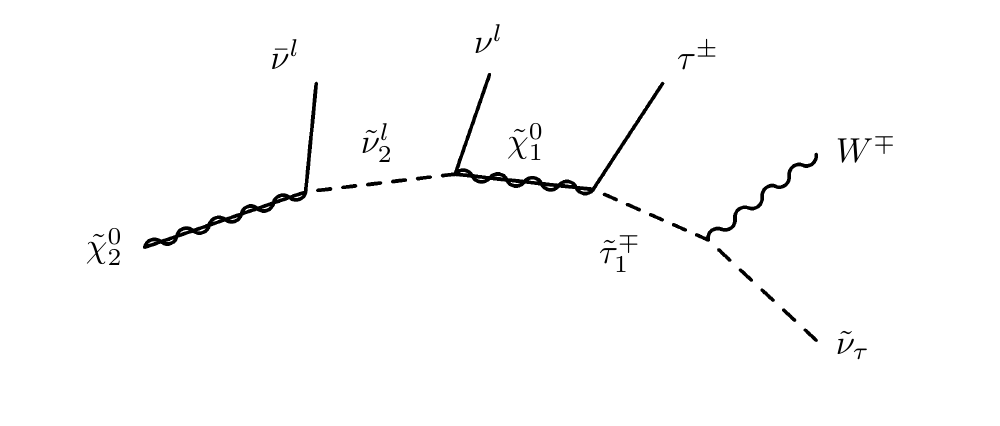}
  \vspace{-0.5cm}
  \label{fig:feyn-multileptons}
  \caption{Typical chargino and neutralino decay chains for the Higgs resonance region.}
\end{figure}
Notice that we consider the decay of $\tilde{\nu}_2$ to $\tilde{\chi}_1^0\ \nu$, which is shown in figure~\ref{fig:feyn-multileptons}. There is however another possibility, $\tilde{\nu}_2$ decaying to Higgs and $\tilde{\nu}_1$. This possibility is interesting since the coupling $h-\snu-\snu$ is also the relevant one in the sneutrino
annihilation in the early Universe. The study of this second possibility goes
beyond the scope of this work, since it will not give rise to a distinguishable leptonic signature. 

The process shown in~\ref{eq:multilepton} contains two $W$ bosons at the end of the
decay chain. Considering the $W$ leptonic decay to electrons and muons, the final
state is given by three leptons not correlated in sign and flavor and two taus.  To
study this signature in more detail we use benchmark $B3$ described in
table~\ref{tab:benchmark}, with relevant masses and couplings
\begin{eqnarray}
  \label{eq:bench3}
  m_{\tilde{\chi}_1^{\pm}} = 781.1\ \mathrm{GeV},&&
  m_{\tilde{\chi}_2^{0}} = 780.02\ \mathrm{GeV},\\ \nonumber
  m_{\tilde{\nu}_2^{l(\tau)}}=671.1(647.3)\ \mathrm{GeV},&&
  \sin{\theta_{\tilde{\nu}^{l(\tau)}}} = 0.007\,,\\ \nonumber
  m_{\tilde{\tau}_1}=240.3\ \mathrm{GeV},&&
  \sin{\theta_{\tilde{\tau}}}= -0.09\,.
\end{eqnarray}
and branching ratios summarized in table~\ref{tab:BR3}.
\begin{table}[t!]
\caption{Relevant branching ratios for decays in benchmark $B3$ for the three uncorrelated lepton signature.}
  \centering
  \begin{tabular}[h]{r c l | c |  r c l| c}
    \hline
       &   Process  &  & BR  & &  Process  &  & BR\\
    \hline
    $\tilde{\chi}_1^+ $ & $\rightarrow$ & $e^+\ \tilde{\nu}_2\ $ & $\ 15\%$ &
    $\tilde{\chi}_2^0 $ & $\rightarrow$ & $\nu\ \tilde{\nu_2}\ $ & $\ 48\% $\\ 
    & & $ \mu^+ \tilde{\nu}_2\ $ & $\ 15\%$  & && $\tilde{\l}_L\
    l\ $ & $\ 28\% $   \\ 
    && $ \tau^+ \tilde{\nu}_2\ $ & $\ 21\%$  &&& & \\
    \hline
    $\tilde{\chi}_1^0 $ & $\rightarrow$ & $\tau^+ \tilde{\tau}^-_1$ & $\ 90\%$ &
     $\tilde{\nu}_2 $ & $\rightarrow$ & $\tilde{\chi}_1^0\ \nu\ $ & $\ 98\%$ \\
    \hline
    $\tilde{\tau}_1^{\pm} $ & $\rightarrow$ & $W^{\pm}\ \tilde{\nu}_1 $& $\ 100\% $ \\
    \hline
  \end{tabular}
  \label{tab:BR3}
\end{table}

In order to single out the most distinctive signature from the final state in process~\ref{eq:multilepton}, we  require three electrons or muons
but neglect events with opposite sign same flavor leptons. This condition could be relaxed allowing opposite sign same flavor leptons but forbidding the
ones with invariant mass close to the $Z$ boson mass.

For the background we consider $W Z\rightarrow W l^+\ l^-$ and $t\bar{t} W$. As in the previous case, $WZ$ is simulated with MG5 and \texttt{Pythia 8} and $t\ \bar{t} W$ using MG5 and \texttt{Pythia 6}.

Selected events are required to pass the following cuts
\begin{enumerate}
\item Three leptons with $p_T>20$ GeV and $\eta<2.5$;
\item Events with opposite sign same flavor (OSSF) leptons are forbidden;
\item When OSSF events are allowed we require them to have
  $|M_\mathrm{inv}-M_{Z}| < 10$ GeV;
\item At least one lepton with $p_T > 25$ GeV;
\item $E_T^{\mathrm{miss}}>100$ GeV.
\end{enumerate}
\begin{figure}[t]
  \centering
  \includegraphics[width=0.45\linewidth]{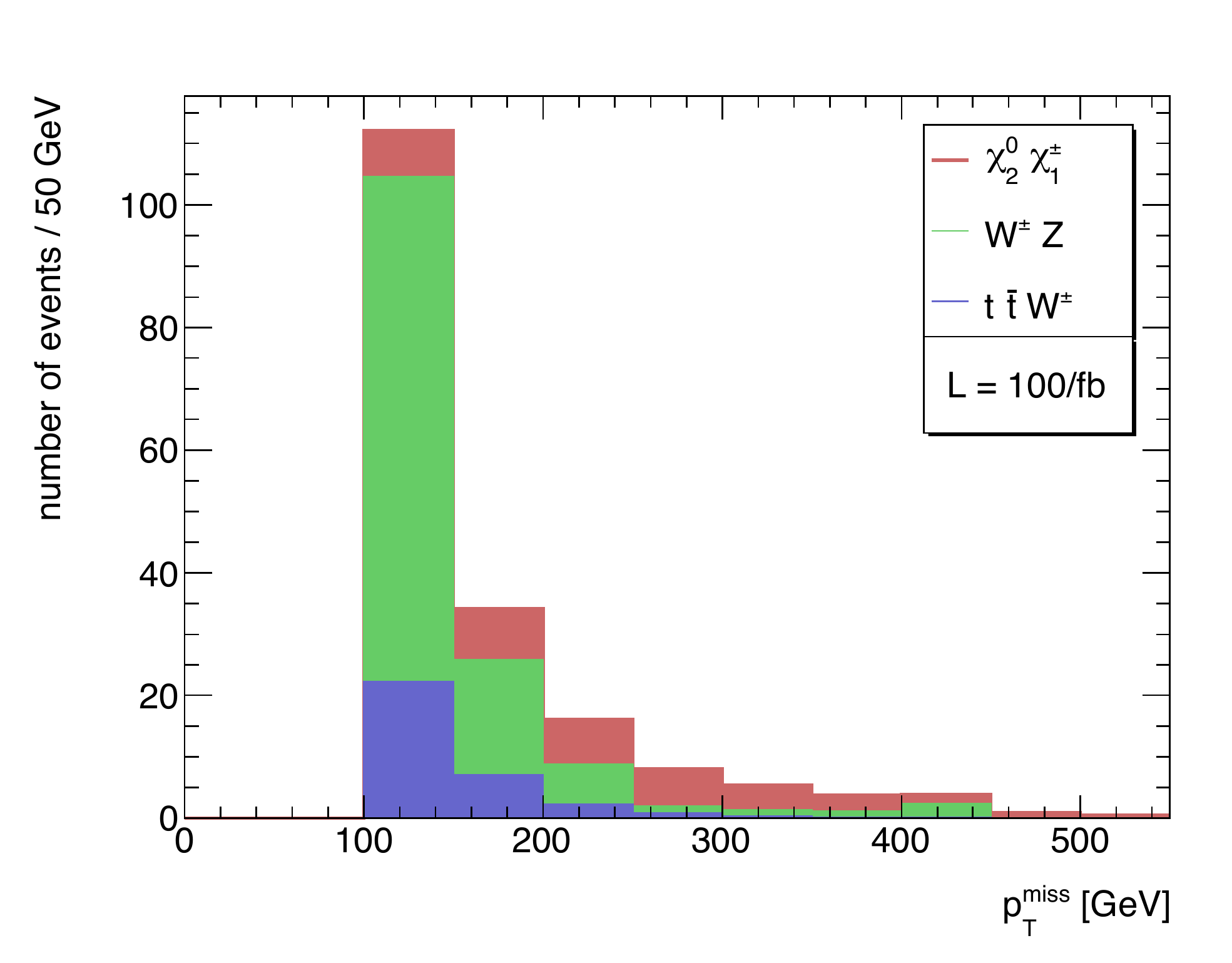}
  \includegraphics[width=0.45\linewidth]{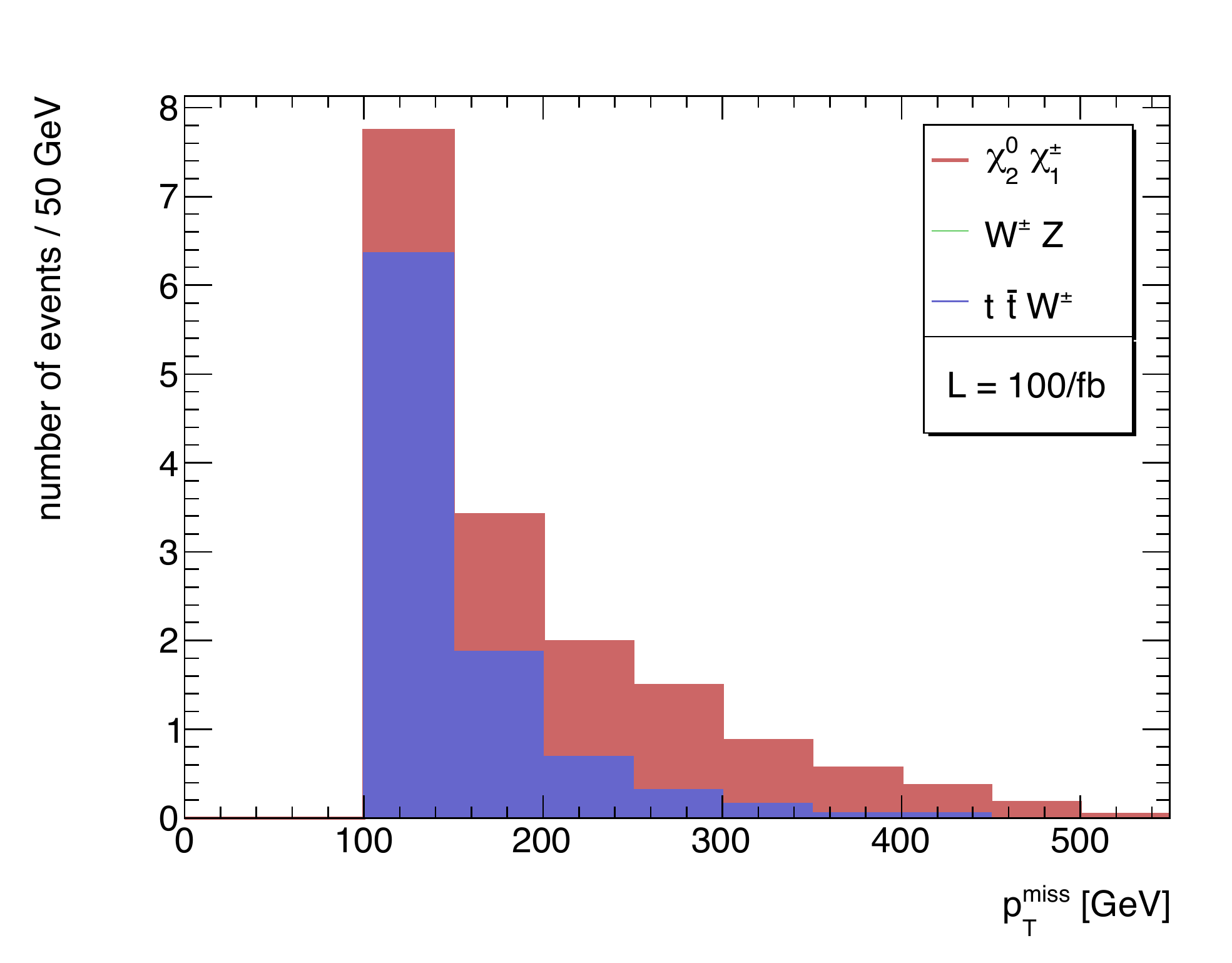}\\
    \includegraphics[width=0.45\linewidth]{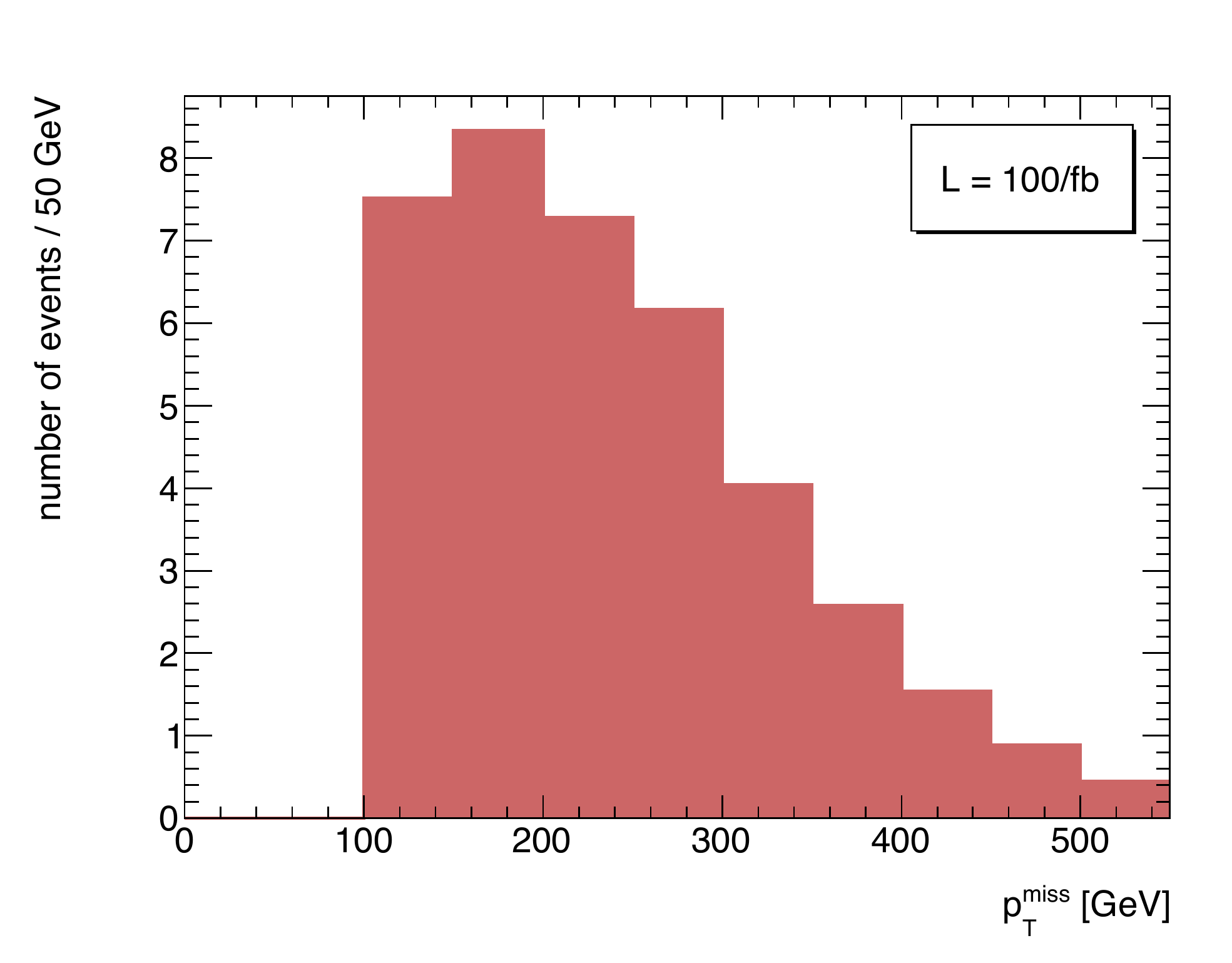}
  \includegraphics[width=0.45\linewidth]{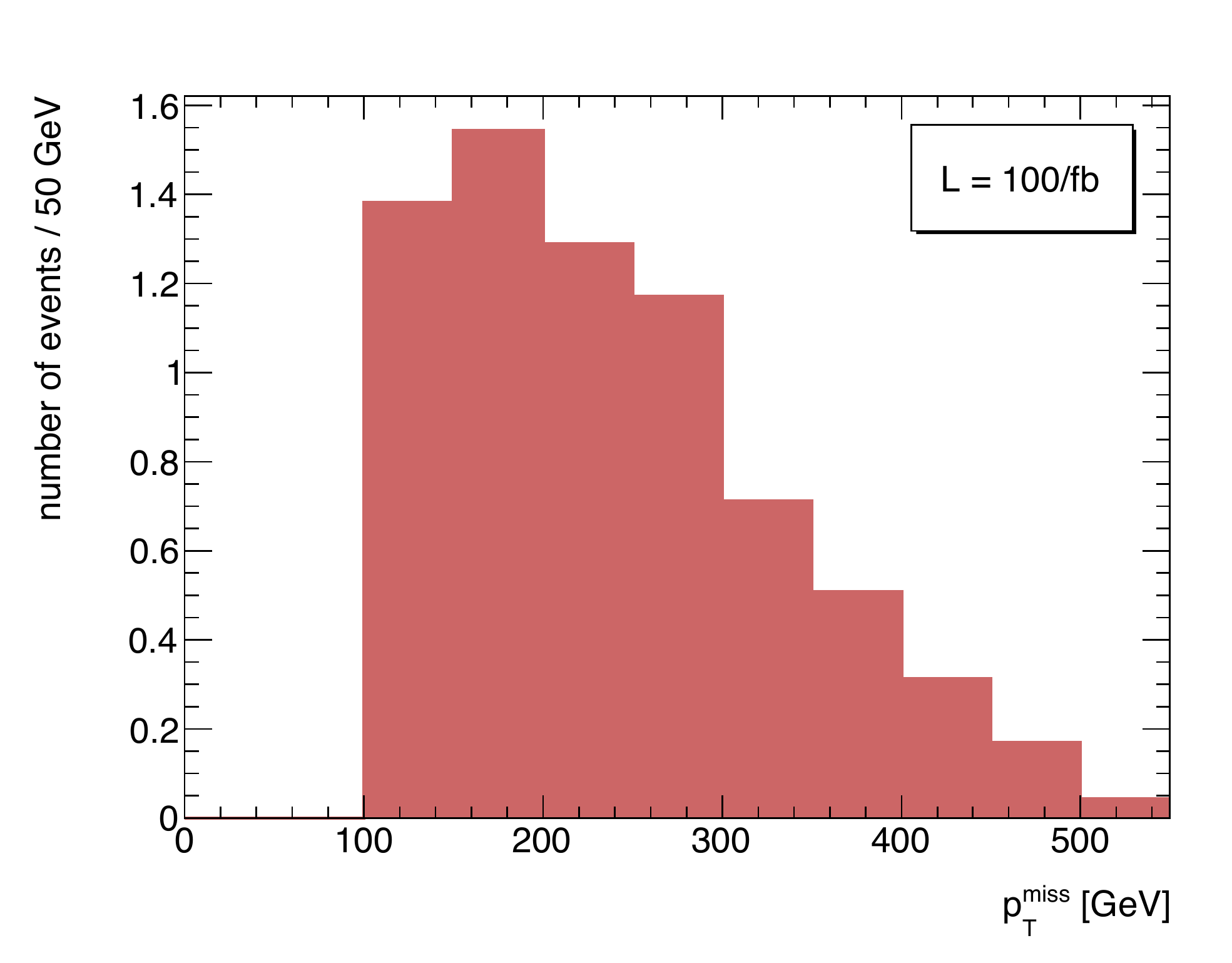}\\
  \includegraphics[width=0.45\linewidth]{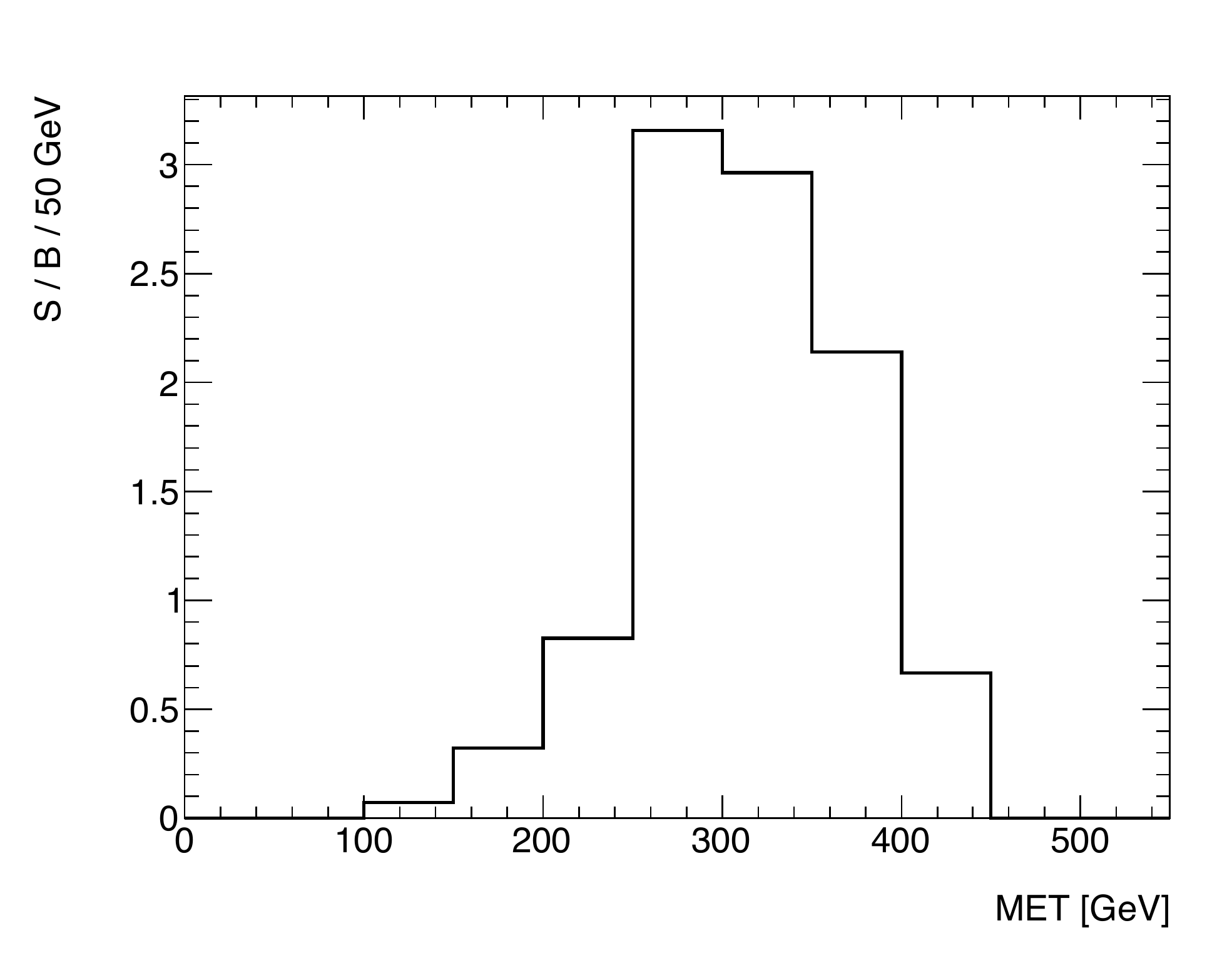}
  \includegraphics[width=0.45\linewidth]{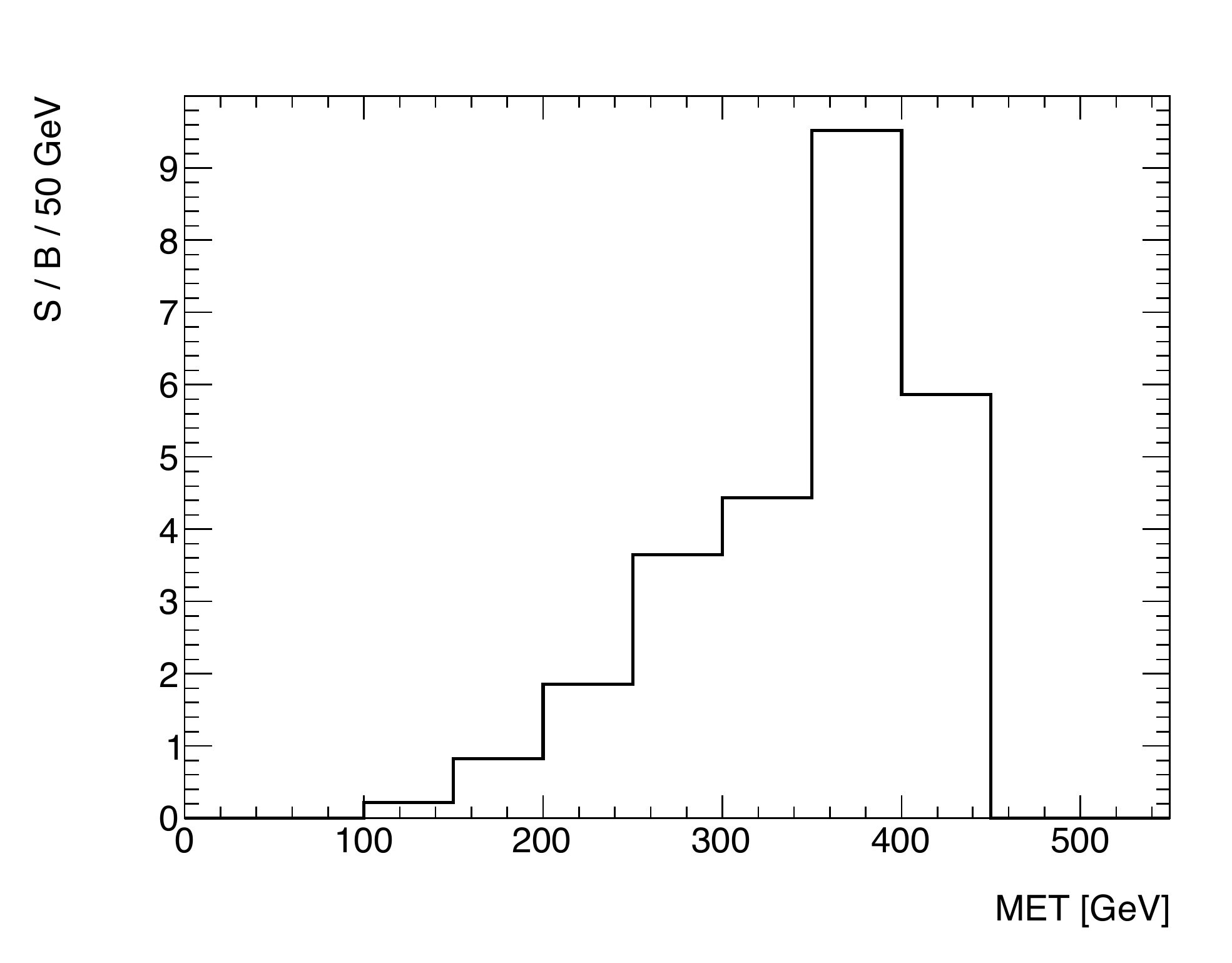}
  \label{fig:multileptons}
  \caption{{\it Top left:} Missing transverse momentum
        distribution of the signal allowing opposite sign same flavor
        leptons. {\it Top right:} Missing transverse momentum distribution of
        the signal neglecting events with opposite sign same flavor
        leptons. {\it Center (Bottom) left and right:} Same as top but for the signal distribution only (signal over background ratio).}
\end{figure}
Figure \ref{fig:multileptons} shows the missing
  transverse momentum distribution in the case when we allow (left) and forbid
  (right) OSSF leptons. As one can see, the ratio between signal and
  background is remarkable good (lower panels). Indeed the background for three uncorrelated leptons is very small and the signal stands well above it and is in the full reach of LHC at 14 TeV.

\subsection{Direct chargino production}
\label{sec:ddp}

In the previous subsection we have shown that the sneutrino as LSP and the sleptons as NLSPs could have quite
particular signatures of leptons without correlation of sign and flavor. However there is a significant region in the parameter space where sleptons
are heavier than some of the neutralinos and charginos (see figure~\ref{fig:sleptonMasses}). These regions have more
``traditional'' signatures but still exhibit some particularities with respect to
the MSSM. Direct chargino production could be a window to access these
regions. The difference between the MSSM+RN with respect to the MSSM is that the chargino
decay chain could be dominantly into two-body ($\tilde{\chi}_1^\pm \rightarrow l^\pm
\tilde{\nu}_l$) instead of three-body ($\tilde{\chi}_1^\pm \rightarrow W^\pm
\tilde{\chi}_1^0 \rightarrow f'\ \bar{f} \tilde{\chi}_1^0$), producing a
sharper distribution in the signal.
\begin{figure}[t]
  \centering
  \includegraphics[width=0.5\linewidth]{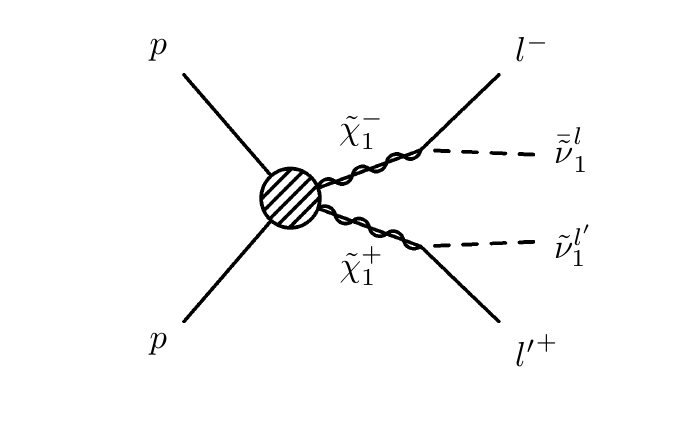}
  \vspace{-0.5cm}
  \caption{Chargino production relevant for the two uncorrelated lepton signal discussed in section~\ref{sec:ddp}.}
  \label{fig:CCproduction}
\end{figure}

We focus on direct chargino production, depicted in figure~\ref{fig:CCproduction}, with a final state of two leptons and
missing transverse momentum
\begin{eqnarray}
p p \rightarrow \tilde{\chi}_1^+\ \tilde{\chi}_1^- \rightarrow l^+\ {l'}^-\ \tilde{\nu}_1^{l}\ \tilde{\nu}_{1}^{l'}\,.
\end{eqnarray}
This signal (two opposite sign leptons and two invisible particles ) also
exists in the MSSM, arising from direct production of slepton
($\tilde{l}^{\pm}\rightarrow l^{\pm}\ \tilde{\chi}_1^0$) but with smaller
production cross-section. This search will access a large portion of the
parameter space, specially if the lightest neutralino is higgsino or wino and
therefore quasi-degenerated with the chargino, but also in most
of the cases of bino-like lightest neutralino region. In order to get good
efficiency in these searches we require a large enough splitting in mass
between the $\tilde{\chi}_1^0$ and the $\tilde{\nu}_1$ to be able to detect
the decay of the chargino. This condition is recovered in the region where the
sneutrino annihilates efficiently through the $Z$ boson to satisfy dark matter
constraints (gray points) and in the region where $\tilde{\nu}_1$ is
degenerated with the neutralino but not with $\tilde{\chi}_1^\pm$ (magenta
points).

For the study of this signature we consider the benchmark $B4$ with parameters
at EW scale described in table~\ref{tab:benchmark}. The chargino is mostly
wino and the relevant masses and mixing angles are given by
\begin{eqnarray}
  \label{eq:bench2}
  m_{\tilde{\chi}_1^{\pm}} = 440.8\ \mathrm{GeV},\
  m_{\tilde{\nu}_1^{l(\tau)}}=125.6(124.1)\ \mathrm{GeV},\
  \sin{\theta_{\tilde{\nu}_{l(\tau)}}} = 0.038(0.042)\,.
\end{eqnarray}
The branching ratios are shown in table~\ref{table:BRsBench4}: notice that the decay into the LSP has the largest branching ratio, and this is a general situation for most of the sneutrino parameter space we consider (gray points, with sizable left-handed component, see figure~\ref{fig:goodDM}).
\begin{table}[t]
  \caption{Relevant branching ratios for chargino decay in benchmark $B4$ for the signature of two opposite sign leptons and two invisible particles.}
  \centering
  \begin{tabular}[t]{c c c | c}
  \hline
&   Process   & & BR \\
  	\hline
    $\cha_1$ & $\rightarrow$ & $W^+\ \neu_1$ & $\ 18.1\%$ \\
    & & $e^+\ \tilde{\nu}_1^e$ & $\ 25.4\%$ \\
    & & $\mu^+\ \tilde{\nu}_1^\mu$ & $\ 25.4\%$\\
    & & $\tau^+\ \tilde{\nu}_1^\tau$ & $\ 31.1\%$\\
    \hline
  \end{tabular}
  \label{table:BRsBench4}
\end{table}

For background simulation we consider $W^+W^-$ and $W Z$ production at
leading order including also the case where one of the gauge bosons is
off-shell. 

In the analysis we use two kinematical variables, $m_{T2}$~\cite{Barr:2003rg}
\begin{eqnarray}
  \label{eq:mT2}
  m_{T2} = \mathrm{min}_{p_1+p_2=p_T^\mathrm{miss}}\left\{\mathrm{max}[ M_T(p_{l_1},p_1),M_T(p_{l_2},p_2) ]\right\}\,,
\end{eqnarray}
where $l_1$ and $l_2$ correspond to the two leptons we require in this
analysis and $M_T$ is the transverse mass. We also use the effective
transverse energy~\cite{Cabrera:2012cj},
\begin{eqnarray}
  \label{eq:ETeff}
  {\cal E}_T^{\rm eff} = \sqrt{(M_\mathrm{inv}^{ll})^2 + (p_T^{ll})^2} + 2 |{p}_T^{\ \mathrm{miss}}|\ ,
\end{eqnarray}
where $M_\mathrm{inv}^{ll}$ and $p_T^{ll}$ are the invariant mass and
transverse momentum of the two selected leptons. Keep in mind that $m_{T2}$
and $ {\cal E}_T^{\rm eff}$ are expected to have a distribution with a maximum
at the mass of the chargino for $m_{T2}$ and twice this value for $ {\cal
  E}_T^{\rm eff}$.

\begin{figure}[t]
  \centering
  \includegraphics[width=0.49\linewidth]{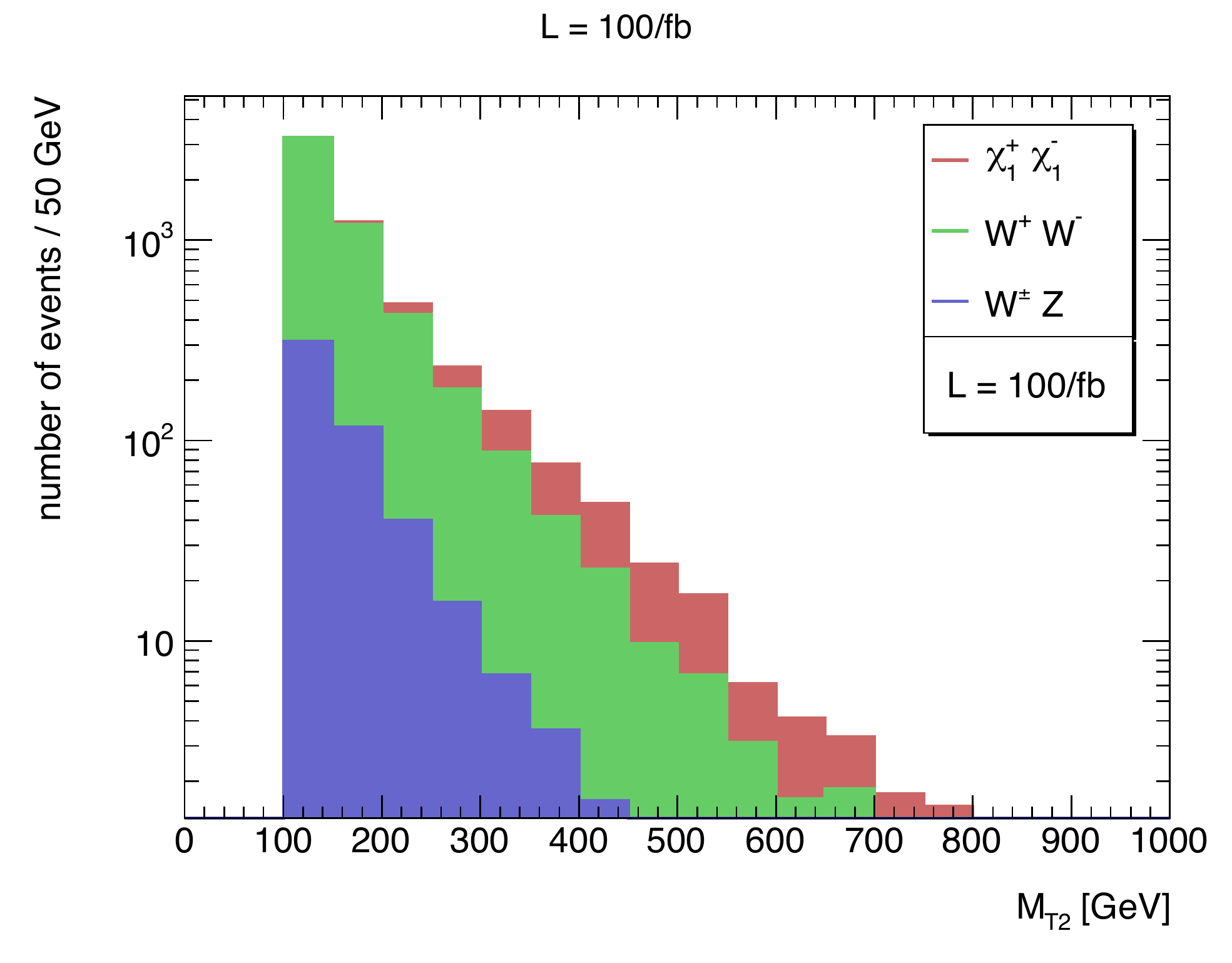}
  \includegraphics[width=0.49\linewidth]{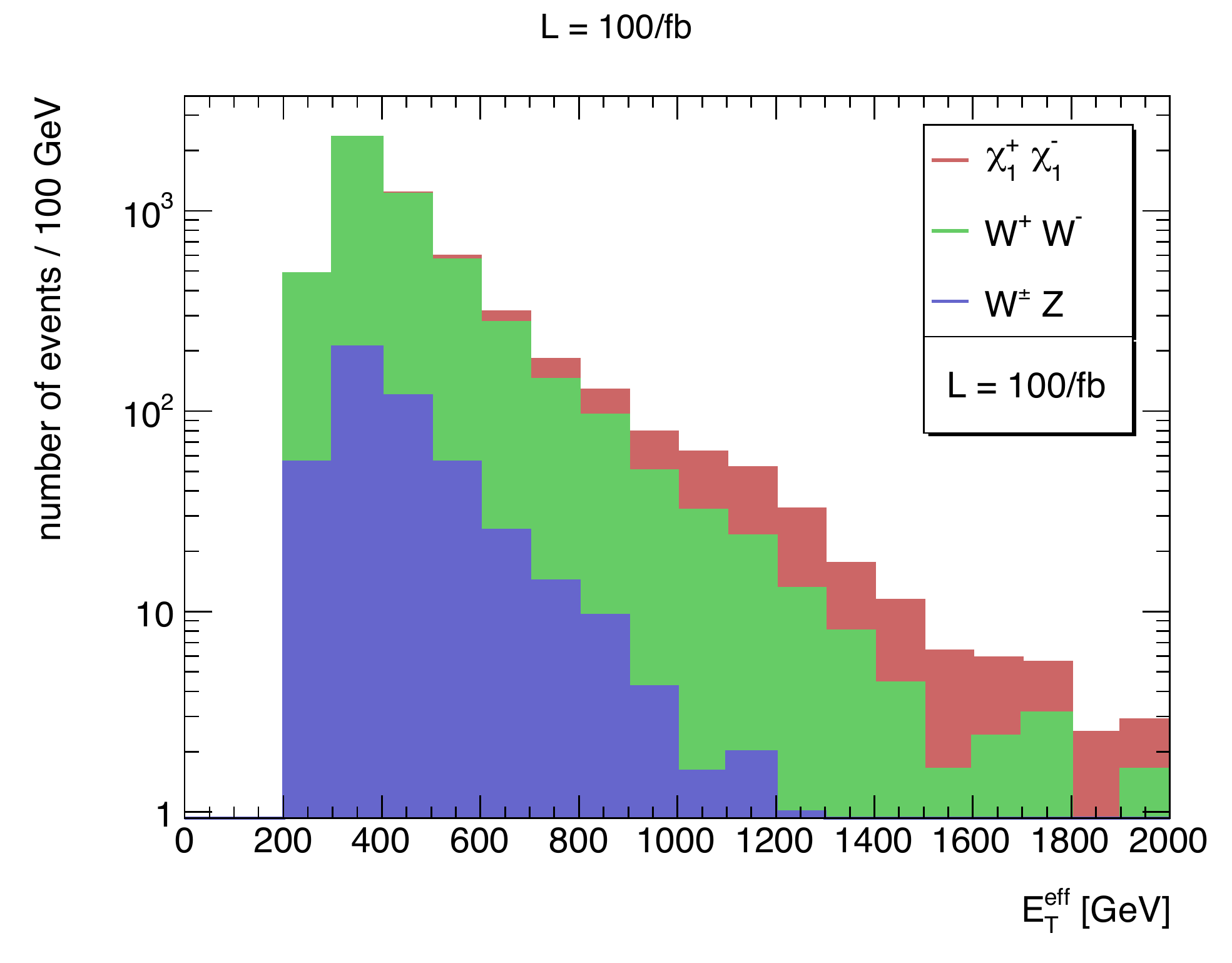}\\
    \includegraphics[width=0.49\linewidth]{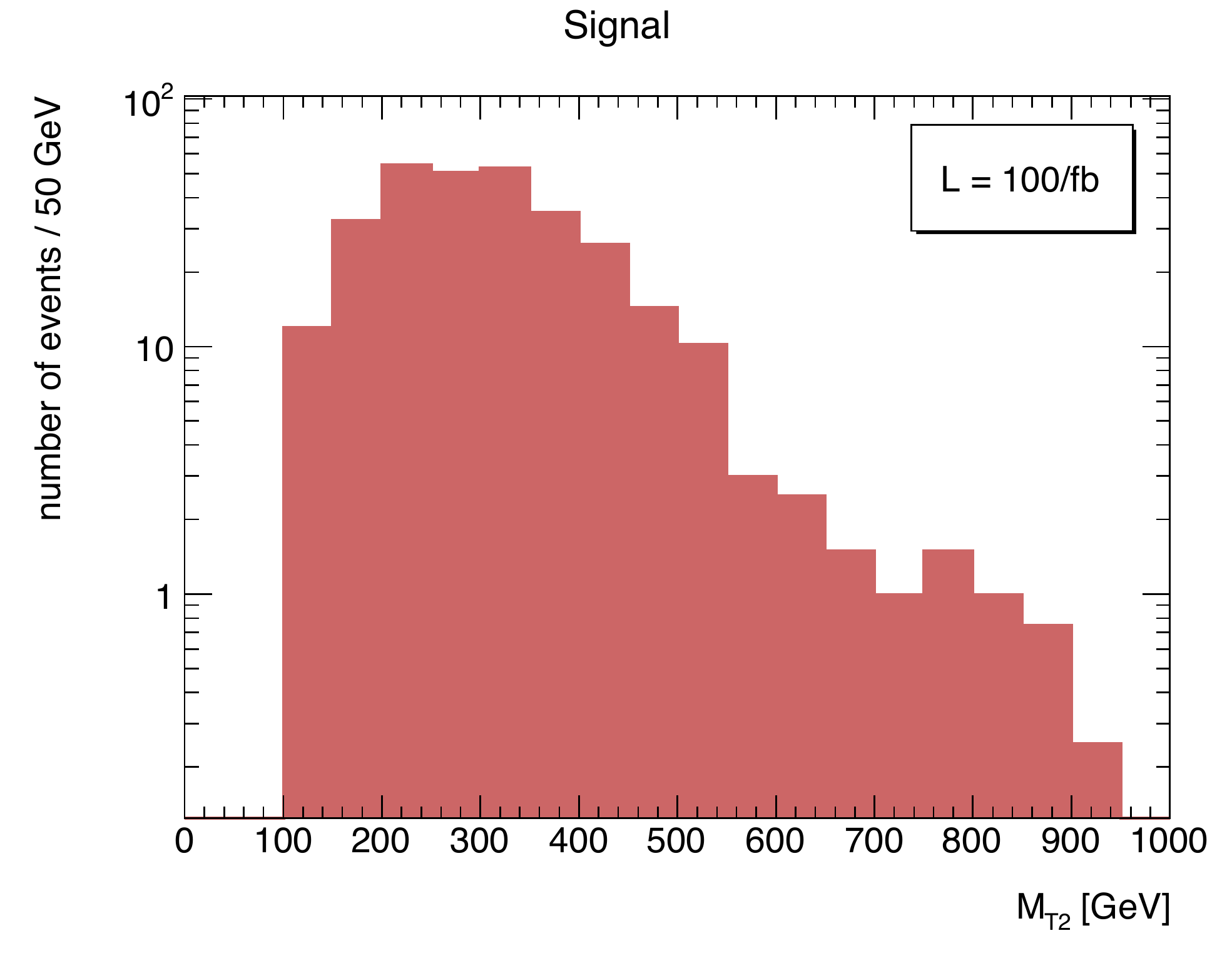}
  \includegraphics[width=0.49\linewidth]{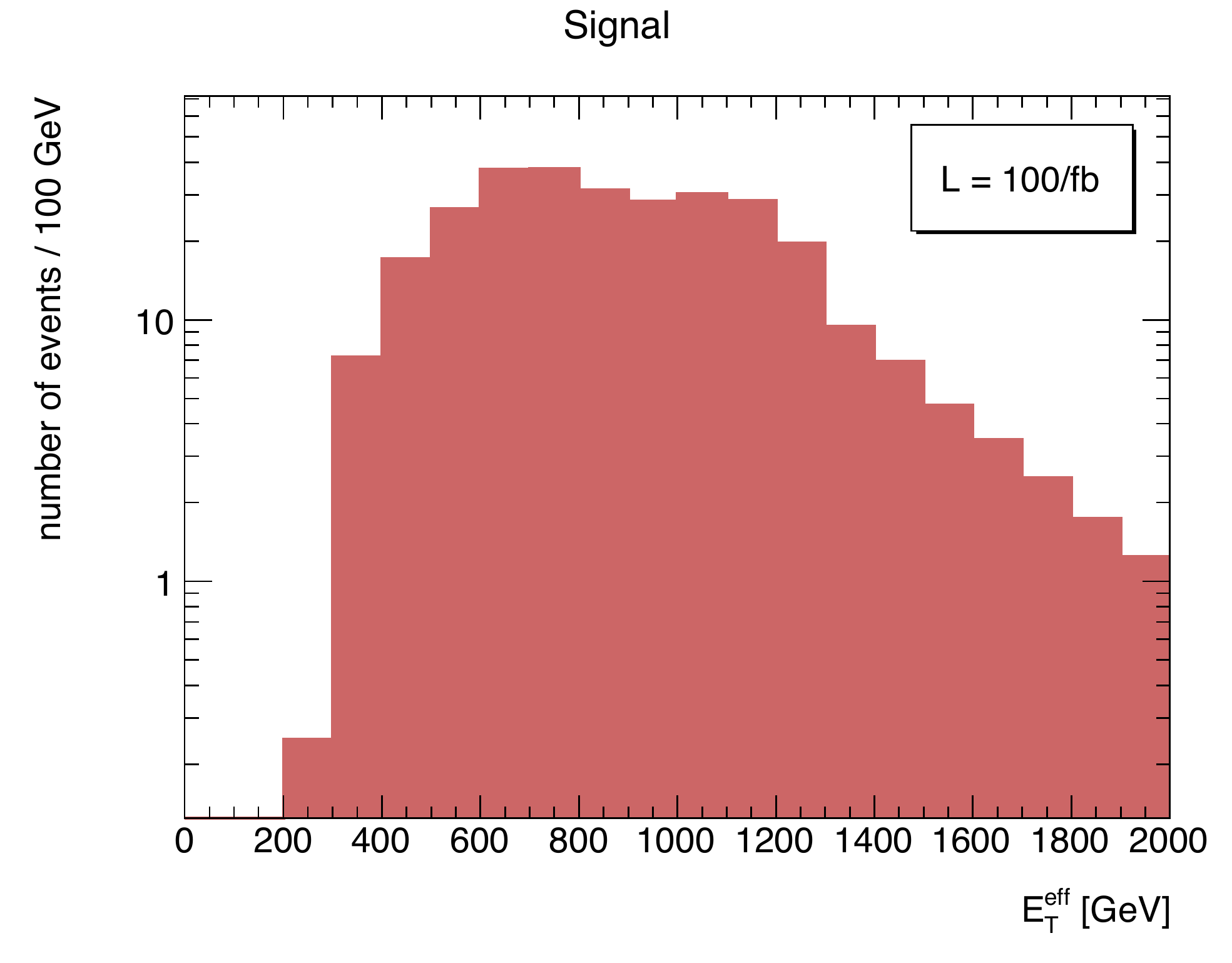}\\
  \includegraphics[width=0.49\linewidth]{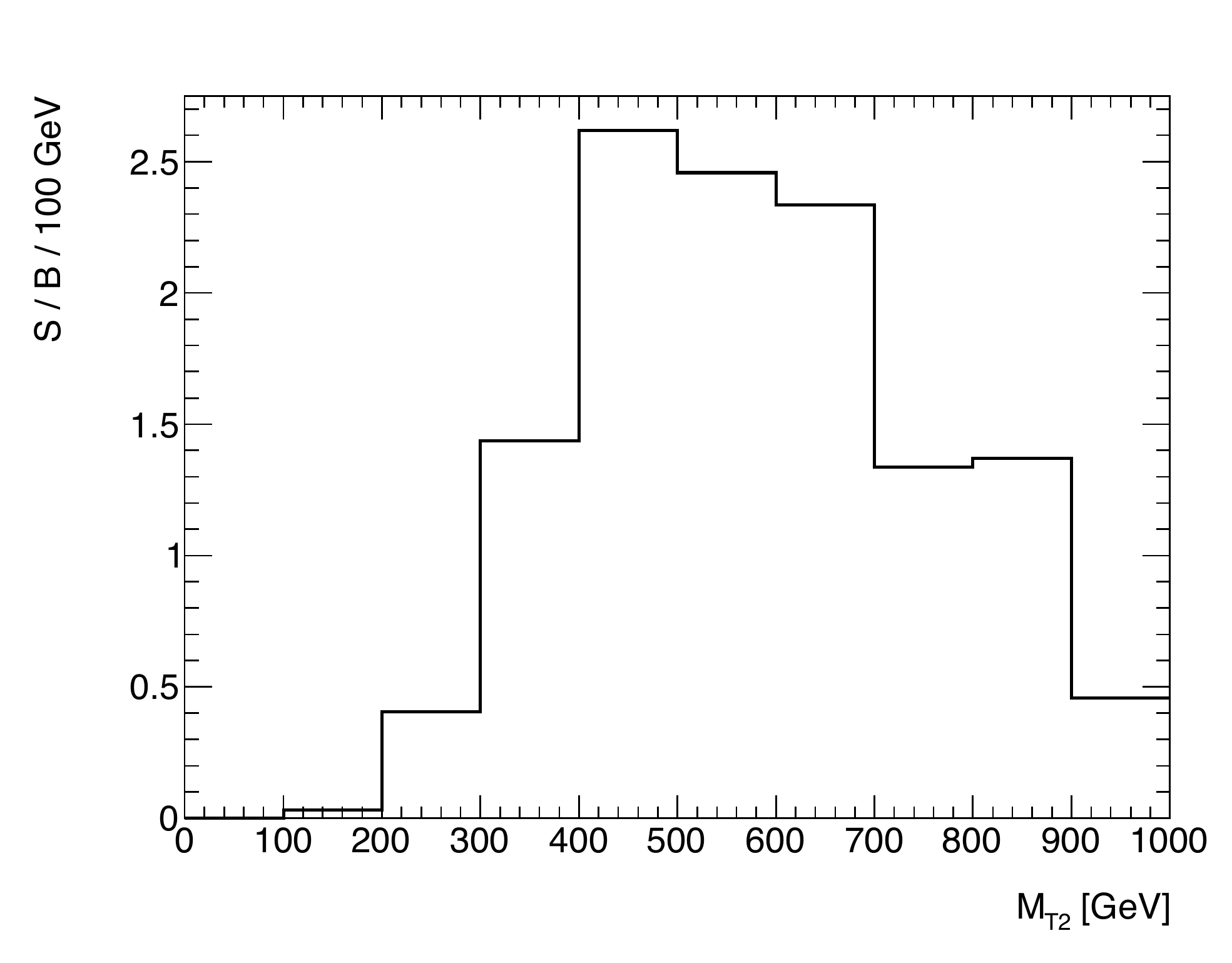}
  \includegraphics[width=0.49\linewidth]{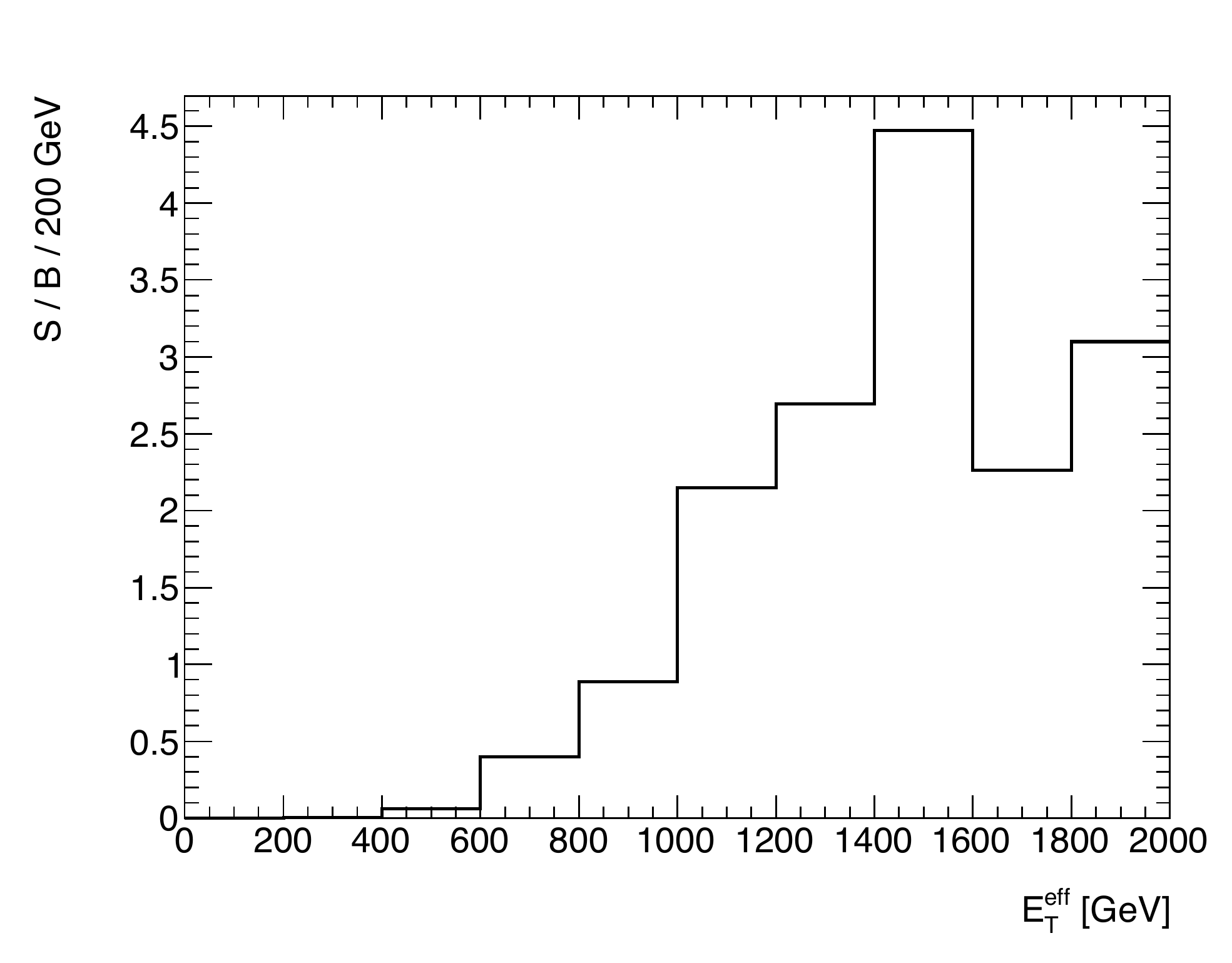}
  \caption{{\it Top left:} The $M_{T2}$ distribution is shown, for chargino
    production to two opposite sign leptons and missing energy. The signal and background are labelled in the figure. {\it Top right:} Same as left for the variable $E_T^{\mathrm{eff}}$. {\it Center (Bottom) left and right:} Same as above for the signal distribution only (signal over background ratio).} 
  \label{fig:C1prod}
\end{figure}

Taking~\cite{TheATLAScollaboration:2013hha} as a reference, we consider the following cuts
\begin{enumerate}
\item Two opposite sign leptons (electrons and muons);
\item $Z$ veto ( $|m_{ll}-m_Z|>10$ GeV );
\item $m_{T2}>110$ GeV;
\item $p_T^\mathrm{miss}>40$ GeV;
\item Second hardest jet with $p_T<50$ GeV.
\end{enumerate}
Figure~\ref{fig:C1prod} shows the signal superposed with the background for
$m_{T2}$ and $E_T^\mathrm{eff}$, central panels show the signal only. Notice that the maximum of the signal
    is located around 400 GeV for $m_{T2}$ (and 800 GeV for
    $E_T^\mathrm{eff}$) where the background has decreased significantly,
    allowing us to disentangle the signal from the background. This is confirmed by the signal over background ratio in the lower panels. In case LHC
    measures this kind of signal it would be possible to estimate the size of
    the supersymmetric masses when combining $m_{T2}$ and
    $E_T^\mathrm{eff}$.

Colored particles communicate with the sneutrinos through neutralinos and charginos, making this signature general and very interesting. In this case however the signal will be associated to multi-jets as well. Hence the gluino and squark production is less promising for this type of search because of the huge expected background.

\section{Discussion on other potential signatures}\label{sec:disc}

\subsection{Multi-leptons from $\tilde{l}_R$ three-body decay}\label{sec:gp}

\begin{figure}[t]
  \centering
  \includegraphics[width=0.5\linewidth]{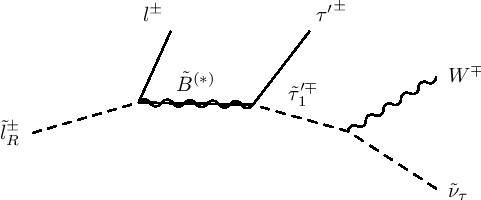}
  \caption{Right-handed slepton decay chain.}
  \label{fig:slR}
\end{figure}

An interesting phenomenological region is denoted by right-handed sleptons lighter than both the left-handed sleptons and the lightest neutralino $\neu_1$. Staus are typically lighter that selectrons and smuons, as discussed before. The right-handed selectrons and smuons will decay through three-body
\begin{eqnarray}
  \label{eq:slR}
  \tilde{l}_R &\rightarrow& l \ \nu_{l'}\ \tilde{\nu}_{l'}\,,\\ \nonumber
  \tilde{l}_R &\rightarrow& l \ \tau\ \tilde{\tau}_1\,.
\end{eqnarray}
The second case of~\ref{eq:slR}, where $\tilde{l}_R$ decays to $\tilde{\tau}_1$, is a particular signature, as the final
state will lead to three uncorrelated leptons in flavor, depicted in figure~\ref{fig:slR}.  We do not simulate this signature since in our data sample right-handed sleptons lighter that neutralinos
tend to be heavier than 700 GeV (gray points). Hence this signal will be suppressed because the cross-section for direct
production falls down steeply for increasing slepton mass and is beyond LHC reach. However we point out that this could be a very
interesting signature when associated with production of colored particles
decaying to neutralino and the neutralino consequently into slepton right. 

\subsection{Same signatures as the MSSM}\label{sec:4l}

Until now we have discussed LHC signatures arising from the Higgs resonance region, from the $Z$ boson region and when neutralinos (mostly bino like) and sneutrinos are degenerate. The blue region (see figure~\ref{fig:goodDM}) is characterized by lightest neutralinos and charginos which are either wino or higgsino and are degenerate with the sneutrino LSP. Hence the typical configuration for the SUSY mass spectrum is essentially MSSM like, the only exception being that the LSP is now the $\snu_1$. In this case we do not expect signatures that differ significantly from the MSSM predictions. If the decay chain ends up with a neutralino, the process $\neu_1 \to \snu_1 \nu^{\tau}$ will be completely invisible, hence the nature of the dark matter can not be determined.

Another possibility to disentangle the MSSM and MSSM+RN scenarios is given by the chargino decay. There are scenarios in which the $\neu_1$ and $\cha$ are degenerate, due to their composition, and the splitting in mass can be smaller than few \%, such that charginos can be long-lived particles, {\it e.g.} in MSSM frameworks as anomaly mediated~\cite{ATLAS:2012ab}. On the other hand, when the sneutrino is the dark matter, it is enough for relic density constraint to have charginos within 10\% in mass degenerate with the sneutrino, so typically the charginos will not be long-lived.

\section{Conclusions}\label{sec:concl}

In this paper we have investigated distinctive leptonic signatures at colliders of the simplest extension of the MSSM in which neutrinos have Dirac masses (MSSM+RN), motivated by the fact that neutrino are massive.
The connection with neutrino masses has a significant impact on the nature of the LSP and dark matter candidate. With the addition of right-handed neutrino superfield, the phenomenology of the scalar neutrino is modified as well: the left-handed component and right-handed component can substantially mix because of large trilinear scalar coupling $A_{\snu}$ (which are not related to the small neutrino Yukawa coupling) and becomes the dark matter candidate.

Assuming the SUSY parameters unified at GUT scale, we revise the status of sneutrino dark matter, finding that it is a viable candidate for masses above the Higgs pole, and that a large portion of the parameter space is compatible with the exclusion limit of LUX and can be probed by the future direct detection experiment XENON1T. 
We have found that there is a correlation between the annihilation channels that fix the LSP relic density and the signatures at LHC. In some regions, as the Higgs pole or when bino-neutralino and sneutrino are degenerate, the sleptons might be lighter than the electroweak fermions, leading to interesting features. The most promising signatures of sneutrino dark matter are (i) decay into two leptons of opposite sign but uncorrelated flavor and (ii) three uncorrelated leptons, which have a negligible standard model background. A higher number of leptons ($\sim 4l$) in the final state is also expected, even though such signature is suppressed by the long decay chain.
 Simulated Monte Carlo events for both signals and backgrounds have been used to assess the experimental sensitivity to specific benchmark points, representative of generic configurations arising in the MSSM+RN. We have pointed out that the signal is in the reach of LHC at 14 TeV of center of mass energy and $100/\rm fb$ of luminosity. Interestingly, some configurations of the MSSM+RN parameter space lead to long-lived staus, which can be detected by next LHC run and can as well provide a hint on the nature of the dark matter, if ever detected. Indeed the life-time of the $\stau$ in the MSSM and MSSM+RN has a different behavior as a function of the mass splitting with the LSP.

The anomalous production of events with 4 leptons recently observed by the CMS collaboration~\cite{CMS:2013jfa} has started to increase the interest towards multi-lepton signatures~\cite{D'Hondt:2013ula}. In this paper we have proposed interesting leptonic signatures from a motivated extension of the MSSM, which can be probed by future LHC run with the appropriate search strategies and by future astroparticle experiments in a complementary way.

\acknowledgments
The authors gratefully thank D. Cerdeno and V. Martin-Lozano for useful discussions and P. Artoisenet for comments on technical details of the \texttt{MadGraph5} software. CA and MEC acknowledge the support of the European Research Council through the ERC starting grant {\it WIMPs Kairos}, PI G. Bertone. CA acknowledges the support of the ERC project 267117 (DARK) hosted by Universit\'e Pierre et Marie Curie - Paris 6, PI J. Silk.

\appendix
\section{Sensitivity analysis to the choice of priors in the MSSM+RN}\label{sec:appA}
Below we discuss the dependence on the prior choice of the dark matter phenomenology. In the analysis we use flat priors for the slepton parameters and for the gaugino masses. Here we run an additional chain with log prior for the parameters $M_1,M_2,m_N,m_L,m_R,A_L$ and $A_{\snu}$. The prior range is the same as reported in table~\ref{tab:priors}.

\begin{figure}[t]
\begin{minipage}[t]{0.32\textwidth}
\centering
\includegraphics[width=1.\columnwidth,trim=25mm 15mm 60mm 120mm, clip]{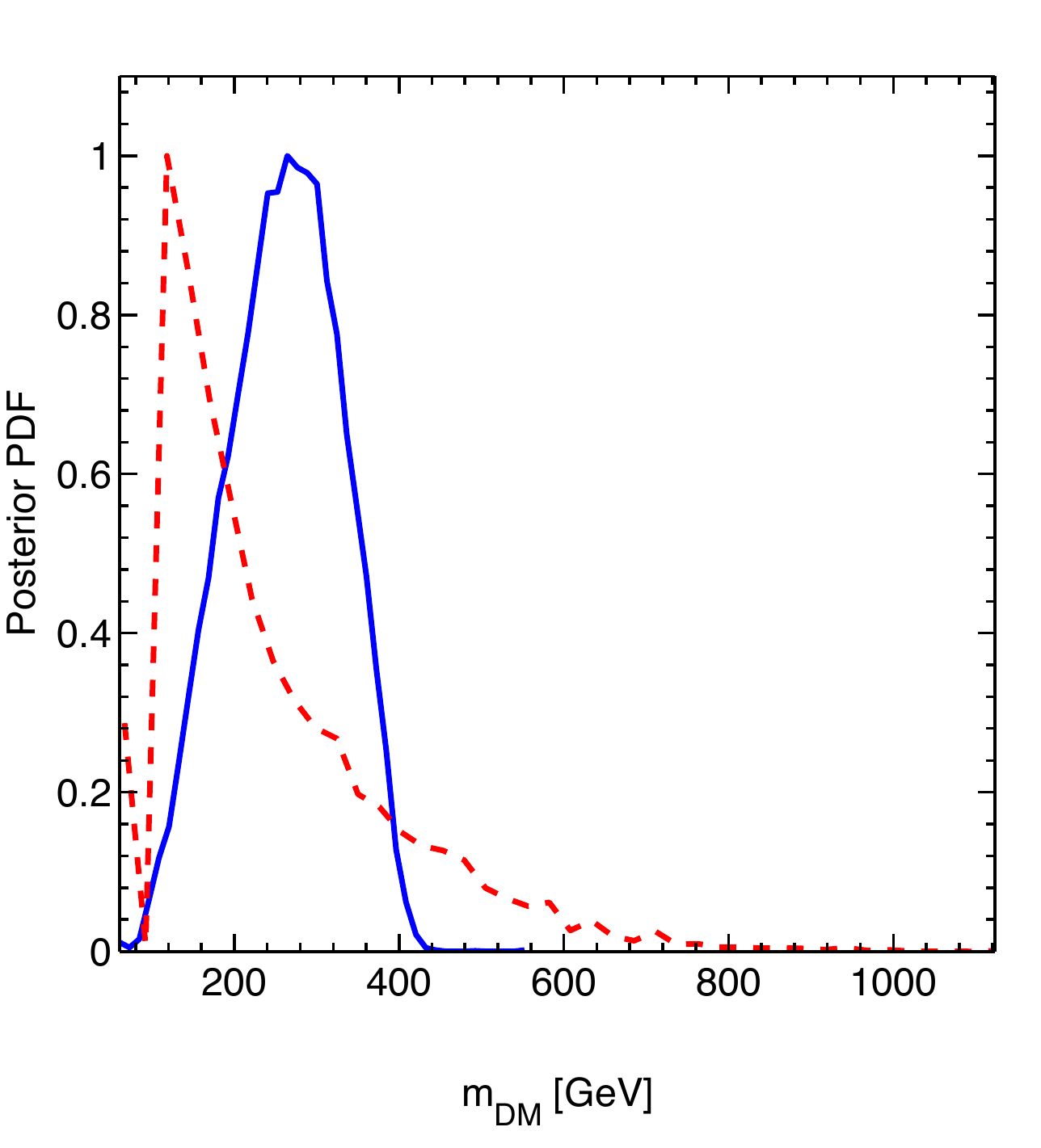}
\end{minipage}
\begin{minipage}[t]{0.32\textwidth}
\includegraphics[width=1.\columnwidth,trim=25mm 15mm 60mm 120mm, clip]{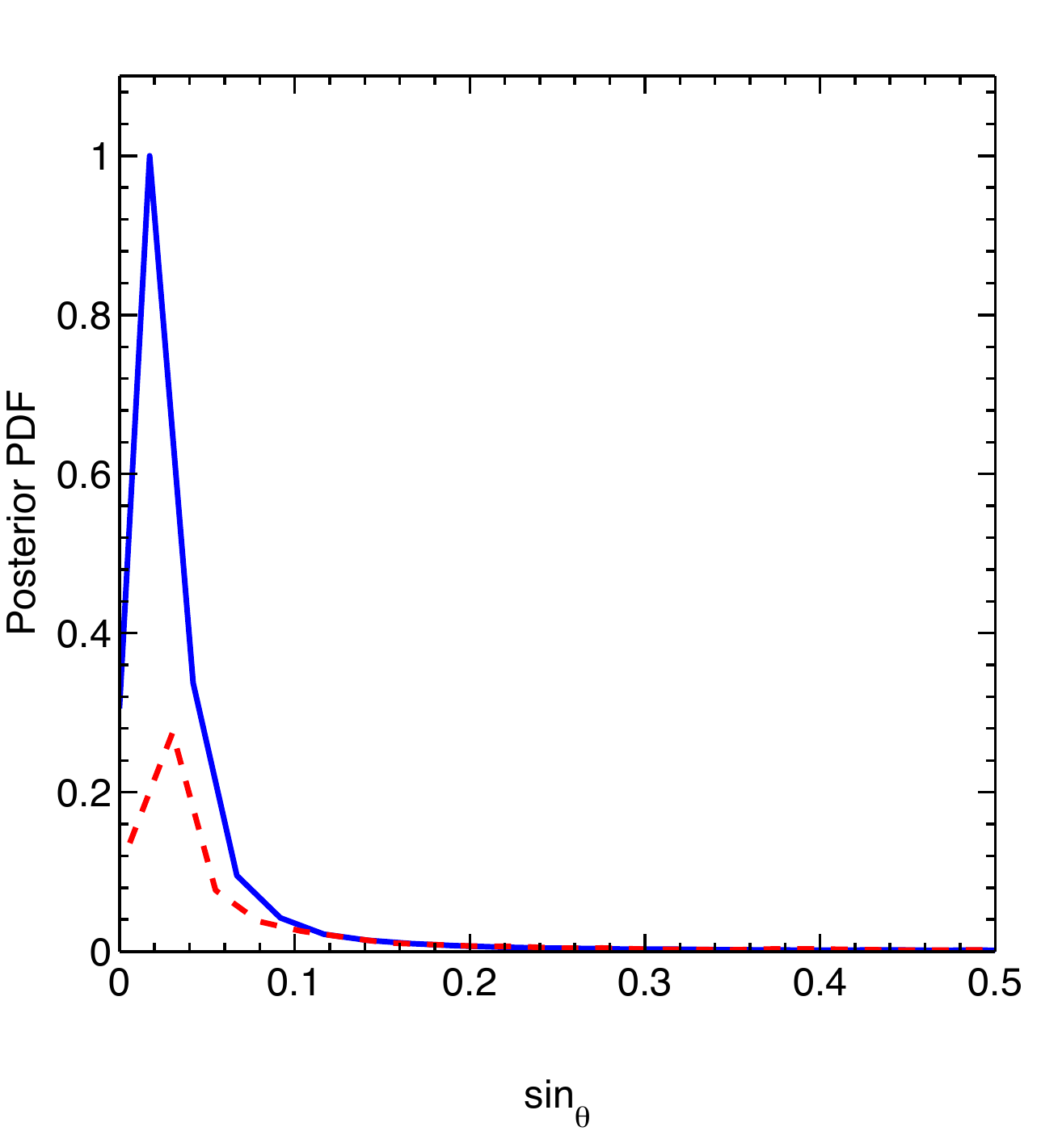}
\end{minipage}
\begin{minipage}[t]{0.32\textwidth}
\centering
\includegraphics[width=1.\columnwidth,trim=25mm 15mm 60mm 120mm, clip]{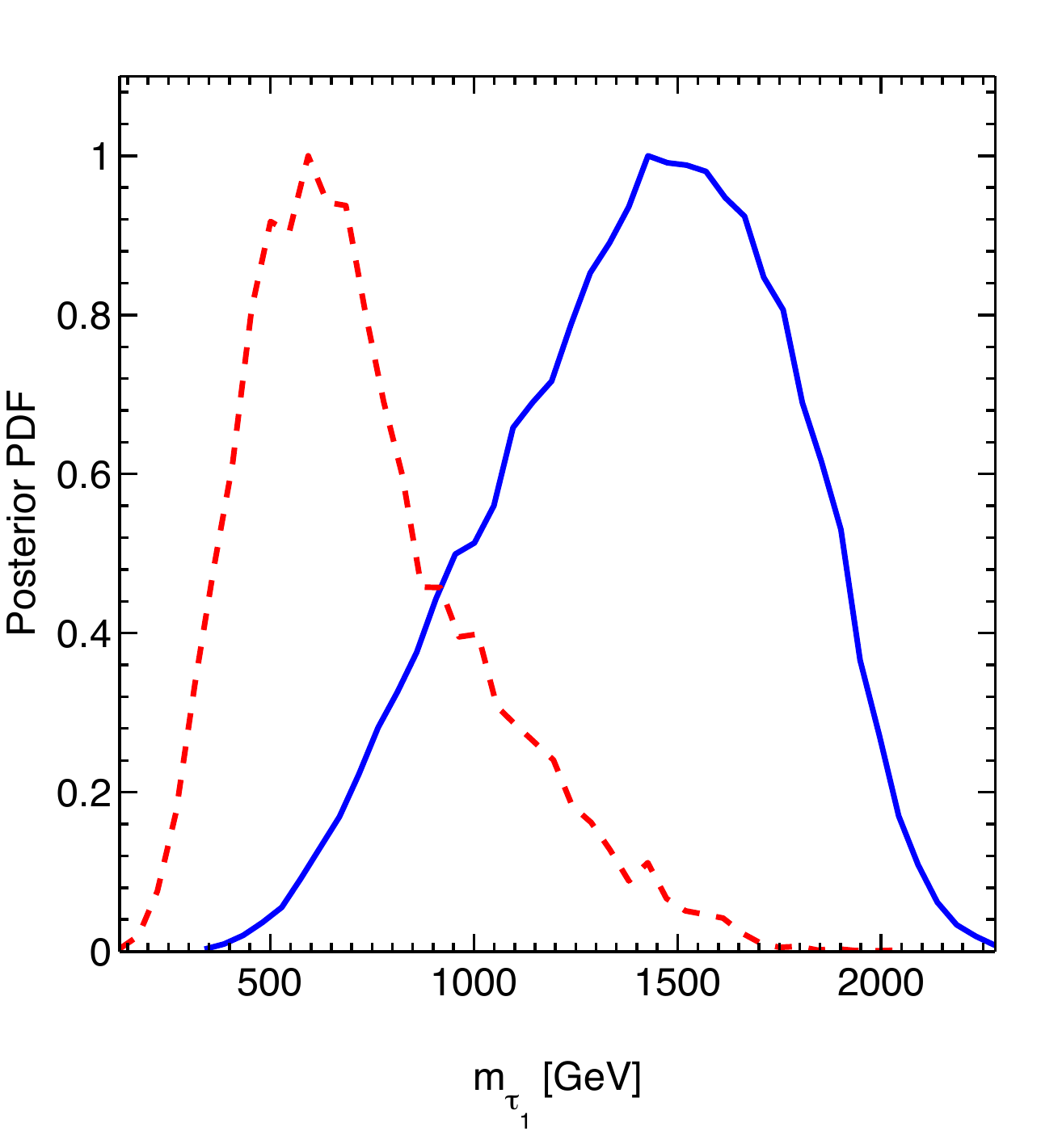}
\end{minipage}
\\
\begin{minipage}[t]{0.32\textwidth}
\centering
\includegraphics[width=1.\columnwidth,trim=25mm 15mm 60mm 120mm, clip]{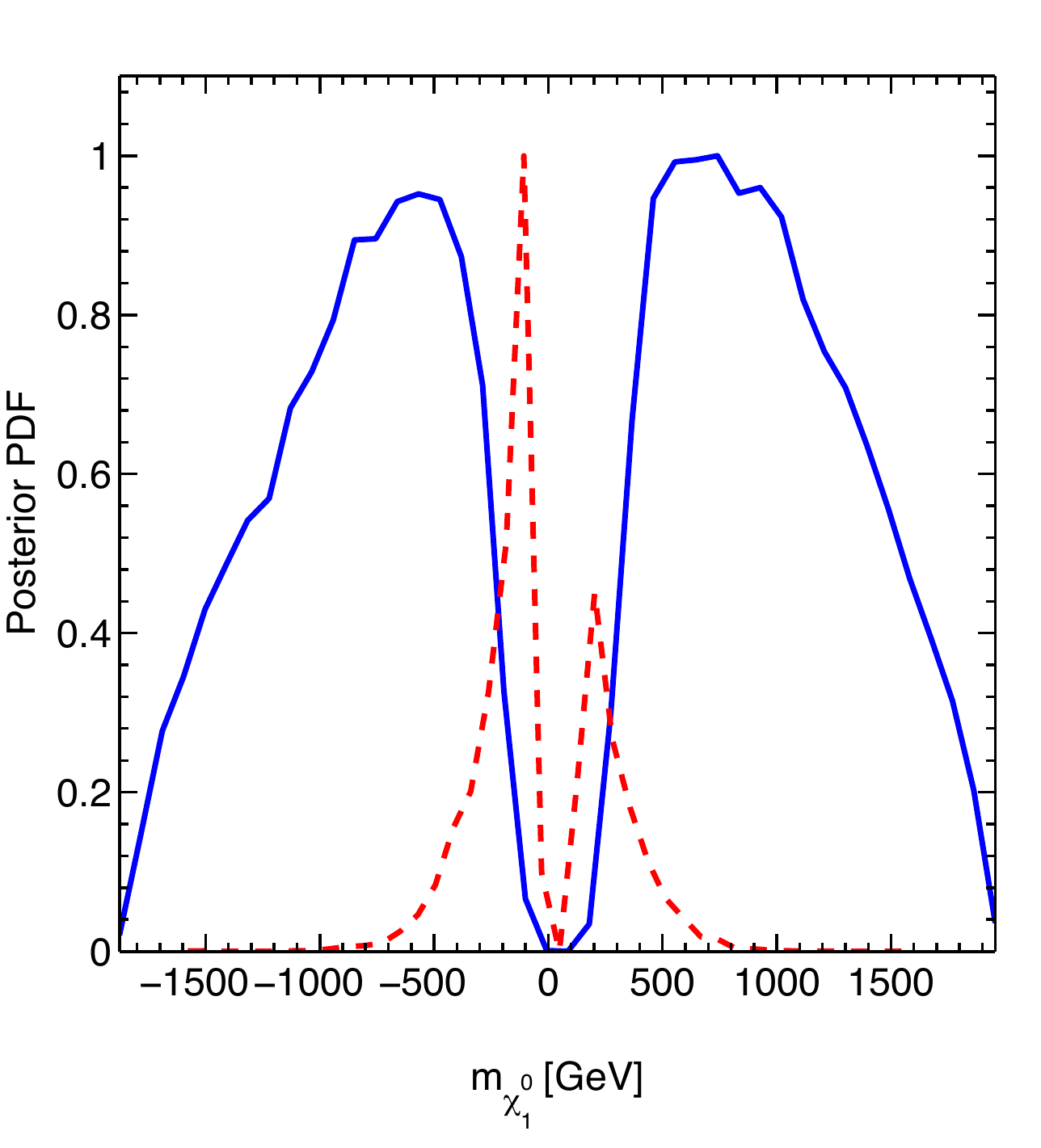}
\end{minipage}
\begin{minipage}[t]{0.32\textwidth}
\includegraphics[width=1.\columnwidth,trim=25mm 15mm 60mm 120mm, clip]{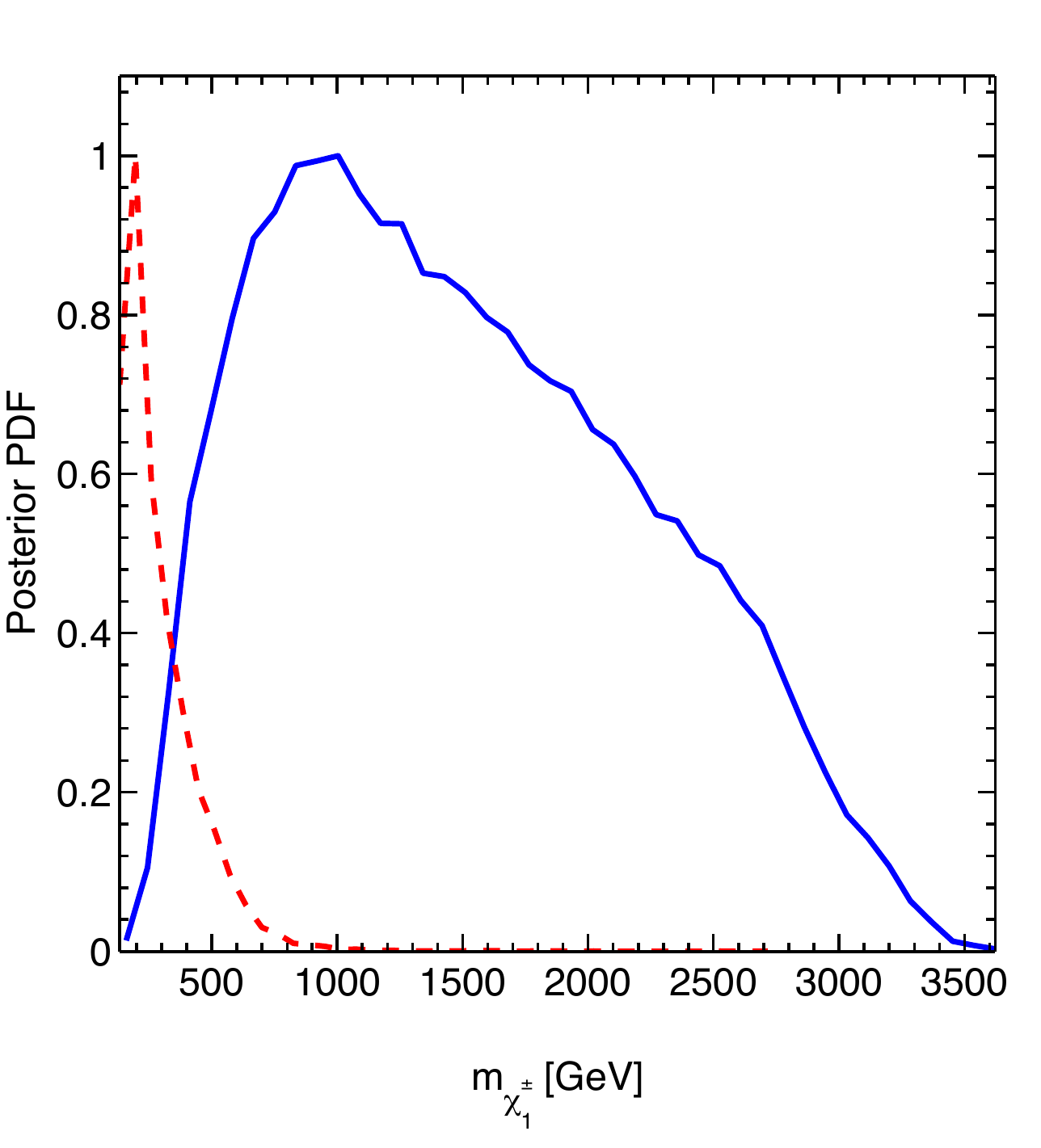}
\end{minipage}
\begin{minipage}[t]{0.32\textwidth}
\centering
\includegraphics[width=1.\columnwidth,trim=25mm 15mm 60mm 120mm, clip]{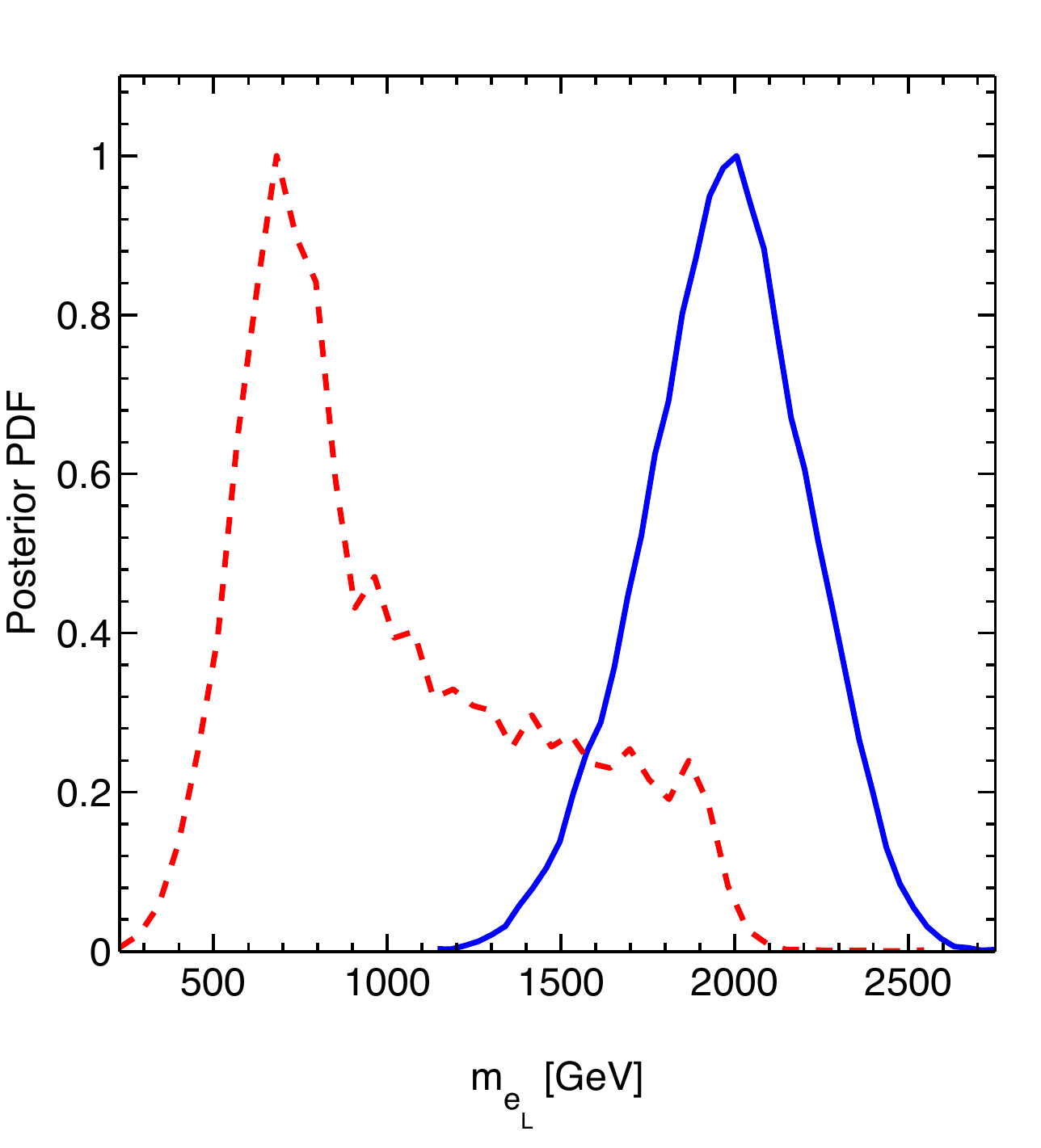}
\end{minipage}
\\
\begin{minipage}[t]{0.32\textwidth}
\centering
\includegraphics[width=1.\columnwidth,trim=25mm 15mm 60mm 120mm, clip]{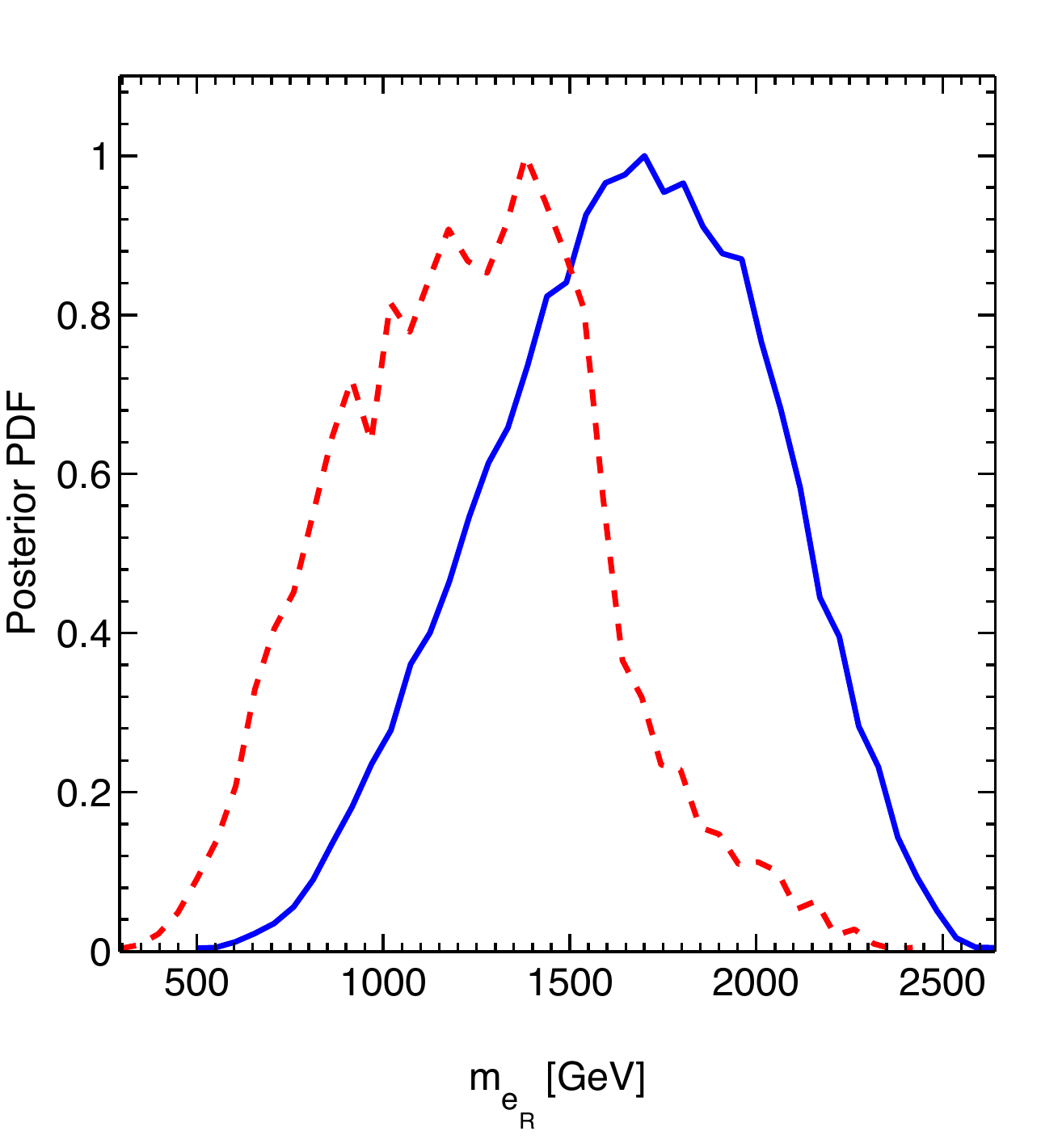}
\end{minipage}
\begin{minipage}[t]{0.32\textwidth}
\includegraphics[width=1.\columnwidth,trim=25mm 15mm 60mm 120mm, clip]{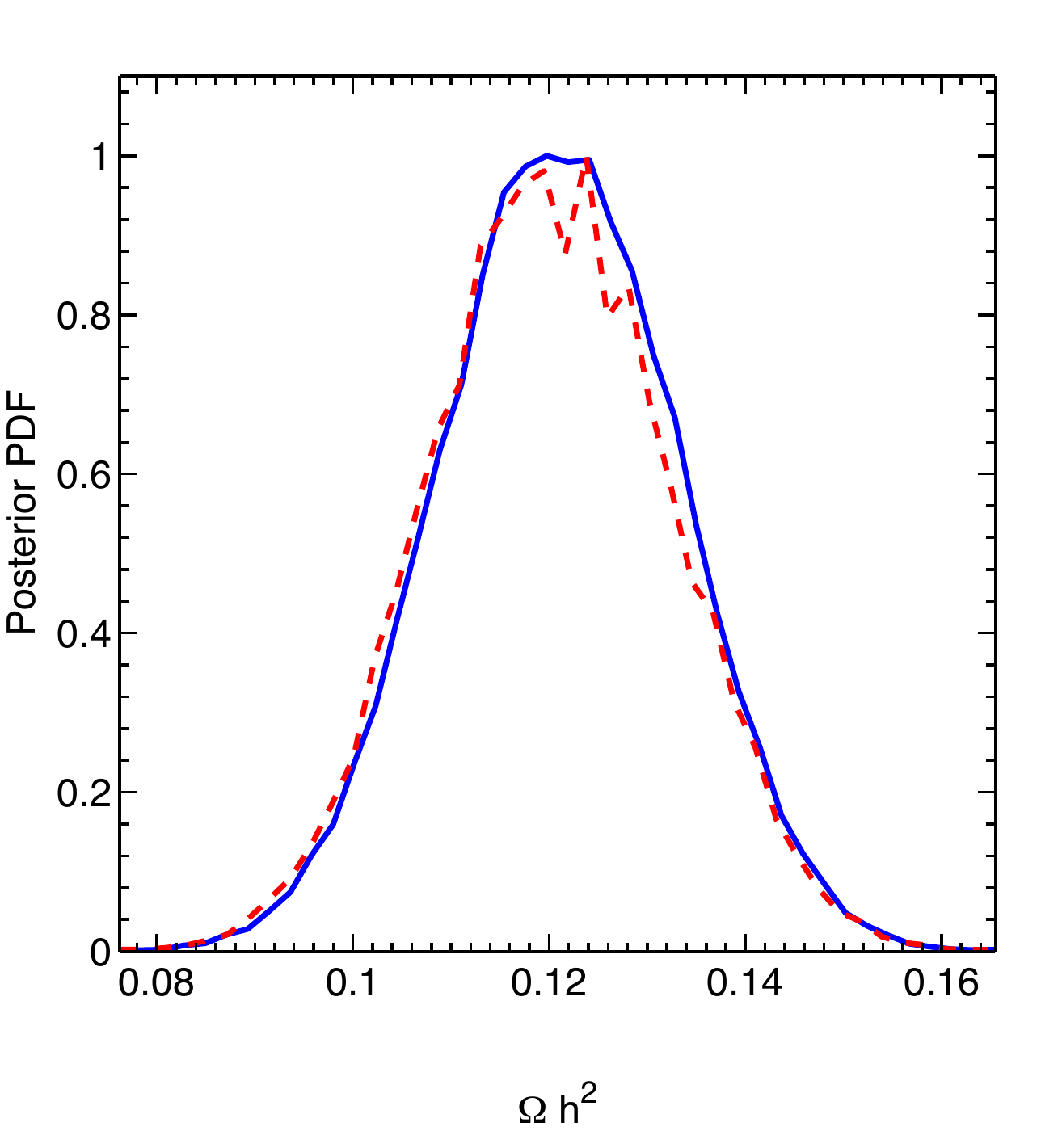}
\end{minipage}
\begin{minipage}[t]{0.32\textwidth}
\centering
\includegraphics[width=1.\columnwidth,trim=25mm 15mm 60mm 120mm, clip]{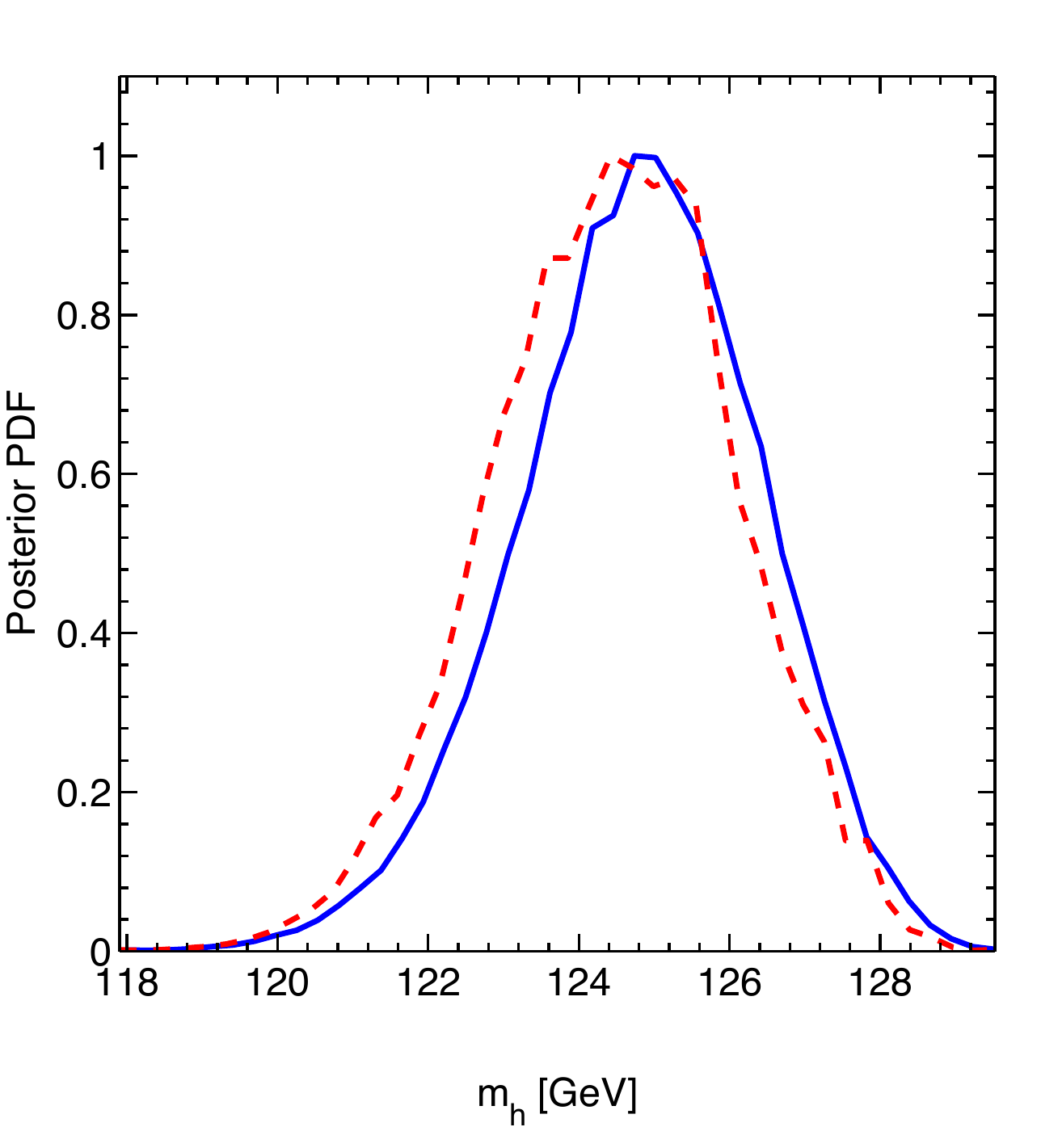}
\end{minipage}
\caption{{\it First panel (top left):} 1D marginalized posterior pdf for $m_{\snu}$. {\it All other panels:} Same as top left for the sneutrino mixing angle and the lightest stau mass (top center, right);  for the lightest neutralino and lightest chargino mass (second raw, left and center); for the mass of the slepton left and of the slepton right (second raw right and bottom left );  for $\Omega_{\rm DM} h^2$ and $m_h$ (bottom center and right). The posterior pdfs are normalized to their maximum,  the blue solid line stands for flat priors and the red dashed line for log priors. \label{fig:1dpost1}}
\end{figure}

\begin{figure}[t]
\begin{minipage}[t]{0.49\textwidth}
\centering
\includegraphics[width=1.\columnwidth,trim=0mm 0mm 0mm 0mm, clip]{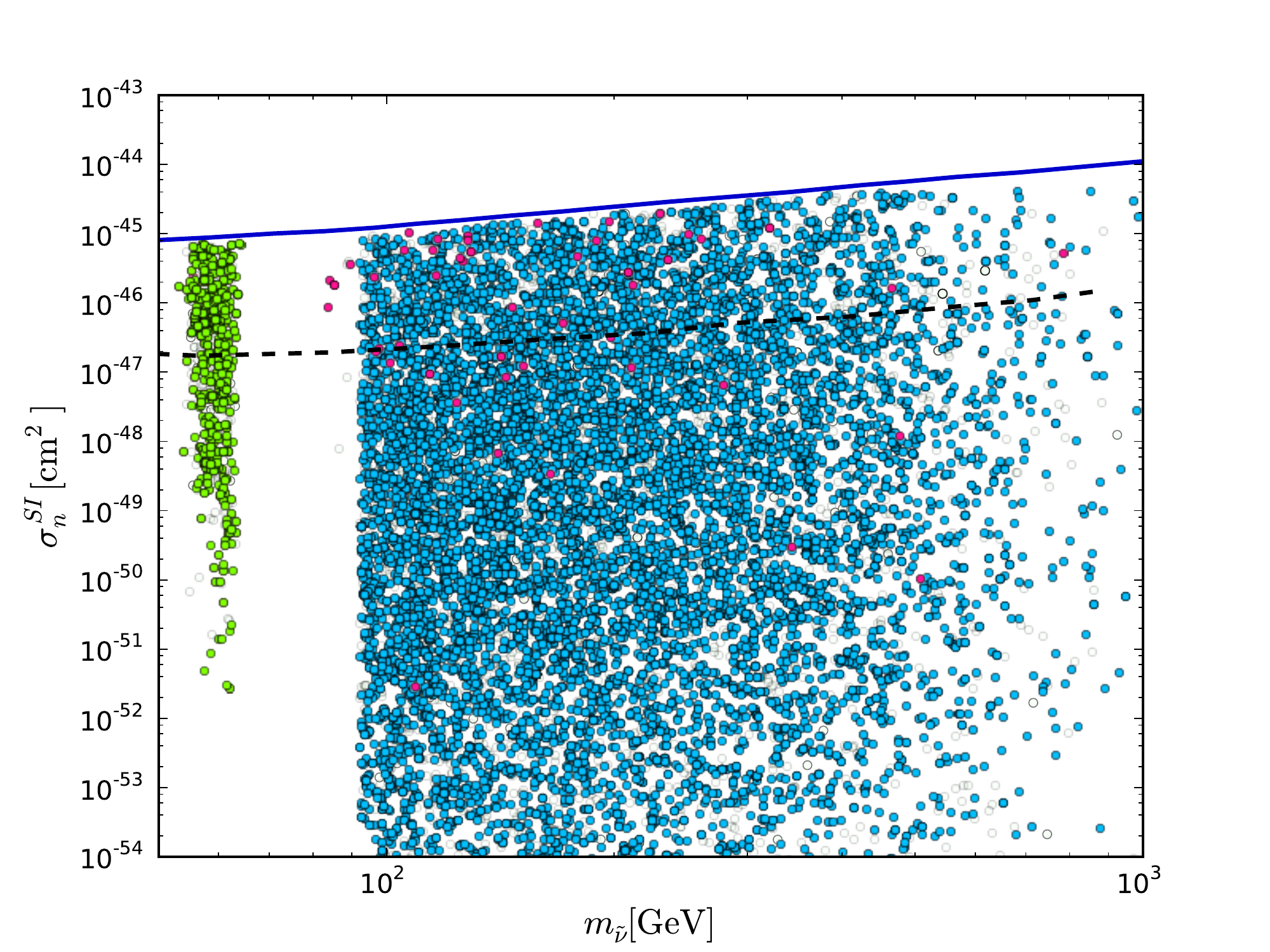}
\end{minipage}
\hspace*{0.2cm}
\begin{minipage}[t]{0.49\textwidth}
\includegraphics[width=1.\columnwidth,trim=0mm 0mm 0mm 0mm, clip]{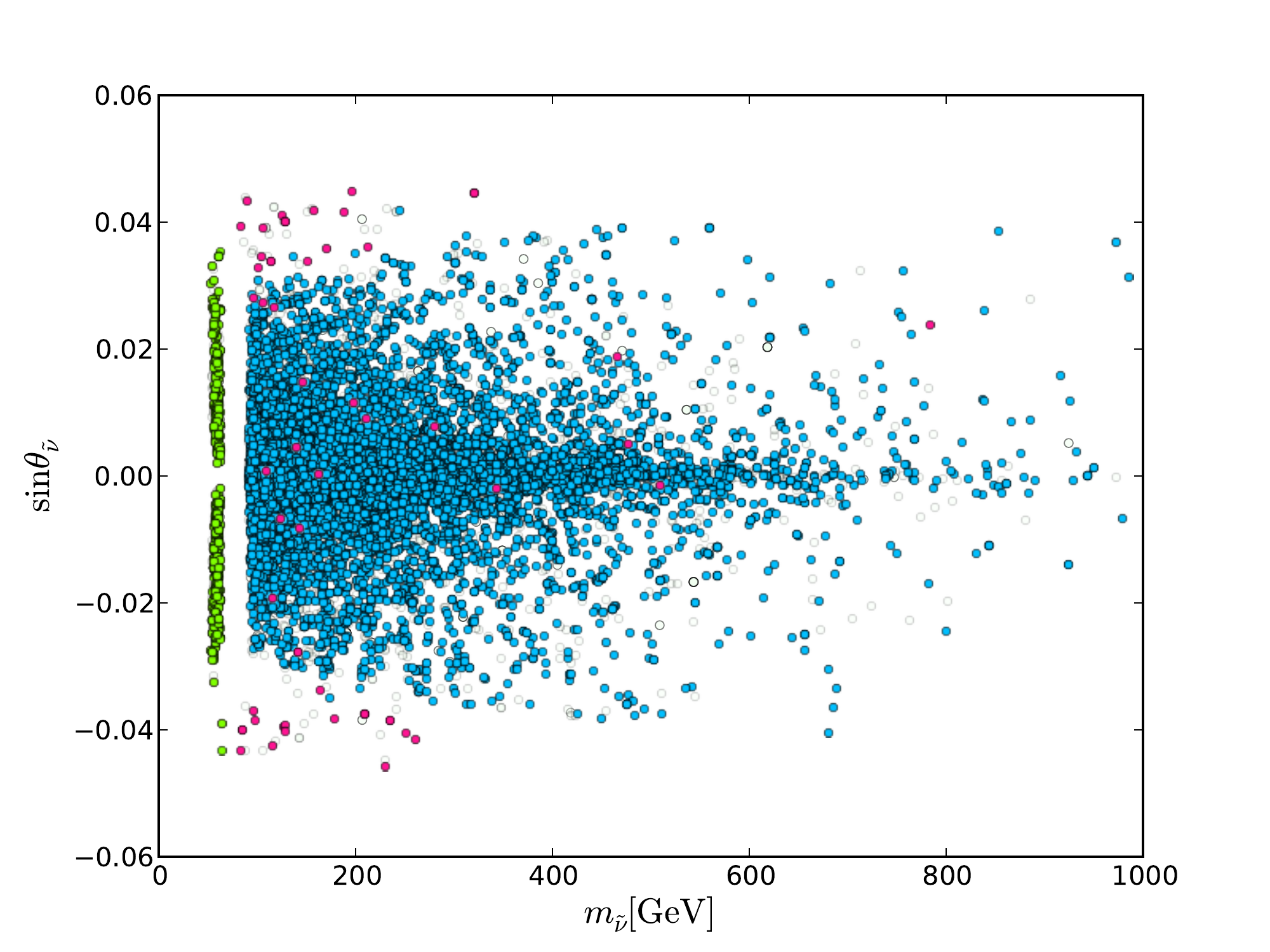}
\end{minipage}
\caption{{\it Left:} Equal weight points from \texttt{MultiNest} chains plotted as a function of the sneutrino mass $m_{\snu}$ and the scattering cross-section $\sigma^{\rm SI}_n$ for a log prior choice. {\it Right:} Same as left in the $\{ \snu-\sin\theta_{\snu} \}$-plane. The color code is as in figure~\ref{fig:goodDM}.\label{fig:figsnulp}}
\end{figure}

As mentioned before, we run \texttt{MultiNest} in the mode which provides a good determination of the posterior pdf. In figure~\ref{fig:1dpost1} we show the 1D posterior pdfs for the quantities relevant for the dark matter phenomenology: first the sneutrino mass itself and its mixing angle. Then the lightest neutralino and lightest chargino as they are relevant for setting the DM relic density; the $\tilde{\tau}_1$ mass pattern is relevant  for both the relic density and for the signatures at LHC. Finally we show the spectrum of the left-handed and right-handed slepton, relevant for LHC signatures. All the 1D posterior pdfs are marginalized over the hidden directions, namely these are integrated out. 
As expected the log prior (red dashed) tends to prefer low values of the masses with respect to the case of flat priors (blue solid). The impact of changing priors is relevant in particular for the gaugino sector: $\chi^0_1$ and $\chi^\pm_1$ tends to be lighter and close in mass. This means that the lightest neutralino is mostly wino and degenerate with the chargino, hence with this choice of priors the most common phenomenology we retrieve is the one denoted by the blue points. Unfortunately this is the case where the model is indistinguishable from the MSSM at LHC. This is confirmed by figure~\ref{fig:figsnulp}, where we see that there is a large portion of the parameter space where the sneutrino is almost sterile. 

Figure~\ref{fig:1dpost1} shows as well the marginalized one dimensional posterior pdf for two relevant observables: the dark matter relic density and the Higgs mass, again for flat and log priors. The change in priors does not affect significantly those observables. These plots confirm as well that all points in figures~\ref{fig:goodDM},~\ref{fig:sleptonMasses} and~\ref{fig:figsnulp} satisfy the constraints from Planck and the Higgs mass.

\bibliographystyle{JHEP}
\bibliography{biblio}

\end{document}